\documentclass{mythesis}
\usepackage[fleqn]{myeq}
\usepackage{amssymb}
\usepackage{graphicx}
\usepackage{calrsfs}
\usepackage{cite}
\usepackage{varioref,xr-hyper}
\usepackage{color}
\usepackage{epsfig}
\usepackage{makeidx}
\usepackage{amssymb,latexsym}

\newcommand{\be}{\begin{equation}}
\newcommand{\ee}{\end{equation}} 
\newcommand{\bcenter}{\begin{center}}
\newcommand{\ecenter}{\end{center}} 
\newcommand{\bfig}{\begin{figure}}
\newcommand{\efig}{\end{figure}}

\title{Phenomenological aspects of dark energy dominated cosmologies} 
\author{Pier Stefano Corasaniti} 
\qualification{Doctor of Philosophy}

\begin{document}

\include{cover}

\maketitle

\pagenumbering{roman}


\begin{declaration}
The work presented in this thesis is the result of a number of 
collaborations with 
my supervisor Ed Copeland and
Bruce Bassett and Carlo Ungarelli at the Institute of Cosmology
of Portsmouth University. Some of the work has already been 
published in \cite{CORAS1},\cite{CORAS2},\cite{CORAS3},\cite{CORAS4}.

\vspace{2cm}

I hereby declare that this thesis describes my own original work,
except where explicitly stated. No part of this work has previously
been submitted, either in the same or different form, to this or any
other University in connection with a higher degree or qualification.

\vspace{3cm}
\noindent{Signed{\tt .....................................}
\hspace{1cm}
Date{\tt .....................}}\\

\noindent{Pier Stefano Corasaniti}

\end{declaration} 
\begin{acknowledgements}

This thesis is the final result of almost three years of hard work
and God only knows how much this has coasted me in terms of personal
affections and sacrifices, but I would do it again.
It has been a really great experience
and has given me the opportunity to realise myself as the theoretical
cosmologist that I always wanted to be.
Living in UK was not easy at the beginning, but now I am perfectly
integrated with this country that I consider my second home.
I have learned not to complain about the weather and the food 
since there are other things that make life worth living. 
At the end of this adventure I have discovered that all the joys and 
sorrows I went through have made me a much better person.
I really have to thank my supervisor Ed Copeland who gave me this chance
without knowing me,
a chance that my home country refused to offer to me. 
I am particularly thankful to him for allowing me to develop
my own ideas about research projects, ideas that we improved together
through open minded discussions. I also thank the Institute of 
Cosmology of Portsmouth for their welcoming hospitality
during these many years. It has been a pleasure to start a fruitful
collaboration with Bruce Bassett and Carlo Ungarelli and if 
this has been possible I have
to thank my dear friend Cristiano Germani with whom I established a strong
brotherhood. I thank Luca Amendola for the hospitality during
my visit at the Astronomical Observatory of Rome and
the Physics Department of Dartmouth College (New Hampshire)
where a large part of this thesis was 
written. In particular it is a pleasure to thank Marcelo Gleiser, Robert
Caldwell and Michael Doran for the welcoming atmosphere I found in Dartmouth,
for the useful and interesting physics discussions
and for bearing the visit of a crazy roman.
I am particularly grateful to Andrew Liddle, Nelson Nunes, Nicola Bartolo,
Martin Kunz and Michael Malquarti for the many cosmological
discussions and all the members of the physics and astronomy group,
in particular Mark Hindmarsh, Andre Lukas, Peter Schroder, 
Beatriz de Carlos and David Bailin,
for the pleasant and friendly environment of the Sussex group.
I am glad to have met and mention an enormous number of friends: 
first of all my friend Gonzalo
Alvarez, also Lys Geherls, Jane Hunter, Gemma Dalton, Electra Lambiri,
 Isaac Neumann, Asher
Weinberg, my flat mates Stephen Morris and particularly
Neil McNair for booking concert tickets so many times ,
Matteo and Chiara Santin for their kind and nice friendship, Fernando Santoro,
James Gray, Malcom Fairbairn, Seung-Joo Lee, Maria Angulo, Liam 
O'Connell for is effort to explain to me the rules of
cricket during the world cup
football matches, James Fisher, Jon Roberts,
Neil Bevis and in particular Yaiza Schmohe. I will never forget
the amazing Sweet Charity Band and its conductor Chloe Nicholson who
gave me the opportunity to play my clarinet on the stage again.
They made me remember who I truly am deep in my soul, simply a clarinet
player. A particularly express my thanks to my parents for their moral support
and for paying the flight tickets to Italy.
This thesis is dedicated to the memory of Franco Occhionero, he was
an extraordinary scientist and one of the most lovely people I have ever met.
Always generous in helping and guiding his students he paid attention 
to teaching and researching as much as to the popularization of science.
I will never forget his public lectures, his kindness and the
strength and genuine humility of his personality.
My approach to Cosmology is the result
of the daily interaction with such a great man and
I am proud of being one of his many pupils. 
Finally I would like to conclude with a little bit of the sarcasm that
plays a fundamental role in my life. It may be that all this work is entirely
wrong, I do not think it is since I am responsible for it, but in the remote
eventuality I would like to remind the reader of Francis Bacon's words:
{\em `The truth arises more from the errors rather than from the confusion'}.

\end{acknowledgements}

\begin{abstract}
Cosmological observations suggest that the Universe
is undergoing an accelerated phase of expansion driven by
an unknown form of matter called dark energy.
In the minimal standard cosmological model that best fits the
observational data the dark energy is provided by a
cosmological constant term. However there is currently no convincing 
theoretical
explanation for the origin and the nature of such an exotic component.
In an attempt to justify the existence of the dark energy within
the framework of particle physics theories, several scenarios have
been considered. In this thesis we present
and discuss the phenomenological aspects of some of these dark energy models.
We start by reviewing the cosmological measurements that give direct
and indirect evidence for the dark energy. Then we focus on a class of
theoretical models where the role of the dark energy is played
by a minimally coupled scalar field called quintessence.
For the sake of simplicity we place special emphasis on two of these models,
the Inverse Power Law times an Exponential potential and the Two
Exponential potential. We then
consider the effect of scalar field fluctuations on the structure
formation process in these two models.
By making use of the cosmological distance measurements such as the 
supernova luminosity distance and the position of the acoustic peaks in
the CMB power spectrum, we constrain the shape of a general parameterized
quintessence potential.
We find that by the present time the scalar field is evolving in a very flat
region or close to a minimum of its potential. In such a situation
we are still unable to distinguish between a dynamical model of dark energy
and the cosmological constant scenario. 
Going beyond constraining specific classes of models,
we develop a model independent approach that allows us to determine the 
physical properties of the dark energy without the need to refer to 
a particular model. We introduce a parameterization of the dark energy
equation of state that can account for most of the proposed quintessence
models and for more general cases as well.
Then we study the imprint that dark energy leaves on the CMB anisotropy
power spectrum. We find that dynamical models of dark energy produce
a distinctive signature by means of the integrated Sachs-Wolfe effect.
However only models characterized by a rapid transition of
their equation of state can  most likely be distinguished 
from the cosmological constant case. By using a formalism to model
localized non-Gaussian CMB anisotropies, we compute analytical
formulae for the spectrum and the bispectrum. The use of these formulae
in specific cases such as the SZ signature of clusters of galaxies provide
an alternative cosmological test.
\end{abstract}

\begin{flushright}

{\small {\sl To the memory of \\ Prof. Franco Occhionero }}

\end{flushright}

\tableofcontents

\starttext

\pagenumbering{arabic}

\chapter*{Introduction} \label{intro}

The set of astrophysical observations collected in the past 
decades and the theoretical and experimental developments in high
energy physics have provided the natural framework that defines
Cosmology as a scientific discipline.
The identification of the `Hot Big-Bang'
scenario as a paradigm has been of crucial importance for the 
beginning of Cosmology as a modern science. In fact it has allowed us
to address a number of questions about the nature and the evolution of
the Universe that otherwise would have been the subject
of investigation of philosophers and theologians.
Since this paradigm has been accepted by the majority of the scientific
community, more specific and detailed studies have been undertaken in order
to extend the validity of the paradigm to a wider class of phenomena
such as the formation of the structures we observe in the Universe.
As result of this intense activity, 
that the philosopher of science T.S.~Kuhn would define as
{\em normal science investigation} \cite{KUHN}, 
the initial paradigm of Cosmology has been extended in order to include
the inflationary mechanism and dark matter, two necessary
ingredients to explain a number of issues that arise within the `Hot Big-Bang'
scenario. This extended paradigm is extremely successful
and recent measurements in observational cosmology have widely confirmed
its prediction.
It is very remarkable that long before the recent developments
of the cosmological science, T.S.~Kuhn identified 
the basic steps that a scientific discipline follows in its evolution,
steps that applies to modern Cosmology too. In particular he has pointed
out that there are phenomena
which evading an explanation of the paradigm are the subject
of `extraordinary' investigations, in opposition to the `ordinary'
normal science activity. Most of the time these studies lead to 
a crisis of the underlying paradigm and trigger what he has called
a `scientific revolution'. In this light the
discovery that the Universe is dominated by a dark energy component
that accounts for $70\%$ of the total matter budget  
belongs to this class of phenomena. 
This is the subject of this thesis. In Chapter~\ref{chap1} we will describe
the observational evidence of dark energy and in Chapter~\ref{chap2}
we will review some of proposed explanations. 
From Kuhn's point of view these explanations would represent the attempt
to force Nature to fit within the
`Hot Big-Bang' paradigm. In fact these models fail to succeed since
they manifest inconsistencies with the expectations of 
particle physics theories that are included in the standard 
cosmological paradigm. The solution to this difficulty most probably
will need a new paradigm. This necessity will be more urgent 
if the observational data will indicate a time dependence of the 
dark energy properties.
In this perspective the aim of this thesis is to investigate some of
the phenomenological aspects of minimally coupled quintessence
scalar field scenarios. 
In Chapter~\ref{chap3} we describe the evolution of fluctuations
in the quintessence field. In Chapter~\ref{chap4} we discuss the 
constraints on a general class of quintessence potentials obtained
from the analysis of the position of the acoustic peaks in the Cosmic
Microwave Background (CMB) anisotropy power spectrum and the Sn Ia data.
In Chapter~\ref{chap5} we describe some of the
methods used to constrain the dark energy. We then introduce
a model independent approach that allows us to study the full impact
a general dark energy fluid has had in Cosmology.
In fact this fluid description, based on a very general parameterization
of the dark energy equation of state, allows us to infer the dark
energy properties from cosmological observations, instead of constraining
specific classes of dark energy models.
In Chapter~\ref{chap6} we study the effects dark energy produces in the 
CMB anisotropy power spectrum and show that clustering of dark 
energy leaves a distinguishable signature only for a specific class
of models. In Chapter~\ref{chap7} we present
the results of preliminary work that aims to develop alternative cosmological
tests using the non-gaussianity produced by localized sources of CMB 
anisotropies. 
We hope that the work reviewed in this thesis will provide the basis for
those `extraordinary' investigations that will help us in developing
the new paradigm that modern Cosmology needs.

\chapter{The dark side of the Universe} 
\label{chap1}

The recent results obtained in different areas of observational cosmology provide an
astonishing picture about the present matter content of the Universe.  The
accurate measurements of the Cosmic Microwave Background radiation (CMB) give
strong evidence that the curvature of the space-time is nearly flat.  On the
other hand the analysis of large scale structure surveys shows that the amount
of clustered matter in baryonic and non-baryonic form can account only for
thirty per cent of the critical energy density of the Universe.  In order to be
consistent these two independent analyses require the existence of
an exotic form of matter, that we call {\em dark energy}. More direct evidence
is provided by the Hubble diagram of type Ia
supernova (Sn Ia) at high redshifts.  It suggests that the Universe is
undergoing an accelerated expansion sourced by this dark energy
component that is characterized by a negative value of its equation of state.
A lot of criticism has been levelled to these measurements.  In fact the physics of Sn Ia
is still a matter of debate and consequently their use as standard candles has not
yet convinced the whole astronomical community.  In spite of such an important
issue it is worth underlining that the combination of CMB and supernova data
constrains the amount of clustered matter in the Universe to a value that is
consistent with the large scale structure observations.  As we shall review in
this chapter, indirect evidence for a dominant dark energy contribution comes
mainly from the combination of CMB results and the constraints on the density of
baryons and cold dark matter.  
The simplest explanation for the dark energy would be the presence
for all time of a cosmological constant term $\Lambda$ 
in Einstein's equations of General Relativity.
However evidence for a non vanishing 
value of $\Lambda$ raises a fundamental problem for theoretical physics, (for
a review of the subject \cite{WEINB,CARR1,SAHNI1,CARR2,STRAUM,DOLGOV}). In
fact, as we shall see later, it is rather difficult to explain the small observed
value of $\Lambda$ from the particle physics point of view.
In this chapter we will briefly review some historical 
developments of the dark energy problem. We will introduce the 
standard cosmological model and the equations describing the expansion
of the Universe. Then we will discuss the build up of observational evidence
for dark energy.

\section{A historical introduction}
The cosmological constant was initially introduced by Einstein in the equations
of General Relativity (GR) as a term that could provide static cosmological solutions
\cite{EINSTEIN}.
Motivated by the observed low velocities of the stars, he assumed that
the large scale structure of the Universe is static. We should remind ourselves that
the notion of the existence of other galaxies had been established only a few years
later. Besides, Einstein believed that the GR equations had to be compatible with the 
Mach's principle, which in a few words states that the metric of the space-time is uniquely 
fixed by the energy momentum tensor describing the matter in the Universe.
For this reason he thought the Universe had to be closed. These two
assumptions were, however, not compatible with the original form of the GR equations.
In fact a matter dominated closed Universe is not static, therefore he
needed to introduce a term leading to a repulsive force that could counterbalance
the gravity. In such a case he found a
static and closed solution that preserved Mach's principle. But
in the same year, 1917, de Sitter discovered an apparent static solution
that incorporated the cosmological constant but contained no matter \cite{DESITTER}.
As pointed out by Weyl, it was an {\em anti-Machian} model with an
interesting feature: test bodies are not at rest and an
emitting source would manifest a linear redshift distance relation.
Such an argument was used by Eddington to interpret Slipher's observations
of the redshift of spiral nebula (galaxies).
Subsequently expanding matter dominated cosmological solutions without
a cosmological constant were found by Friedman \cite{FRIED1,FRIED2} and Hubble's
discovery of the linear redshift distance relation \cite{HUBBLE0} 
made these models the standard cosmological framework. 
The cosmological constant was then abandoned.
For some time a non vanishing $\Lambda$ was proposed to solve
an `{\em age problem}'. Eddington pointed out the Hubble
time scale obtained by using the measured Hubble constant was only 2 billion
years in contrast with the estimated age of the Earth, stars and stellar systems.
In the 1950s the revised values of the Hubble 
parameter and the improved constraints on the age of stellar objects resolved
the controversy and the cosmological constant became unnecessary.
However in 1967 it was again invoked to explain a peak in the number count
of quasars at redshift $z=2$. It was argued that the quasars were born during
a hesitation era \citation{PETROSIAN,SHKLOVSKY,KARDASHEV}, at the transition between the matter and the 
$\Lambda$ dominated era. More observational data confirmed the existence
of this peak and allowed for a 
correct interpretation as simply an evolutionary effect of active galactic nuclei,
with no necessity for $\Lambda$.
We shall discuss the theoretical implications of the cosmological constant
in Chapter 2, here we would like to stress that in the past few decades the $\Lambda$
term has played the role of a fitting parameter necessary to reconcile theory and
observations which has been discarded every time systematic effects were considered. 
It is therefore natural to 
ask the question if today we are facing a similar situation.
In what follows we will try to show that this turns not
to be the case and the dark energy is indeed most likely to be present in our Universe.

\section{Standard cosmology}
The Einstein field equations are:

\begin{equation}
R_{\mu\nu}-\frac{1}{2}g_{\mu\nu}R=g_{\mu\nu}\Lambda+\frac{8\pi G}{3}T_{\mu\nu},
\label{ein}
\end{equation}
where $R_{\mu\nu}$ is the Ricci tensor, $R$ is the Ricci scalar, $\Lambda$
is the cosmological constant term and $T_{\mu\nu}$
is the matter energy momentum tensor which determines the dynamics of the 
Universe. When different
non interacting sources are present the energy momentum tensor is
the sum of the energy momentum tensor of each of the sources. 
Assuming an isotropic and homogeneous space-time the large scale
geometry can be described by the Friedman-Robertoson-Walker
metric:

\begin{equation}
ds^2=dt^2-a^2(t)\left(\frac{dr^2}{1-kr^2}+r^2d\theta^2+r^2sin^2\theta d\phi^2 \right),
\end{equation}
where $a(t)$ is a function of time (called the {\em scale factor}) and $k=0,\pm1$
sets a flat, open (-1) or close (+1) geometry. Spatial homogeneity and isotropy implies that
the energy momentum tensor of each component is diagonal:

\begin{equation}
T_{\mu \nu}^i=diag(\rho_i(t),p_i(t),p_i(t),p_i(t)),
\end{equation}
where $\rho_{i}(t)$ is the energy density and $p_i(t)$ is the pressure of the {\em i}-th matter component 
(radiation, baryons, cold dark matter, etc..). 
In the FRW metric the Einstein equations (\ref{ein}) with a mixture of different matter components
are the Friedman equations:

\begin{equation}
H^2=\left(\frac{\dot{a}}{a}\right)^2=\frac{8\pi G}{3}\sum_i\rho_{i}+\frac{\Lambda}{3}-\frac{k}{a^2},
\label{friedmann1}
\end{equation}

\begin{equation}
\frac{\ddot{a}}{a}=-\frac{4\pi G}{3}\sum_i(\rho_{i}+3p_i)+\frac{\Lambda}{3}.
\label{friedmann2}
\end{equation}
We define the density parameters  $\Omega_i=\rho_i/\rho_c$,
$\Omega_{\Lambda}=\Lambda/3\rho_c$ and $\Omega_k=-k/H^2 a^2$ where
$\rho_c=3 H^2/8 \pi G$ is the critical energy density. Then the Friedman equation
Eq.~(\ref{friedmann1}) can be rewritten as:

\begin{equation}
1-\Omega_k=\sum_i \Omega_i=\Omega_{tot},
\end{equation}
showing that the spatial curvature is fixed by the total matter content.
Since the different components do not interact
with each other, their energy momentum tensor must satisfy the energy conservation
equation $T_{\mu;\nu}^{\nu}=0$. Hence in addition to the Friedman equations
the evolution of the energy density of each matter component is given by:

\begin{equation}
\dot{\rho_i}=-3H(\rho_i+p_i).
\label{encons}
\end{equation}
It is worth remarking that, as has been stressed 
by T. Padmanabhan \cite{PADMA}, {\em `absolutely no progress in
cosmology can be made until a relationship between $\rho_i$ and $p_i$ is
provided in the form of the functions $w_i(a)$'}. In fact once these relations
are known, we can solve the dynamical equations and make predictions about
the evolution of the Universe, that can be tested by cosmological observations.
If the matter components consist of normal laboratory matter, then the knowledge
of how the matter equation of state $w$ evolves at different energy
scales is provided by particle physics. At present the behaviour of
matter has been tested up to about $100$ GeV, in this domain the
relation between energy density and pressure can be taken to be that of
an ideal fluid, $p_i=w_i\rho_i$, with $w=0$ for non relativistic matter
and $w=1/3$ for relativistic matter and radiation.
However if a cosmological model based on conventional matter components
fails to account for cosmological observations, we could interpret this fact as
a failure of the cosmological model or as a signal for the existence of a
source not seen in laboratories. For instance the cosmological constant
term behaves as a perfect fluid with negative pressure. This can be seen
rewriting the $\Lambda$ term in 
Eq.~(\ref{friedmann1}) and Eq.~(\ref{friedmann2}) as an energy density and
a pressure term. Then one finds
$p_{\Lambda}=-\rho_{\Lambda}=-\Lambda/(8 \pi G)$. 
The effect of such a component on
the expansion of the Universe can be seen
from Eq.~(\ref{friedmann2}), and the value of deceleration parameter today is
\begin{equation}
q_0\equiv-H_0^{-2}\left(\frac{\ddot{a}}{a}\right)_0=\frac{\Omega_m}{2}-\Omega_{\Lambda},
\end{equation}
and we have neglected the radiation. For
$\Omega_{\Lambda}>\Omega_m/2$ the expansion of the Universe is accelerated
since $q_0<0$. Hence in a Universe dominated by the cosmological constant the
expansion is eternally accelerating. 
In summary the dynamics of our Universe is observationally determined
by two geometrical quantities,
the Hubble parameter $H_0$, which provides us with a measure of the observable
size of the Universe and its age, and the deceleration parameter $q_0$ which probes
the equation of state of matter and the cosmological density parameter.

\section{Observational evidence}
Different cosmological tests can be used to constrain the geometry and
the matter content of the Universe. We shall
briefly review the latest limits on $\Omega_m$ and $\Omega_{\Lambda}$ obtained by
recent experiments in cosmology.

\subsection{CMB anisotropies}
During the last few years an avalanche of balloon and ground experiments, together
with the most recent WMAP satellite observatory have
measured the small angular temperature fluctuations
of the Cosmic Microwave Background Radiation. 
Such measurements have detected a series of acoustic peaks in the
anisotropy power spectrum and confirmed early predictions about the
evolution of pressure waves in the primordial photon-baryon plasma \cite{SZ,PEBYU}. 
The specific features of such peaks are sensitive to the value of the cosmological
parameters, in particular to $\Omega_{tot}$, $\Omega_{b}$ and the scalar spectral index $n$.
The sensitivity to the curvature of the Universe however does not allow us to
constrain independently $\Omega_m$ and $\Omega_{\Lambda}$, that are consequently
degenerates.
The earlier analysis of the Boomerang experiment \cite{DEB1,LANGE,NETTE}
found $\Omega_k \sim 0$ and the latest data released constrain the total energy
density to be $\Omega_{tot}=1.04\pm_{0.04}^{0.06}$ \cite{RUHL}. Such a result is consistent with 
the ones found by other CMB experiments. For instance, the data from MAXIMA-1, another balloon experiment,
when combined with the COBE-DMR data suggest $\Omega_{tot}=1.00\pm_{0.30}^{0.15}$ \cite{BALBI1}.
Similarly the two ground experiments, DASI and CBI provide
$\Omega_{tot}=1.04\pm{0.06}$ \cite{PRYKE} and $\Omega_{tot}=0.99\pm0.12$ \cite{SIEV} respectively.
Recently three more groups, ARCHEOPS 
\cite{BENOIT}, VSA \cite{GRAINGE} and ACBAR \cite{GOLD} have released their data
finding similar results.
The constraints on the baryon density are in good agreement with the prediction
of the Big-Bang Nucleosynthesis (BBN) and the scalar spectral index is found to
be of order unity, as predicted by generic inflationary paradigms.
However CMB alone poorly determines $\Omega_{\Lambda}$ and
a vanishing cosmological constant cannot be excluded at $2\sigma$.
Nonetheless due to the strong constraint on the curvature of the Universe, it is
reasonable to
restrict the data analysis 
to the flat cosmological models ($\Omega_{tot}=1$). In this case,
 assuming the so called `HST prior' on the value of the Hubble constant, $h=0.71\pm0.076$ \cite{FREE},
then all the CMB data constrain the cosmological constant density parameter to be
$\Omega_{\Lambda}=0.69\pm_{0.06}^{0.03}$ \cite{SIEV}.
The WMAP satellite provided CMB data with such an high level of accuracy that is worth mentioning
a part. The experiment has measured CMB anisotropies
in different frequency bands,
allowing for an efficient removal of the foreground emissions. The measurements
mapped the full sky in the unpolarized and polarized components providing an accurate
determination of the temperature power spectrum (TT) and
the temperature-polarization cross-correlation spectrum (TE) \cite{Bennet}.
The position of the first peak in
the TT spectrum constrain the curvature to be $\Omega_k=0.030\pm^{0.026}_{0.025}$ \cite{PAGE}.
The combination of WMAP data with ACBAR and CBI, 2dF measurements and Lyman $\alpha$
forest data find the best fit cosmological parameters: $h=0.71\pm^{0.04}_{0.03}$, the baryon density 
$\Omega_b h^2=0.0224\pm0.0009$, the dark matter density $\Omega_m h^2=0.0135\pm^{0.008}_{0.009}$,
the optical depth $\tau=0.17\pm0.04$, the scalar spectral index $n=0.93\pm0.03$ and the
amplitude of the fluctuations $\sigma_8=0.84\pm0.04$ \cite{Spergel}. The value of $\tau$ comes from an
excess of power on the large angular scales of the TE spectrum \cite{kogut}. This signal cannot be
explained by systematic effects or foreground emissions and has a natural interpretation
as the signature of early reionization, most probably occurred at redshift $z\approx 20$. This
conflicts with the measurements of the Gunn-Peterson absorption trough in spectra, which
indicate the presence of neutral hydrogen at redshift $z\approx 6$ \cite{Fan}. Therefore we have evidence for
a complex ionization history of the Universe, which most probably underwent two reionization
phases, an early and a late one. Of particular interest is the running of the scalar spectral index
that provides a better fit to the data when WMAP is combined with small angular scale measurements
such as ACBAR, CBI, 2dF galaxy survey and Lyman $\alpha$. Another interesting finding of the WMAP TT spectrum is
the lack of power at low multipole. In particular the quadrupole and the octupole are suppressed
compared to the expectation of the best fit $\Lambda$CDM model. It has been claimed that such
suppression could be the signature of new physics \cite{Contaldi,Efcur}.

\subsection{Clustering of matter}
The cosmological structures we observe today have been formed by the gravitational amplification of
small density perturbations. The amount of such inhomogeneities at
different cosmological scales is measured by the matter power spectrum. This is estimated from
the statistical analysis of a large sample of galaxies and provides a measurement of the
amount of clustered
matter in the Universe. Recently two large galaxy surveys, the 2dF Galaxy Redshift Survey \cite{PERCI}
and the Sloan Digital Sky Survey \cite{DOD}, have probed intermediate scales ($10-100$ Mpc). The fit
to the power spectrum data of the 2dF yields $\Omega_m h=0.20\pm0.03$ and the baryon fraction 
$\Omega_b/\Omega_m=0.15\pm0.07$ \cite{PERCI}. Such a low value of $\Omega_m$ gives
indirect evidence for a large non vanishing cosmological constant contribution when
this LSS data is combined with the CMB. A joint analysis
 of the CMB and 2dF data indicates 
$0.65\lesssim\Omega_{\Lambda}\lesssim 0.85$ at $2\sigma$ \cite{EFS1}. 
An independent estimate of $\Omega_m$ is provided by the
peculiar velocities of galaxies. In fact mass density fluctuations cause galaxy
motion with respect to the Hubble flow. Such a motion reflect the matter
distribution and therefore is sensitive to $\Omega_m$. The
analysis of the Mark III and SFI catalogs constrain $\Omega_m=0.3\pm0.06$ \cite{SILBERMAN}. 
Low values of $\Omega_m$ are also indicated by the studies of cluster of galaxies, where it is
assumed the amount of matter in rich clusters provides a fair sample of the matter content
of the Universe.
Recent surveys have precisely determined the local X-ray luminosity 
function. Using the observed mass-luminosity relation, the cluster mass function 
has been compared with the prediction from numerical simulations. This analysis
constrains $\Omega_m<0.36$ at $1\sigma$ \cite{ALLEN1} (see also \cite{VIA}).
This result is in agreement 
with the limits found by an alternative study from which,
$0.1<\Omega_m<0.5$ at $2\sigma$ \cite{VIKH}. Another
way of estimating the amount of dark matter is to measure the baryon fraction $f_b$ from
X-ray cluster observations. In fact the ratio of the baryonic to total mass in cluster should
closely match the ratio $\Omega_b/\Omega_m$. Therefore a measurement of $f_b$ combined with
accurate determination of $\Omega_b$ from BBN calculation can be used to determine $\Omega_m$.
Using such a method it was found $\Omega_m\thickapprox 0.32$ for $h\sim0.7$ \cite{MOHR}. This
is in agreement also with the value obtained by a study of the redshift dependence of the baryon fraction, that indicates $\Omega_m=0.3\pm_{0.03}^{0.04}$ \cite{ALLEN2}.
Similarly a different analysis based on gravitational lens statistics provides
 $\Omega_m=0.31\pm_{0.24}^{0.39}$ \cite{CHAE}.

\subsection{Age of the Universe}
\begin{figure}[t]
\centerline{
\includegraphics[width=.75\textwidth]{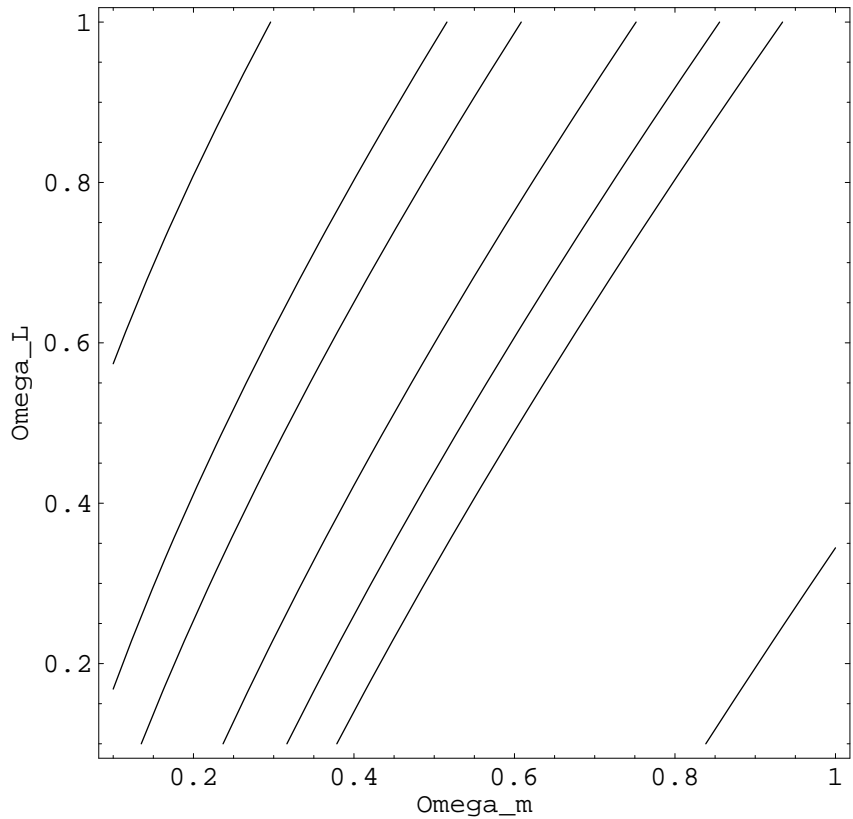}
}
\caption{Lines of constant $H_0 t_0$ in the $\Omega_m-\Omega_{\Lambda}$
plane. From top left to bottom right $H_0 t_0=(1.08,0.94,0.9,0.85,0.82,0.8,0.67)$.}		
\label{f:fig1}			
\end{figure}

The Friedman equation Eq.~(\ref{friedmann1}) can be integrated to obtain the
age of a given cosmological model:
\begin{equation}
H_0 t_0=\int_{0}^{1}\frac{da}{a\sqrt{(1-\Omega_m-\Omega_{\Lambda})/a^2+\Omega_m/a^3+\Omega_{\Lambda}}}.
\end{equation}
The numerical solutions
are shown in figure~\ref{f:fig1}. The solid lines correspond to increasing
value of $H_0 t_0$ in the $\Omega_m-\Omega_{\Lambda}$ plane. It is easy to see
that for fixed values of $\Omega_{m}$ the age of the Universe increases for
larger values of $\Omega_{\Lambda}$. Since matter
dominated cosmological models are younger than globular clusters,
in the past few decades the possibility
of an `age problem' has been a matter of debate.
The presence of a cosmological constant can alleviate such a problem. However a key
role is played by the Hubble parameter, in fact low values of $H_0$ increase
$t_0$. The age of globular clusters is estimated to be
about $11.5\pm1.5$ Gyr \cite{CHAB}, therefore a purely matter dominated
Universe ($\Omega_m=1$ and $\Omega_{\Lambda}=0$) cannot be excluded if
$h<0.54$. However such a low value of $h$ seems to be inconsistent with the
accepted value of $h\approx 0.71\pm0.07$ \cite{FREE}.
It is worth mentioning that the determination of the age of high redshift objects
can be used to 
constrain $\Omega_{\Lambda}$ by studying the 
redshift evolution
of the age of the Universe \cite{ALCA1}.

\subsection{Supernovae Ia and luminosity distance measurements}
Supernovae type Ia are violent stellar explosions, their luminosity
at the peak becomes comparable with the luminosity of the whole hosting galaxy.
For such a reason they are visible to cosmic distances. 
These supernovae appear to be standard candles
and therefore are used to measure cosmological distances by means of
the magnitude-redshift relation:
\begin{equation}
m-M=5\log_{10}\frac{d_L}{Mpc}+25,
\end{equation}
where $m$ is the apparent magnitude, $M$ is absolute magnitude and $d_L$ is the luminosity distance
which depends upon the geometry of the space and its matter content.
In the standard scenario a white dwarf accretes mass from a companion star. 
Once the Chandrasekhar mass limit is reached, the burning
of carbon is ignited in the interior of the white dwarf. This process propagates to the exterior layers
leading to a complete destruction of the star. The physics of these objects is not
completely understood yet. It requires the use of numerical simulations from which it appears that
the thermonuclear combustion is highly turbulent. Such theoretical uncertainties prevent us from
having reliable predictions of possible evolutionary effects. It is also matter of debate as to whether
the history of the supernova progenitors can have important effects in the final explosion (see 
\cite{HILL} for a general review).
For these reasons it is a rather unreliable assumption that supernova Ia are perfect standard candles.
However the observations show the existence of an empirical
relation between the absolute peak luminosity
and the light curve shapes. There are also correlations with the spectral properties. Using
such relations it is possible to reduce the dispersion on magnitude of each supernova to within
$0.17$ magnitudes allowing them to be used for cosmological distance measurements  
(see \cite{LEI} and references therein). The Supernova Cosmology Project \cite{PEL1} and the
High-Z Supernova Research Team \cite{RIES1} have observed and calibrated a large sample of
supernovae at low and high redshifts. The result of their analysis \cite{PEL2,RIES1} 
shows that distant supernovae are on the average about 0.20 magnitudes fainter than
would be expected in a Milne
universe (empty). The likelihood analysis, due to the degeneracy of the luminosity distance
with the values of $\Omega_{\Lambda}$ and $\Omega_m$, constrains these parameters in 
a region approximated by $0.8\Omega_{m}-0.6\Omega_{\Lambda}\approx-0.2\pm0.1$. The data give 
evidence for a non vanishing cosmological constant.
Including the farthest supernova Sn 1997f with $z\approx1.7$ \cite{RIES2}, the
data analysis shows that at redshift $z\sim1.2$ the Universe was in a decelerating phase. 
However the presence of possible systematic uncertainties has attracted some criticism. In particular
there could be a dimming of the light coming from the supernovae due to intergalactic dust.
Moreover the Sn Ia might have an evolution over the cosmic time, due to changes in characteristics
of the progenitors so as to make their use as standard candles unreliable.
The argument against the extinction is that high-redshift supernovae suffer little reddening.
While the fact that their spectra appear similar
to those at low-redshift seems to exclude the possibility of evolutionary effects in the
data. For instance in \cite{Sul}, the Hubble diagram of distant type Ia supernovae segregated
according to the type of host galaxy has been analysed. The results shows that
host galaxy extinction is unlikely to systematically affect the luminosity of Sn Ia in
a manner that would produce a spurious cosmological constant. 
In reality only a theoretical prediction, not available at the present time, would convince the
entire community. 
Nevertheless none of these systematic errors can reconcile the data with a vanishing $\Lambda$.

\section{Cosmic complementarity}
\begin{figure}[t]
\centerline{
\includegraphics[width=.75\textwidth,angle=270]{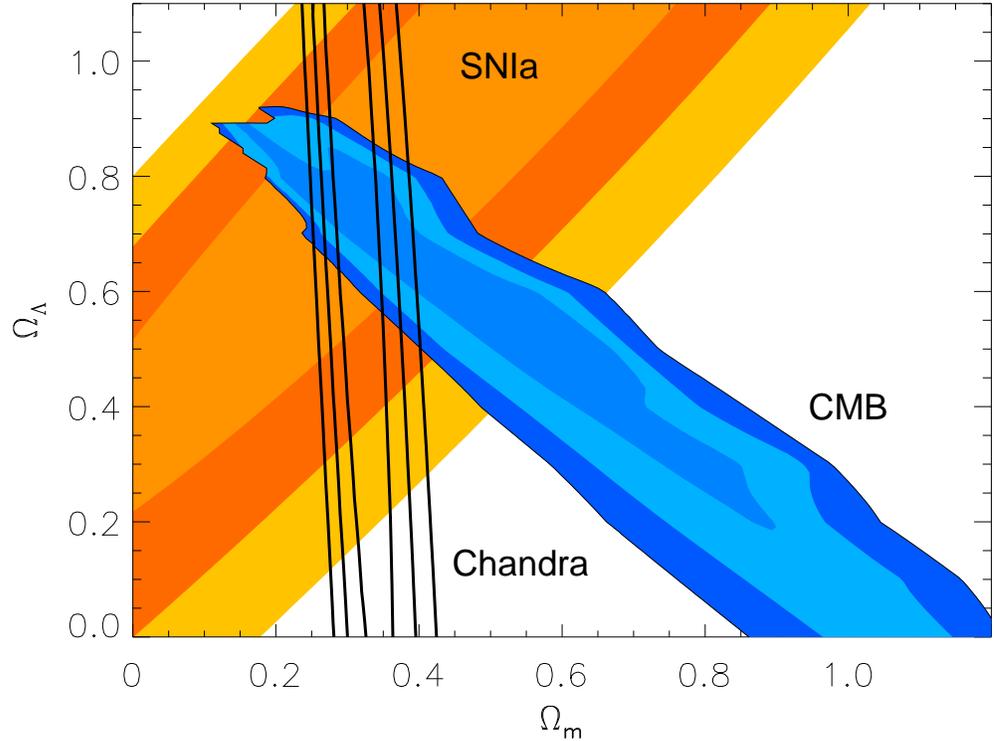}
}
\caption{1, 2 and 3 $\sigma$ confidence contours on
$\Omega_{m}-\Omega_{\Lambda}$ plane determined from Sn Ia, CMB and Chandra $f_{\rm gas}$(z) data
(from \cite{ALLEN2})}		
\label{f:fig2}			
\end{figure}

Figure~\ref{f:fig2} shows the region of the $\Omega_{m}-\Omega_{\Lambda}$ constrained
by different cosmological observations.
It appears evident that the joint analysis of
the recent CMB, large scale structure and Sn Ia observations, previously
reviewed, indicate a region of the parameter space where they are consistent \cite{BAHC,TEG1,TEG2}.
In particular the likelihood contours of Sn Ia and CMB are orthogonal and therefore
their combination breaks the geometric degeneracy between $\Omega_m$ and $\Omega_{\Lambda}$.
Such consistency tells us that the Universe is nearly flat,
the structures we observe today are the result of the growth of initial density fluctuations 
characterized by a nearly scale invariant spectrum as predicted by inflationary scenarios.
The matter content of the Universe
consist of baryons ($3\%$), while most of the clustered matter 
consists of cold dark particles
that account for only $30\%$ of the total energy density.
About $67\%$ of the matter is in a `dark energy' form and is responsible for the present
accelerated expansion. In spite of any astronomical uncertainties, the limits on
$\Omega_{\Lambda}$ impose that $\Lambda\lesssim10^{-47}$ GeV$^4$.
Understanding the origin and the smallness of this term is the challenge of modern theoretical
physics. In principle nothing prevents the existence of a $\Lambda$-term
in the Einstein equations. However if $\Lambda$ appears on the left hand side, the gravitational
part of the Einstein-Hilbert action will depend on two fundamental constants, $G$ and $\Lambda$,
which differ widely in scale. For instance the dimensionless combination of fundamental constants
$(G \hbar/c^3)^2\Lambda\lesssim10^{-124}$. On the other hand we know that several independent 
phenomena can contribute as an effective $\Lambda$ 
term on the right hand side of the Einstein equations \cite{WEINB}.
However, as we shall see in the next chapter, in order to reproduce the small observed value,
these terms have to be fine tuned with bizarre
accuracy. Therefore the solution to such an enigma may well lead us to the discovery
of new physics.

\chapter{An explanation for the dark energy?} 
\label{chap2}

The cosmological constant can be naturally interpreted as the energy 
contribution of the vacuum. However its measured value turns out to be
extremely small compared to the particle physics expectations.
Therefore alternative candidates for
the dark energy component have been considered. In particular a light scalar
field rolling down its self-interacting potential can provide the missing
energy in the Universe and drive a late time phase of accelerated expansion.
A lot
of effort has gone into justifying the existence of this field,
called
{\em quintessence}, within the context of particle theories 
beyond the Standard
Model. Different versions of this original idea have been 
developed in the literature. In this Chapter we will review
the vacuum energy problem. Then we will discuss the application of the
anthropic principle to the solution of the 'coincidence problem'. 
We will describe the main
characteristic of a minimally coupled
quintessence scenario. At the end
we will analyse the dynamics
of two specific scalar field models.

\section{Vacuum energy}
It was initially
pointed out by Y.B. Zeldovich \cite{ZEL} that in Minkowski space-time,
Lorentz invariance constrains the energy momentum tensor of zero 
point vacuum fluctuations to be proportional to the Minkowski metric, {\em i.e.} 
$T_{\mu\nu}^{vac}=const.\times diag(1,-1,-1,-1)$. This relation can be generalized
to the case of a curved space-time with metric $g_{\mu\nu}$. 
The principle of general covariance 
requires that $T_{\mu\nu}^{vac}\propto g_{\mu\nu}$, which has the form 
of a cosmological constant. This implies that in General Relativity, 
since the gravitational field couples through the Einstein
equations with all kinds of energy, the vacuum energy contributes to the
total curvature of space-time. The vacuum state of a collection
of quantum fields, that describes the known forces and particles,
is defined to be the lowest energy density state. 
If we think of the fields as a set of harmonic oscillators
the total zero point energy is given by:

\begin{equation}
\rho_{vac}=\frac{1}{2}\sum_{\bf k}=\frac{1}{4\pi^2}\int_0^{\infty}\sqrt{k^2+m^2}k^2dk,
\label{vac}
\end{equation}
that diverges as $k^4$ (ultraviolet divergence). However
any quantum field theory is valid up to a limiting cut-off scale,
beyond which it is
necessary to formulate a more fundamental description. 
Consequently the integral
Eq.~(\ref{vac}) can be regularized imposing a cut-off $k_{max}$,
and we obtain

\begin{equation}
\rho_{vac}=\frac{k_{max}^4}{16\pi^2}.
\end{equation}
If we set the cut-off $k_{max}$ at the Planck scale, the energy
density of the vacuum is
$\rho_{vac}\approx (10^{19}$ GeV$)^4$ which is about $120$ orders of magnitude larger
than the observed value of $\rho_{\Lambda}$. Fixing the cut-off
scale at the QCD phase transition, $k_{max}=\Lambda_{QCD}$,
we find $\rho_{vac}^{QCD}\approx10^{-3}$ GeV$^4$ which is still 44 orders
of magnitude above the expected one. On the other hand if Supersymmetry is realized
in nature, the cosmological constant vanishes because the vacuum energy 
contribution of the bosonic degree of freedom exactly cancels that of
the fermionic ones. However, because we do not observe super-particles,
Supersymmetry must be broken at low energy. This implies that
the cosmological constant vanishes in the early Universe and reappears
later after SUSY breaking. Assuming that Supersymmetry
is broken at $M_{SUSY}\approx 1 TeV$ the resulting $\rho_{vac}$ is about
60 orders of magnitude larger then the observational upper bounds.
Hence any cancellation mechanism will require a bizarre
fine tuning in order to explain the huge discrepancy between 
$\rho_{vac}$ and $\rho_{\Lambda}$. By the present time we do not have any
theoretical explanation for this cosmological constant problem. Moreover such a
tiny value presents an other intriguing aspect. In fact we could ask why
$\Lambda$ has been fixed at very early time with such an extraordinary accuracy
that today it becomes the dominant component of the Universe.
In other words we should explain why the time when $\Lambda$ starts dominating
nearly coincides with the epoch of galaxy formation. This is the so called
coincidence or 'why now' problem. The solution to the cosmological constant
problem will provide an explanation also for this cosmic coincidence. 
On the other hand it could be easier to justify a vanishing cosmological
constant assuming the existence of some unknown symmetry coming 
from quantum gravity or string theory \cite{WITTEN0}. 
As we shall see in the following sections, alternative scenarios of dark energy
formulate the initial condition problem and
the coincidence problem in a different way. 

\section{Anthropic solutions}
The use of anthropic arguments in cosmology has been often seen as
an anti-scientific approach. However a different use of the `Anthropic
Principle' has been recently proposed in the literature and for a
review of the subject we refer to \cite{GARRIGA}.
We should always have in mind
that at the speculative level our Universe can be one particular
realization of possible universes. Therefore the fact that we live in this
Universe makes us privileged observers, since under other
circumstances we would not be here. For instance several authors
pointed out that not all values of $\Lambda$ are consistent with
the existence of conscious observers \cite{DAVIES,BARROW,WEINBERG}.
The reason is that in a flat space-time the gravitational collapse of structure 
stops at the time $t\sim t_{\Lambda}$, as consequence universes with large 
values of $\Lambda$ will not have galaxies formed at all. This
argument can be used to put an anthropic bound on $\rho_{\Lambda}$
by requiring that it does not dominate before the redshift $z_{max}$ 
when the earliest galaxy formed. In \cite{WEINBERG}
assuming $z_{max}=4$ it was found $\rho_{\Lambda}\lesssim \rho_m^0$.
However it was suggested in \cite{VK,EFA} that observers are in
galaxies and therefore there is a conditional probability to observe
a given value of $\Lambda$. In particular this value will be the one 
that maximizes the number of galaxies. In such a case the probability
distribution can be written as
\begin{equation}
d\mathcal{P}(\rho_{\Lambda})=\mathcal{P}_*(\rho_{\Lambda})
\nu(\rho_{\Lambda})d\rho_{\Lambda},
\end{equation}
where $\mathcal{P}_*(\rho_{\Lambda})$ is the {\em a priori} probability density distribution
and $\nu(\rho_{\Lambda})$ is the average number of galaxies that form
per unit volume with a given value of $\rho_{\Lambda}$. The calculation
of $\nu(\rho_{\Lambda})$ can be done using the Press-Schechter formalism.
Assuming a flat {\em a priori} probability density distribution the authors of \cite{MARTEL}
found that the peak of $\mathcal{P}(\rho_{\Lambda})$ is close to the 
observed value of $\Lambda$.
This anthropic solution to the cosmological constant problem would be incomplete
without an underlying theory that allows $\Lambda$ to take different values and predicts
a flat $\mathcal{P}_*(\rho_{\Lambda})$. The recent developments in string/M theory seems
to provide a natural framework where such issues can be addressed (see \cite{POLCH}). 

\section{Quintessence}
A non-anthropic solution to the cosmic coincidence problem would be 
an exotic form of matter playing the role of dark energy. 
The existence of such a component should be the prediction of some fundamental
theory of particle physics. For instance it was initially suggested that
a network of topological defects could
provide such a form of energy \cite{TURNER1,CHIBA1,CHIBA2}. In fact
topological defects are characterized by a negative value of
the equation of state $w_X\lesssim -1/3$ and lead to an accelerated
expansion if they dominate the energy budget of the Universe.
However these models are ruled out by current cosmological observations.
On the other hand, long before the time of Sn Ia measurements,
it was considered that an
evolving scalar field, called quintessence,
could take into account for the missing energy of the Universe 
\cite{WETT1,RATRA1,FRIEMAN,COBLE,CALDWELL1,FERRE,WETT2}. In this scenario the
cosmic coincidence problem is formulated in a different way. In fact the evolution
of the quintessence is determined by the initial conditions and by the scalar field potential.
Consequently there would be no coincidence problem only if
the quintessence becomes the dominant component today independently of the initial conditions,
that have been set at very early time. It was pointed
out by Zlatev, Wang and Steinhardt \cite{ZLATEV,STEIN1} that viable quintessence
potentials are those manifesting `tracking' properties. In these cases, for a wide range
of initial conditions, the scalar field evolves towards an attractor solution
such that at late time it dominates over the other matter components. However such a time
will depend on the energy scale of the potential and is fixed in way such that $\rho_{Q}$ reproduces 
the observed amount of dark energy. In other words the tracker quintessence solves the initial
conditions problem, but the `why now' problem is related to the energy scale of the model.
If such a scale is consistent with the high energy physics scales there is no fine tunning
and the fact that the acceleration starts only by the present time does not have any
particular meaning. On the contrary the coincidence problem would result if such a scale is much 
smaller than any particle physics scale, because this will require a fine tuning similar to the
cosmological constant case. As we shall see consistent quintessence model
building is a difficult challenge \cite{KOLDA}. Cosmic coincidence is absent
in quintessence models where the scalar field is non-minimally coupled to the cold dark matter
\cite{LUCA1,WANDS,LUCA2,BEAN1}.
Coupling with baryons is strongly constrained by tests of the equivalence principle, however
a coupling with cold dark matter cannot be excluded. In this case the coupling will naturally
produce the gravitational collapsing time scale of the order of the time when the $Q$ field
starts dominating, $t_G\sim t_Q$. Moreover in these models
structure formation can occur even during the accelerated phase of expansion and consequently
no coincidence have to be explained \cite{LUCA3}. It may be argued that for this class of models
to fully succeed what has to be explained is the strength of the couplings. However the non universality
of the couplings may arise in the context of brane models, where dark energy and dark matter belong
to an hidden sector.  

\subsection{Scalar field dynamics} 
\begin{figure}[t]
\centerline{
\includegraphics[scale=0.7]{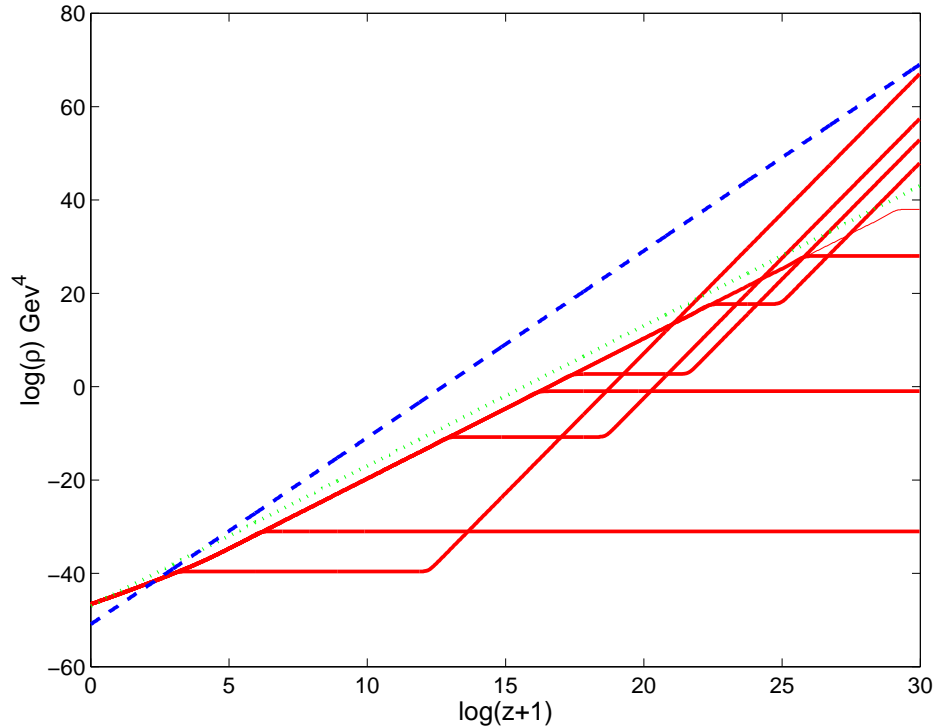}
}
\caption{Energy density versus redshift for radiation (blue dash line),
matter (green dot line) and quintessence (red solid line). As we may notice
for a large range of initial conditions the quintessence energy density converges
to the tracker solution.}		
\label{f:fig3}			
\end{figure}

\begin{figure}
\centerline{
\includegraphics[scale=0.6]{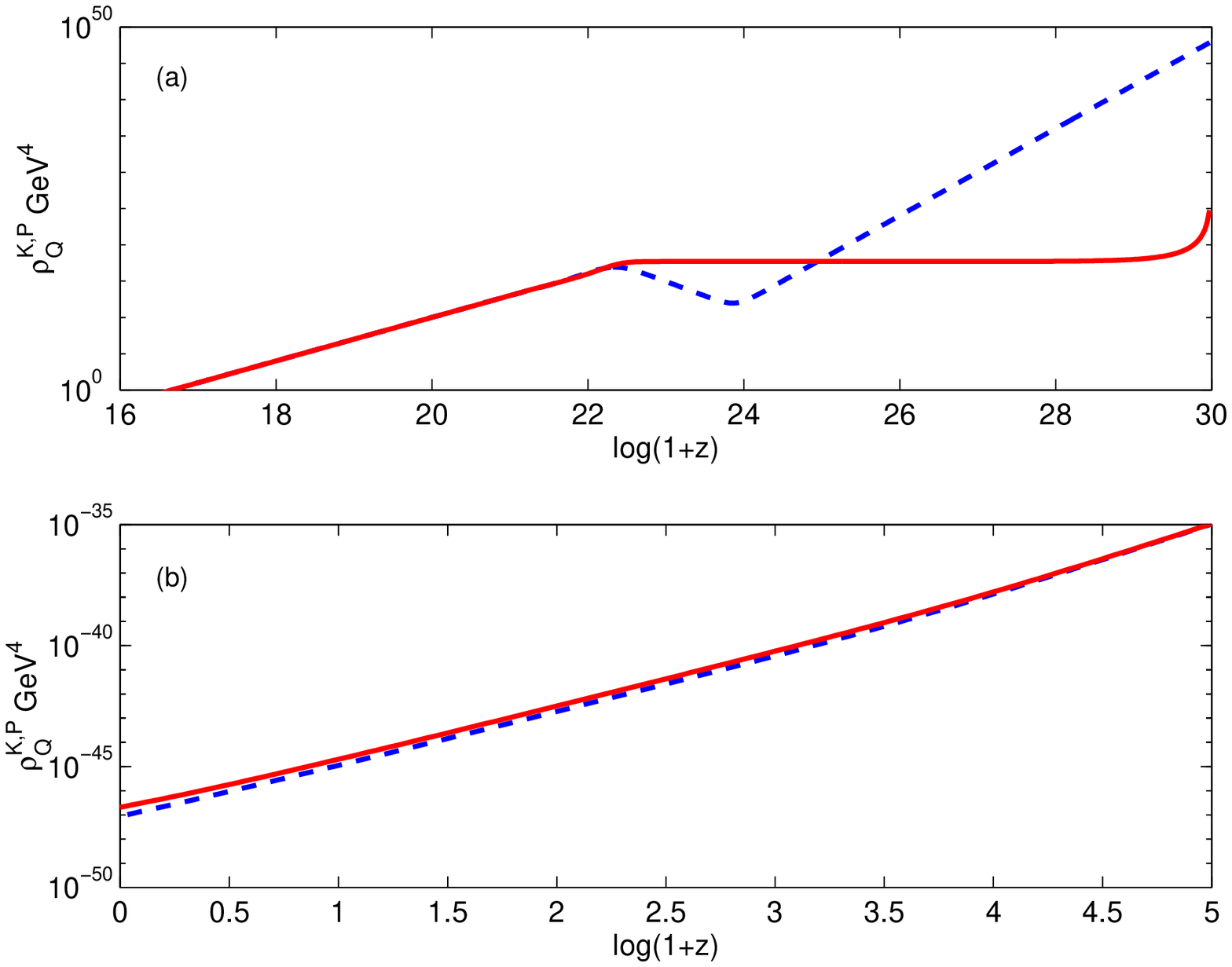}
}
\caption{Evolution of the kinetic (blue dash line) and potential energy 
density (red solid line) at early times (a) and after matter-radiation
equality (b).}		
\label{f:fig4}			
\end{figure}

\begin{figure}
\centerline{
\includegraphics[scale=0.6]{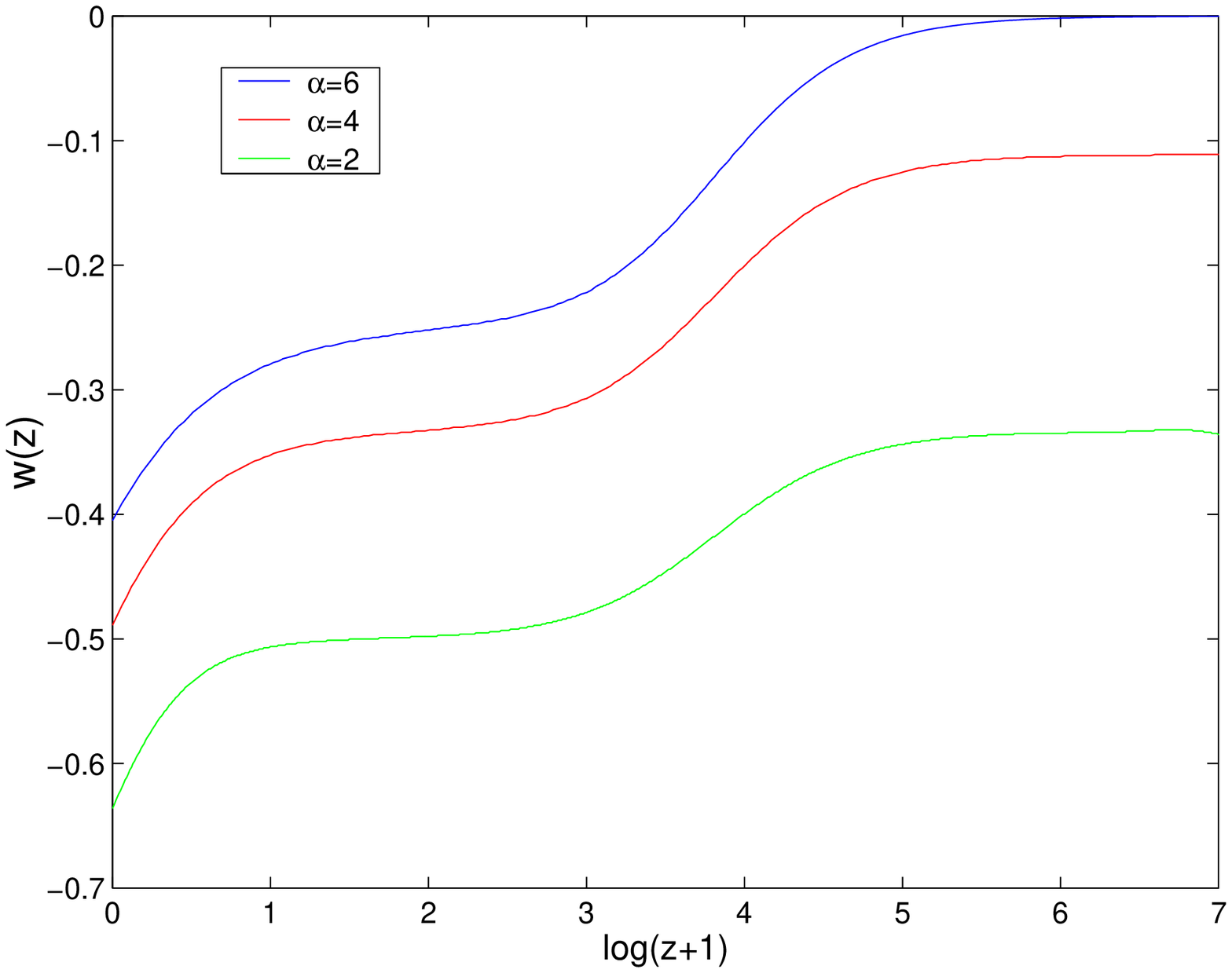}
}
\caption{Equation of state versus redshift for an Inverse Power Law potential with
$\alpha=6$ (blue solid line) $\alpha=4$ (red solid line) and $\alpha=2$ (green solid line).}	
\label{f:fig5}			
\end{figure}

A multiple fluid system consisting of a scalar field, pressureless matter and radiation
interacting through the gravitational field is described by the action:

\begin{equation}
S = -\frac{1}{16\pi G} \int d^4x \sqrt{-g} R + \int d^4x \sqrt{-g} ({\cal L}_Q + {\cal L}_m),
\end{equation}
where $R$ is the Ricci scalar and ${\cal L}_m$ Lagrangian density of
the matter and radiation and ${\cal L}_Q$ is the
Lagrangian density of the quintessence field which is given by:

\begin{equation}
{\cal L}_{Q} = \frac{1}{2} \partial^{\mu}Q\partial_{\mu}Q
- V(Q).
\end{equation}
The scalar field energy-momentum tensor then reads as
\begin{equation}
{T_{Q}}_{\mu\nu} = \partial_{\mu}Q\partial_{\nu}Q -
g_{\mu\nu} \left( \frac{1}{2} \partial^{\alpha}Q\partial_{\alpha}Q 
-V(Q) \right).
\end{equation}
In a FRW flat Universe for a nearly homogeneous scalar field,
the quintessence pressure and energy density are
$p_{Q} = \dot{Q}^2/2 - V$ and $\rho_{Q} = \dot{Q}^2/2 + V$. The quintessence
behaves as perfect fluid with a time dependent equation of state which is given by:
\begin{equation}
w=\frac{\dot{Q}^2/2-V(Q)}{\dot{Q}^2/2+V(Q)}.
\label{eqstato}
\end{equation}
The scalar field evolution is described by the Klein-Gordon equation, 
\begin{equation}
\ddot{Q}+3H\dot{Q}+\frac{dV}{dQ}=0,
\label{klein}
\end{equation}
with 
\begin{equation}
H^2=\frac{8\pi G}{3} \left[ \rho_{m}+\rho_{r}+\frac{\dot{Q}^2}{2}+V(Q) \right] ,
\label{friedmann}
\end{equation}
where $\rho_m$ and $\rho_r$ are the matter and radiation energy densities
and evolve according to the energy conservation equation Eq.~(\ref{encons}).
As an example we analyse the dynamics of this system in the
case of an Inverse Power Law potential \cite{RATRA1,ZLATEV}:
\begin{equation}
V(Q)=\frac{\Lambda^{\alpha+4}}{Q^{\alpha}},
\label{inv}
\end{equation}
where $\alpha=6$ and we set $\Lambda$ such that $\Omega_Q=0.7$.
We solve numerically the equations of motion.
In figure~\ref{f:fig3} we show the evolution with redshift of the
energy density of radiation (blue dash line), matter (green dot line) and quintessence
(red solid line).
Different red lines correspond to different initial conditions,
we may distinguish two distinct behaviours. For initial values of the quintessence
energy density larger than matter energy density, $\rho_Q^{in}>\rho_m^{in}$,
$\rho_Q$ rapidly decreases. This regime is called {\em kination}. In fact, as we can see
in figure~\ref{f:fig4}a, the kinetic energy 
rapidly falls off while the potential energy remains nearly
constant. The overshooting is then followed by the {\em frozen field} phase, where
the energy density is dominated by the potential. 
During this period $\rho_Q$ remains constant
until the kinetic energy becomes comparable with the potential one and 
the field reaches the {\em tracker regime}.
In the tracker solution the kinetic and potential
energies scale with a constant ratio, therefore the equation of state is constant
or slowly varying. For a given potential the existence of a period of tracking is
guaranteed by the condition $\Gamma=V''V/(V')^2>1$ \cite{STEIN1}. Moreover during
this phase the value of the quintessence equation of state mimics the value $w_B$
of the background component according to the relation: 
\begin{equation}
w_Q\approx\frac{w_B-2(\Gamma-1)}{1+2(\Gamma-1)}.
\end{equation}
For inverse power
law potentials $\Gamma=1+\alpha^{-1}$.
At late time the field leaves the tracker solution, the potential energy
starts dominating over the kinetic one (figure~\ref{f:fig4}b)
and the equation of state tends to negative
values. During this final phase the field 
becomes the dominant component of the Universe and drives
the accelerated expansion. The same arguments hold for initial
conditions corresponding to $\rho_Q^{in}<\rho_m^{in}$.
In such a case the quintessence starts its evolution in the frozen regime.
Such a behavior occurs over a range
of initial conditions that covers more
then 100 orders of magnitude, consequently the quintessence dominated
period is obtained with no need of fine tunning of the 
initial conditions.
For tracker models the final value of the equation of state
depends on the parameter $\Lambda$ or equivalently on $\Omega_Q$ and on the
slope of the potential. In general for large values of $\Omega_Q$ or flat potential
$w_Q\rightarrow -1$, in the case of the Inverse Power Law
potential small values of $\alpha$ corresponds to large negative value of $w_Q^0$
(figure~\ref{f:fig5}).

\subsection{Quintessential problems}
\begin{figure}
\centerline{
\includegraphics[scale=0.6]{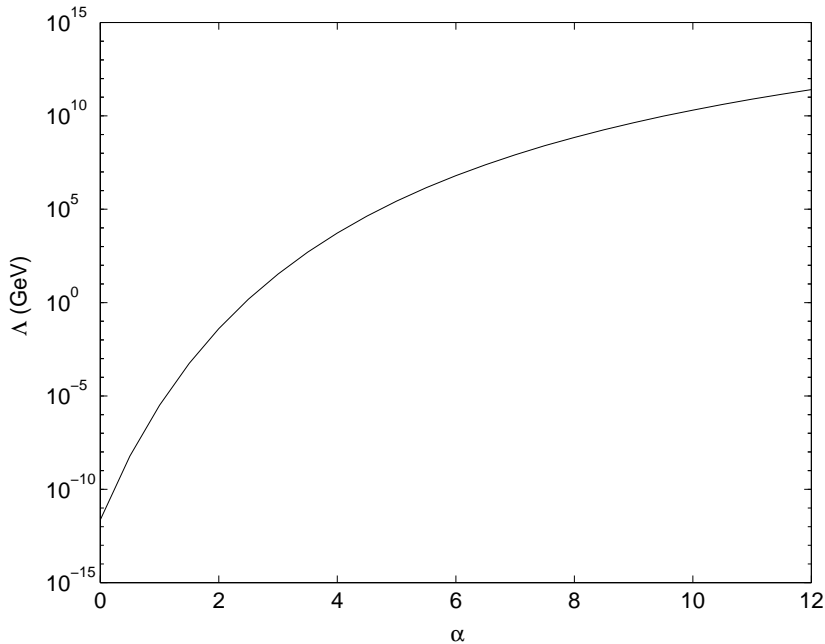}
}
\caption{Energy scale versus $\alpha$ for the quintessence Inverse Power Law potential.}	
\label{f:fig6}			
\end{figure}

The existence of the tracker phase cancels any knowledge of the initial condition
of the scalar field providing an elegant way of solving the coincidence problem.
Nevertheless it is a rather difficult task to build consistent particle physics models
of quintessence. One of the reasons it that for a given tracker potential, 
$V(Q)=\Lambda f(Q)$, the cosmic coincidence is fully solved only if the scale $\Lambda$
is consistent with the energy scale of the underlying particle physics theory that
predicts the shape $f(Q)$ of the quintessence potential. For example let us consider the
Inverse Power Law model. It was shown in \cite{BINE,MASIE} that it can be derived
from a Supersymmetric extension of QCD. 
By the present time the condition $|V'/V|<1$ has to hold in order to guarantee the
Universe is accelerating, this implies that today the scalar field $Q\sim M_{Pl}$.
Since the observations suggest $\rho_Q\approx\rho_c$, for values of the slope
$\alpha\ge 6$ we find that $\Lambda\approx4.8\times10^6$ GeV, a very reasonable
scale from the particle theory point of view. On the other hand for tracker
models the slope of the potential is constrained by measurements of the 
present value of the quintessence equation of state $w_Q^o$ that indicate
a low value of $\alpha$. But, as we can see in figure~\ref{f:fig6}
for values of $\alpha<6$ the energy scale is much smaller than any known particle
physics scale. Therefore the Inverse Power Law seems not to be a viable
quintessence model.
Alternative models have been proposed in the literature, they can be distinguished
into two categories, dilatonic and supersymmetric quintessence.
The former class of models assume that the quintessence field is the dilaton,
this possibility has been studied in \cite{GASPE1,GASPE2}.
The dilaton is predicted by all string theory models
and it couples to all the fields including gravity \cite{WITTEN}. Therefore it 
could be a good candidate for dark energy. However it predicts
a running of the different coupling constants, that are 
strongly constrained by present observations and the violation of the
equivalence principle. Nevertheless these models deserve more investigation and
some of these issues have been recently addressed in \cite{DAMOUR,WETT3}.
It is worth mentioning that the non-minimally coupled scalar field models which
solve the coincidence problem belong to this category 
\cite{LUCA1,WANDS,LUCA2,BEAN1,NICO,PIETRONI}.
In the second class of models the quintessence is one of the scalar fields
predicted by Supersymmetric extensions of the Standard Model of particle physics.
In particular a lot of effort has been recently devoted to the formulation of 
viable quintessence models in the context of Supergravity theory.
In fact it was noticed that Inverse Power Law potentials generated by Supersymmetric
gauge theories are stable against quantum and curvature corrections,
but not against K$\ddot{a}$hlerian corrections \cite{BRAX1,BRAX2,BRAX3}. Since $Q_0\approx M_{Pl}$,
Supergravity (SUGRA) corrections cannot be neglected and therefore any realistic model
of quintessence must be based on SUGRA.
It is argued that also this class of models lead to violation of the equivalence principle,
the quintessence field can in fact mediate a long range fifth force that we do not observe.
However we want to stress that in the context of Supersymmetric theories this is not the
main problem. The reason is that the quintessence can belong to a hidden sector of the theory
that couples only gravitationally to the visible sector.
It was pointed out in \cite{KOLDA} that any supersymmetric inspired model has to address
two specific issues. The first one concerns the case of Supersymmetry breaking quintessence, if
the quintessence field belongs to a sector of the theory that breaks Supersymmetry, because of
the shape of the potential it turns out it cannot be the main source of breaking. In such a case
SUSY can be broken by the presence of an F-term that leads to an intolerably large vacuum energy
contribution that completely spoils the nice properties of the quintessence potential. The other
difficulty arises from the coupling of quintessence to the field responsible for 
the supersymmetry breaking. Such a coupling leads to corrections of the scalar field
potential such that the quintessence
acquires a large mass. Some alternatives have been recently investigated, for instance
a way of avoiding such problems has been considered in \cite{CHOI},
where a Goldstone-type quintessence model in heterotic M-theory has
been proposed. 

\section{Supergravity inspired models}
We now review some of the properties of two models proposed in the context
of Supergravity theories: the Exponential Times Inverse Power Law
potential and the Two Exponential potential.  

\subsection{Exponential Times Inverse Power Law potential}
The authors of \cite{BRAX1} have shown that taking into account 
Supergravity corrections to the Inverse Power Law
potential, the quintessence potential takes the form:
\begin{equation}
V(Q)=\frac{\Lambda^{4+\alpha}}{Q^{\alpha}}e^{\frac{\kappa}{2}Q^2},
\label{SUGRA}
\end{equation}
where $\kappa=1/M_{Pl}^2$.
This potential is an improvement the Inverse Power Law. In fact 
the dynamic remains unchanged during the radiation and
the matter dominated era, while the presence of the exponential
term flatten the shape of the potential in the region corresponding to the late time evolution
of the scalar field. This allows for more negative values of the equation of state today
independently of the slope of the inverse power law. Consequently we can have a reasonable
particle physics energy scale even for large values of $\alpha$. For instance for $\alpha=11$ and
$\Omega_Q=0.7$ we have $\Lambda\approx10^{11}$ GeV and the present value of the
equation of state is $w_Q^0=-0.82$, in better agreement with observational constraints.

\subsection{Two Exponential potential}
\begin{figure}[t]
\centerline{
\includegraphics[scale=0.6]{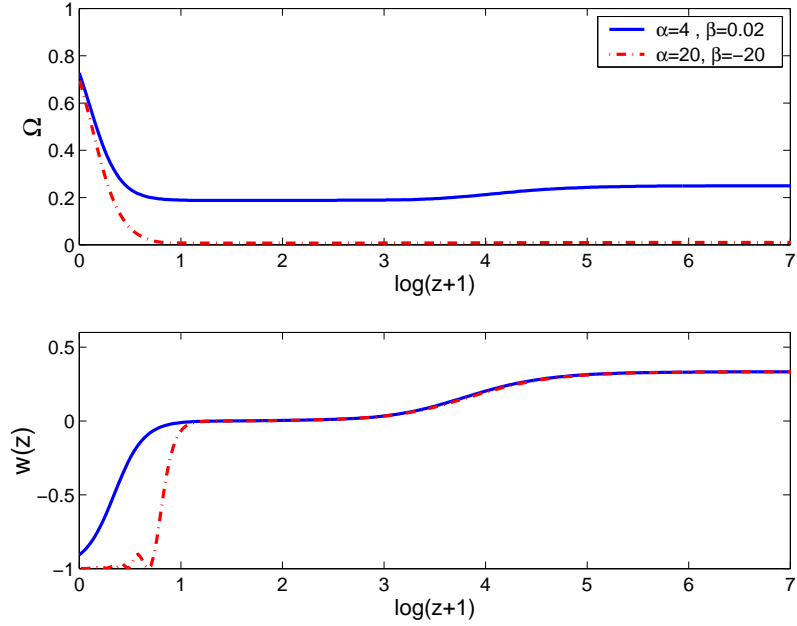}
}
\caption{Evolution of the quintessence energy density and equation of state for parameters
($\alpha$,$\beta$): blue solide line (20,0.5); red dashed line (-20,-20) and $\Omega_Q=0.7$.}	
\label{f:fig8}			
\end{figure}

The dynamics of cosmologically relevant scalar fields with a single
exponential potential has been largely studied
in the literature and within a variety of contexts (see for instance \cite{FERRE}).
The existence of scalar field dominated attractor solutions is well known, however
for this class of models an accelerated phase of expansion can be obtained with an extreme
fine tunning of the initial conditions \cite{CLINE}. It has been shown in \cite{NELSON,ED}
that quintessence Supergravity inspired models predict the scalar field potential to be of the form: 
\begin{equation}
V(Q)=M_{Pl}^4\left(e^{\alpha \sqrt{\kappa}(Q-A)}+e^{\beta \sqrt{\kappa}(Q-B)}\right),
\label{2EXP}
\end{equation}
where $A$ is a free parameter, while $B$ has to be fixed such that $M_{Pl}^4 e^{-\beta B}\sim \rho_Q^0$.
This potential has a number of interesting features. As it has been pointed out by the authors
of \cite{NELSON} in the form given by the Eq.~(\ref{2EXP}) all the parameters are of
the order of the Planck scale. Only $B$ has to be adjusted so that
$M_{Pl}e^{-\beta B}\approx \rho_Q$, it turns out to be $B={\cal L}(100)M_{Pl}$.
For a large range of initial conditions the quintessence field reaches the tracker regime
during which it exactly mimics the evolution of the barotropic fluid and
at some recent epoch it evolves into a quintessence dominated regime.
It is useful to rewrite the two exponential potential as:
\begin{equation}
V(Q)=M^4\left(e^{-\alpha Q/M_{Pl}}+e^{-\beta Q/M_{Pl}}\right),
\end{equation}
where $M$ is the usual energy scale parameter.
For a given value of $\Omega_Q$ the slopes $\alpha,\beta$ fix the final value of the
equation of state. The sign of the slopes distinguish this class of potentials into two
categories: $\alpha>\beta>0$
and $\alpha>0$, $\beta<0$. For both the cases the Q-field initially assumes negative
values and rolls down the
region of the potential dominated by the exponential of $\alpha$.
When it reaches the tracker regime its equation of state exactly reproduces
the value of the background dominant component, $w_Q=w_B$ and the energy
density is given by 
\begin{equation}
\Omega_Q=3(w_B+1)/\alpha^2.
\label{om}
\end{equation}
The late time evolution is determined by the value of $\beta$ that fixes the present value of
the equation of state. In the case of slopes with the same sign $w_Q^0\rightarrow-1$ for small
value of $\beta$, while for $\alpha$ and $\beta$ with opposite sign the scalar field reaches by the
present time the minimum of potential at $Q_{min}/M_{Pl}=ln(-\alpha/\beta)/(\alpha-\beta)$. 
Consequently $w_Q^0\approx-1$ after a series of small damped oscillations. 
This can be seen in figure~\ref{f:fig8} where we plot the evolution of the quintessence energy 
density parameter $\Omega_Q$ and the equation of state $w_Q$ for  
$\alpha=4$, $\beta=0.02$ and $\alpha=20$, $\beta=-20$. 
In the latter model the equation of rapidly drops to $-1$ after few damped oscillations,
while the former shows a more smooth behaviour. It is worth remarking that for the two exponential potential
with same sign of the slopes the accelerated phase can be a transient regime. This can occurs
for large value of $\beta$, in such a case after a short period dominated by the potential energy
the scalar field acquires kinetic energy so that the equation of state can be $w_Q>-1/3$.
As we may note in figure~\ref{f:fig8}, because Eq.~(\ref{om}) holds during the tracker regime,
for small values of $\alpha$ the quintessence energy density assumes non-negligible values at early
times. Such an early contribution during the radiation dominated era is constrained by 
nucleosynthesis bound to be $\Omega_Q(1$ MeV$)<0.13$. This implies that
$\alpha>5.5$. Such a limit has pushed toward larger value by a new analysis
of the Big-Bang nucleosynthesis \cite{RACHEL} that constrains
$\Omega_Q(1$ MeV$)<0.045$ at 2$\sigma$.
In principle this bound can be avoided
if the tracker regime starts after Big-Bang nucleosynthesis. However this can be 
obtained only by tuning the scalar field initial conditions in a restricted range of values.
Moreover a recent analysis of the large scale structure data and CMB measurements strongly constrain
the value of $\Omega_Q$ during the matter dominated era \cite{DOR5}.

\chapter{Quintessence field fluctuations} 
\label{chap3}
The cosmological constant is a smooth component and therefore
does not play any active role during the period of structure formation.
On the contrary a peculiar feature of the quintessence field is that it is
spatially inhomogeneous just as any other scalar field. Therefore 
it is possible 
that the clustering properties of the dark energy can play
a determinant role in revealing the nature of this exotic fluid.
In this Chapter we introduce the scalar field 
fluctuation equations in the Newtonian gauge. 
We derive an analytical solution of the quintessence perturbation in
the tracking regime and describe the behaviour during different cosmological
epochs. We then present the numerical analysis of the perturbations in
a multiple fluids system in the particular cases
of an Inverse Power law potential and the two exponential
potential. 

\section{Quintessence perturbations in Newtonian gauge}
The evolution of minimally coupled scalar field perturbations
has been studied in a number of papers. For instance in
 \cite{RATRA1,FERRE,PERRO,BRAXRIA} the analysis has been done in the
synchronous gauge, while in \cite{VIANA,FABIO,MICHAEL} the authors
have used the Newtonian gauge. In what follows we use the Newtonian
gauge. When compared to the synchronous one it has a number of advantages.
In fact since the gauge freedom is fully fixed there are no gauge modes 
that can lead to misleading conclusions about the evolution of superhorizon
modes. Besides, the metric perturbations play the role of the gravitational
potential in the Newtonian limit.
The equations we need to linearize are the Einstein's equations
\begin{equation}
R_{\mu \nu}-\frac{1}{2}g_{\mu \nu}R=8\pi G T_{\mu \nu},
\label{einste}
\end{equation}
the Klein-Gordon equation
\begin{equation}
\frac{1}{\sqrt{-g}}\partial_{\mu}(\sqrt{-g}g^{\mu \nu}\partial_{\nu}Q)+
\frac{dV}{dQ}=0
\label{kg}
\end{equation}
and the conservation equation of the stress energy tensor of the different
matter components 
\begin{equation}
T_{\mu;\nu}^{\nu}=0,
\label{tmn}
\end{equation}
In the Newtonian gauge the line element in
a spatially flat FRW background reads as
\begin{equation}
ds^2=(1+2\Phi)dt^2-a^2(t)(1-2\Phi)dx^idx_i,
\end{equation} 
where $\Phi$ is the metric perturbation and $t$ is the real time.
We consider a multiple fluid system composed of a scalar field, cold 
dark matter and radiation. 
Expanding the fluid variables at first order around the homogeneous
value we have:
\begin{equation}
\rho_i^{e}=\rho_i(1+\delta_i),
\end{equation}
where $\delta_i$ is the density perturbation of the $i$-th component,
\begin{equation}
Q^{e}=Q+\delta Q,
\end{equation}
where $\delta Q$ is the scalar field fluctuation and $Q$ is the
homogenous part of the quintessence field.
The linearized 
Eq.~(\ref{einste}), Eq.~(\ref{kg}) and Eq.~(\ref{tmn})
provide a set of differential equations for the metric, scalar field,
radiation and cold dark matter perturbations. In Fourier space 
the equations are:
\begin{equation}
\dot{\Phi}+\Phi+\frac{k^2}{3 H^2a^2}\Phi=\frac{4\pi G}{3 H^2}(\delta\rho_Q+
\rho_r\delta_r+\rho_c\delta_c),
\label{eqphi}
\end{equation}
\begin{equation}
\ddot{\delta Q}+3H\dot{\delta Q}+\frac{k^2}{a^2}\delta Q+\frac{d^2V}{dQ^2}\delta Q=4\dot{Q}\dot{\Phi}-2\frac{dV}{dQ}\Phi,
\label{pertq}
\end{equation}
\begin{equation}
\dot{\delta_c}=\frac{k}{a}V_c+3\dot{\Phi},
\end{equation}
\begin{equation}
\dot{V_c}=-\frac{k}{a}\Phi,
\end{equation}
\begin{equation}
\dot{\delta_r}=\frac{4k}{3a}V_r+4\dot{\Phi},
\end{equation}
\begin{equation}
\dot{V_r}=-\frac{k}{4a}\delta_r-\frac{k}{a}\Phi,
\label{eqvr}
\end{equation}
where $\delta_{c}$ and $\delta_r$ are the density perturbations of
cold dark matter and radiation, while $V_{c}$ and $V_{r}$ are the
corresponding velocity perturbations. The perturbed quintessence
energy density and pressure are,
\begin{equation}
\delta \rho_Q\equiv \rho_Q\delta_Q=\dot{Q}\delta\dot{Q}-{\dot{Q}}^2\Phi+\frac{dV}{dQ}\delta Q,
\label{drhoqe}
\end{equation}
\begin{equation}
\delta p_Q=\dot{Q}\delta\dot{Q}-{\dot{Q}}^2\Phi-\frac{dV}{dQ}\delta Q.
\end{equation}

\section{Evolution of perturbations}

\subsection{Analytical solution in the tracker regime}
From Eq.~(\ref{pertq}) we note that
the perturbations of the other fluids feed back onto the scalar field
perturbations through the gravitational potential $\Phi$. 
Deep during the radiation and the matter dominated eras 
the gravitational potential is constant and therefore the first term
on the right-hand-side can be neglected. As a first approximation we 
ignore the second term as well, hence 
Eq.~(\ref{pertq}) becomes
\begin{equation}
\ddot{\delta Q}+3H\dot{\delta Q}+\left(\frac{k^2}{a^2}+\frac{d^2V}{dQ^2}\right)\delta Q=0.
\label{rido}
\end{equation}
We can find an analytical solution during the tracker regime when
\begin{equation}
\frac{d^2V}{dQ^2}\sim A_Q^B H^2,
\label{ddv}
\end{equation}
with $A_Q^B$ a constant that depends on the specifics of the potential and
on the equation of state of the background dominant component.
This can be obtained as follows. Consider the adiabatic definition
of the sound speed associated with the quintessence field:
\begin{eqnarray}
c_Q^2 & \equiv & \frac{\dot{p}_Q}{\dot{\rho}_Q}
=w_Q-\frac{\dot{w}_Q}{3H(1+w_Q)} \nonumber \\
& = & 1+\frac{2V_{,Q}}{3H\dot{Q}}.
\label{cs2qad}
\end{eqnarray}
During the tracker regime the quintessence equation of state
is nearly constant implying that $c_Q^2\approx w_Q = const$. Hence by differentiating
Eq.~(\ref{cs2qad}) with respect to time and dividing by $H$ we have
\begin{eqnarray}
\frac{\dot{c_Q^2}}{H}=\frac{2V_{,QQ}}{3H^2}
+(1-c_Q^2)\left[1+\frac{\dot{H}}{H^2}-\frac{1}{2}(3 c_Q^2+5)\right],
\label{csdo}
\end{eqnarray}
From Eq.~(\ref{csdo}) using the second of the Hubble equations 
$\dot{H}/H^2=-3(1+w_B)/2$, with $w_B$ being the equation of state of the
background dominant fluid, we finally obtain:
\begin{equation}
\frac{V_{,QQ}}{H^2}=\frac{9}{4}(1-c_Q^2)\left(w_B+c_Q^2+2 \right)\equiv A_Q^B.
\end{equation}
It is useful to rewrite Eq.~(\ref{rido}) in conformal time,
which is defined as $d/d\tau\equiv a(t)d/dt$, 
\begin{equation}
\delta Q''+2\mathcal{H}\delta Q'+\left(k^2+a^2 H^2 A_Q^B\right)\delta Q=0,
\label{confordq}
\end{equation}
where a prime denotes the derivative with respect to $\tau$.
In the radiation dominated era the scale factor evolves as $a=\tau/2$,
while the Hubble rate is given by $H=2/\tau$, hence 
Eq.~(\ref{confordq}) becomes:
\begin{equation}
\delta Q''+2 \delta Q'+(k^2+4 A_Q^r)\delta  Q=0,
\label{dqrid}
\end{equation}
with $A_Q^r$ being the value of $A_Q^B$ in radiation dominated era.
During the tracker regime $c_Q^2\approx w_r=1/3$
and $A_Q^r\approx 4$. In such a case the characteristic
roots of Eq.~(\ref{dqrid}) are complex and 
the solutions are of the form:
\begin{equation}
\delta Q=e^{-2\tau}\left(C_1 \cos{\nu_r \tau}+C_2 \sin{\nu_r \tau} \right),
\label{rd}
\end{equation}
where $C_1$ and $C_2$ are integration constants and
\begin{equation}
\nu_r=\sqrt{k^2+4 A_Q^{r}-1}. 
\label{nur}
\end{equation}
Eq.~(\ref{rd}) is a damped oscillatory solution 
with frequency $\nu_r$, On large scales ($k<1$ Mpc$^{-1}$) the frequency of these
oscillations is scale independent and is set by the tracker properties ($A_Q^r$).
On the other hand if $A_Q^r=0$,
Eq.~(\ref{dqrid}) has real roots and the solutions
contain a constant mode and a decaying one. Similar solutions can be found in the
matter dominated era.
The addition of the source term
$\Phi$ on the right-hand-side of Eq.~(\ref{pertq})
leads to an attractor solution for the quintessence
perturbations in the long-wavelength limit, $\delta Q(t)\rightarrow\delta Q_c$. In fact
from Eq.~(\ref{pertq}) we obtain:
\begin{equation}
\delta Q_c\approx -2 \frac{V_{,Q}}{V_{,QQ}} \Phi_c,
\label{dqc}
\end{equation}
that is constant as long as the quintessence is in the tracker regime.
As a consequence this solution does not hold
in the kinetic and particularly in the potential phase when the quintessence
exits away from the tracker.

\subsection{Numerical analysis}
\begin{figure}[t]
\centerline{
\includegraphics[scale=0.6]{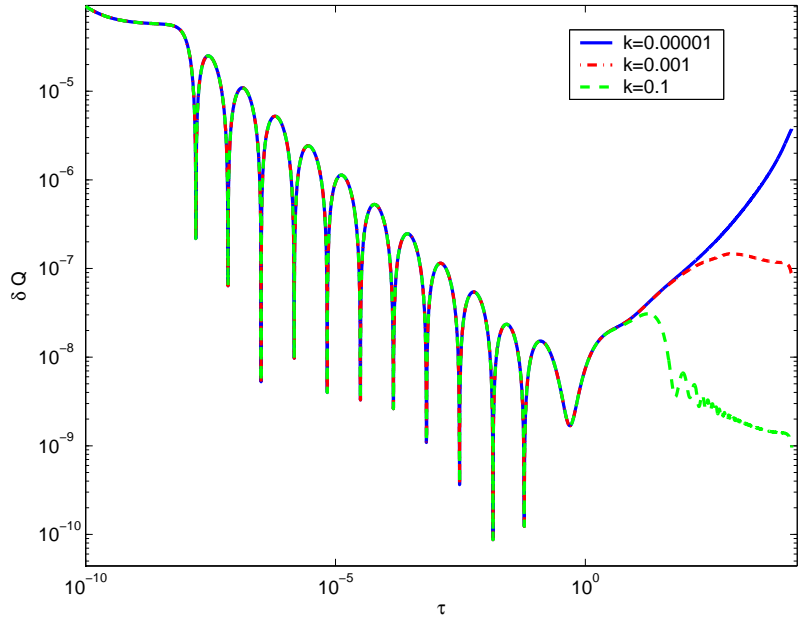}
}
\caption{Evolution of the scalar field perturbation $\delta Q_k$
 for $k=0.00001$ (blue solid line), $k=0.001$ (red dash line) and $k=0.1$ (green dash line).}	
\label{f:dq}			
\end{figure}

\begin{figure}[t]
\centerline{
\includegraphics[scale=0.6]{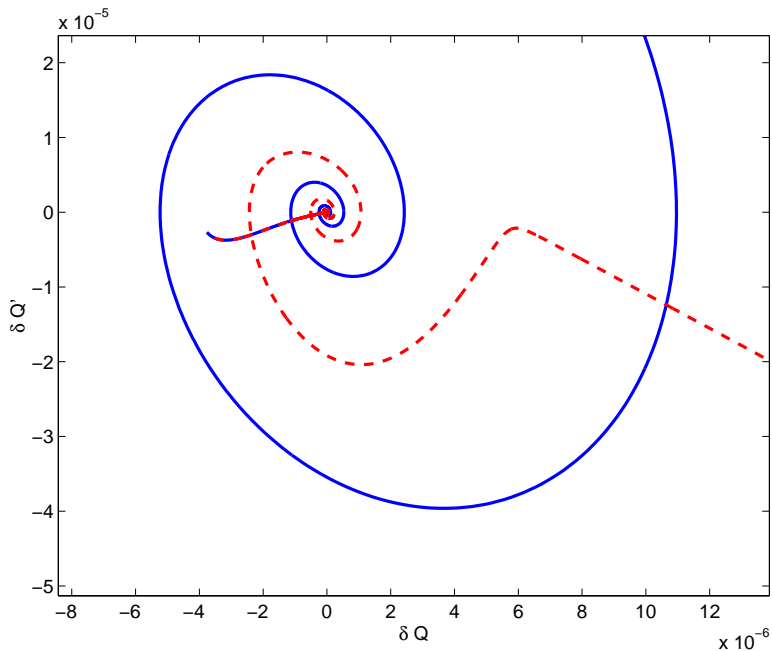}
}
\caption{Phase diagram for long wavelength quintessence 
fluctuations with two different initial conditions. The attractor point
$\delta Q_c$ (Eq.~\ref{dqc}) is a transient attractor.}	
\label{f:dqspa}			
\end{figure}

\begin{figure}[t]
\centerline{
\includegraphics[scale=0.6]{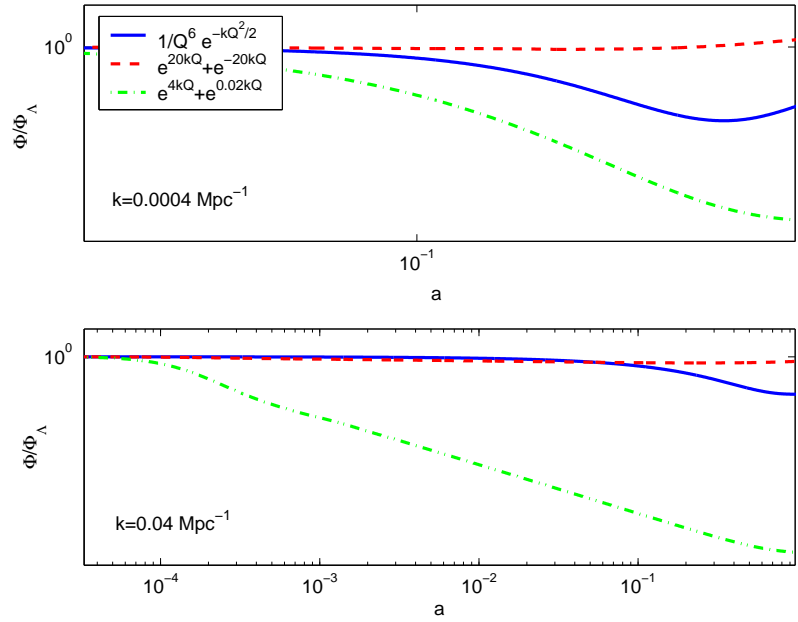}
}
\caption{Evolution of the gravitational potential $\Phi$ normalized to the
cosmological constant case for a SUGRA model (blue solid line), two exponential potentials
with slopes $(-20,20)$ (red dash line) and $(4,0.02)$ (green dash line) respectively.
The top panel shows the mode with $k=0.0004$ and the lower panel shows $k=0.04$.}	
\label{f:phi}			
\end{figure}

In this section we present the analysis of the numerical solution of the system of equations 
Eq.~(\ref{eqphi}-\ref{eqvr}), where we have imposed adiabatic initial conditions. 
We consider a quintessence
SUGRA inspired model described in Chapter~\ref{chap2}. 
The scalar field potential is specified by Eq.~(\ref{SUGRA}) where
we set the slope $\alpha=6$ and the parameter $M$ such that today $\Omega_Q=0.7$. 
The initial conditions for the homogeneous part of the scalar field $Q$ have been
set such that the tracker solution is reached deep
 in the radiation dominated era.
In figure \ref{f:dq} we plot the behaviour
of $\delta Q$ against the conformal time for three different wavenumbers. 
We can see that as the system enters the tracker regime,
decaying oscillations are set for all the three modes with a constant frequency given by Eq.~(\ref{nur}).
Since $k^2$ is negligible in Eq.~(\ref{nur}) the frequency of these oscillations is
the same for all modes.
When the second term in the right-hand-side of Eq.~(\ref{pertq}) becomes comparable
to the term $V_{,QQ}$ the fluctuations evolves onto the attractor solution
given by Eq.~(\ref{dqc}). In the specific case we consider there is not an exact
tracking and consequently the ratio $V_{,Q}/V_{,QQ}$ scales linearly and not
steadily as in the exact tracking case. This explains why $\delta Q$ increases
with time in the long wavelength mode.
As different scales cross the horizon they leave this attractor
solution. For instance the shortest wavelengths ($k=0.1$) enter the horizon after
matter radiation equality. Due to the decay of the gravitational potential
the first term on the right-hand-side
of Eq.~(\ref{pertq}) becomes large leading to a decaying solution with small
damped oscillations whose frequency is set by the 
effective mass term $k$ on the left-hand-side
of Eq.~(\ref{pertq}).
The presence of an attractor solution for the scalar field perturbations is
evident in figure~\ref{f:dqspa}, where we plot the phase space $\delta Q-\delta\dot{Q}$.
As the system evolves in the tracker regime, $\delta Q$ and $\delta \dot{Q}$ spiral toward
the attractor point $\delta Q_c$. As soon as the tracker terminates the
long wavelength fluctuations have converged to the same values
and their evolution is indistinguishable.
In minimally coupled quintessence models the effects of the scalar field
perturbations on the structure formation process are negligible. This is because the
quintessence interacts only gravitationally with the other fluid components and
it is usually a subdominant component
during the epochs that are relevant for the formation of the early structures.
Therefore the clustering of the dark energy can have detectable effects only at late times
through the time evolution of the gravitational potentials.
As we shall see in Chapter~\ref{chap6} this produces a characteristic imprint
in the CMB anisotropy power spectrum.
In figure \ref{f:phi} we plot the behaviour of the gravitational potential $\Phi$ normalized
to the cosmological constant case for
$k=0.0004$ (upper panel) and $k=0.04$ for three different quintessence
models whose dynamics have been described in Chapter~\ref{chap2}:
(model A) the SUGRA potential Eq.~(\ref{SUGRA}) with slope $\alpha=6$ (blue solid line);
(model B) the two exponential potential Eq.~(\ref{2EXP}) with $\alpha=20$ and $\beta=-20$ 
(red dash line) and (model C) $\alpha=4$ and $\beta=0.02$ (blue dash dot line).
We may note a number of features that arises from the presence of dark energy perturbations
and a different expansion rate of the Universe compared to the $\Lambda$CDM model.
Let us consider the mode $k=0.0004$ that
crosses the horizon during the matter era. For model B (red dash line)
 the quintessence energy density is always negligible 
during the matter era and consequently there are no effects due to the perturbations in the
evolution of the gravitational potential. On the other hand because of the scalar
field dynamics the late time expansion rate of the Universe is different
compared to the $\Lambda$ case, hence $\Phi/\Phi_{\Lambda}$ diverges from the unity.
In contrast in models A and C the quintessence energy density dominates earlier,
especially in model C where $\Omega_Q$ is a large fraction of the total energy density 
(see figure~\ref{f:fig8}). In this case we can distinguish two different evolutionary
regimes of the gravitational potential.
At redshift $z>10$ the Universe is matter dominated
and the quintessence is in the tracker solution with an equation of state very close
to the matter value. Therefore during this period the expansion rate is the same
as in the $\Lambda$ model and we would expect $\Phi/\Phi_{\Lambda}\approx1$. 
The fact this ratio deviates from unity is due to the presence of
the dark energy perturbations. On the other hand the evolution of the gravitational 
potential at $z<<10$ is caused by the accelerated phase of expansion. This can be seen
by the change of the slope of $\Phi/\Phi_{\Lambda}$ at a redshift $z\approx2$ when the
acceleration starts. The late time decay of the gravitational potential sources
the formation of CMB anisotropy through the late Integrated Sachs-Wolfe effect (ISW).
The same arguments hold for the mode $k=0.04$ 
that enters the horizon soon after matter-radiation
equality. However it is worth noticing that in the case of model C, the early contribution
of dark energy causes the scalar field perturbations to produce a
 bigger effect on the decay of $\Phi$, leading to a larger early ISW effect.
From this analysis we conclude that different dark energy models will lead 
to a characteristic signature in the CMB power spectrum and will prove this point
in a more general way in Chapter~\ref{chap6}.

\chapter{Constraining the quintessence potential}
\label{chap4}

Cosmological distance measurements test the expansion rate of the
Universe at different redshifts. Therefore they can be used to
constrain the dark energy properties and 
eventually disprove the $\Lambda$CDM model.
A lot of effort has gone into constraining the value of the dark energy
equation of state with the current available data.
In this Chapter we will review some of the constraints obtained in previous
works. Then we will describe the analysis of the Sn Ia data and the measured position
of the CMB peaks that we performed to constrain the shape of a general quintessence
potential \cite{CORAS1}. We will comment on our results and compare them to 
those obtained in other related work.

\section{Upper bounds on the cosmic equation of state}
\begin{figure}[t]
\centerline{
\includegraphics[scale=0.6]{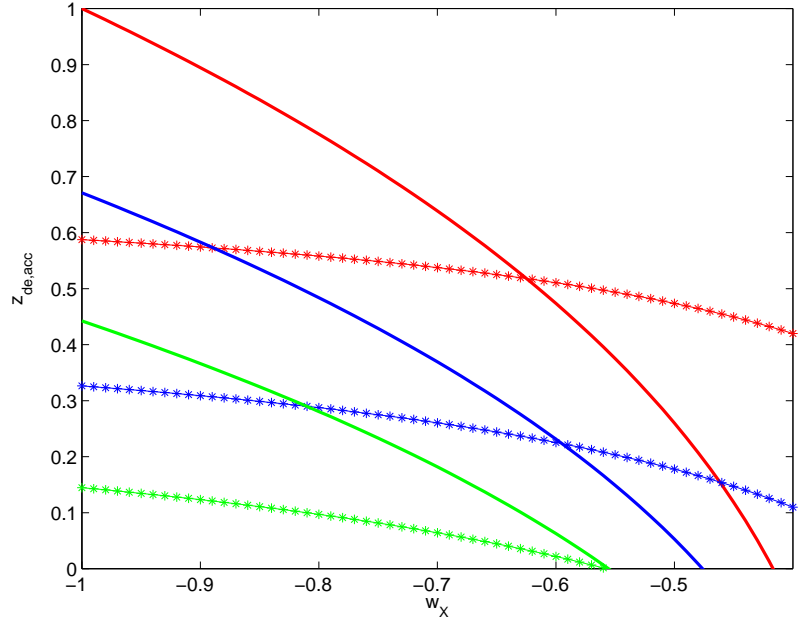}
}
\caption{Redshifts $z_{acc}$ (solid line) and $z_{eq}$ (starred line)
corresponding to the start of the
accelerated phase and the dark energy dominated epoch versus a constant $w_X$ for
$\Omega_X=0.6,0.7,0.8$ (in {\em blue, red} and {\em green} lines respectively).}	
\label{f:fig9}			
\end{figure}

\begin{figure}[t]
\centerline{
\includegraphics[scale=0.6]{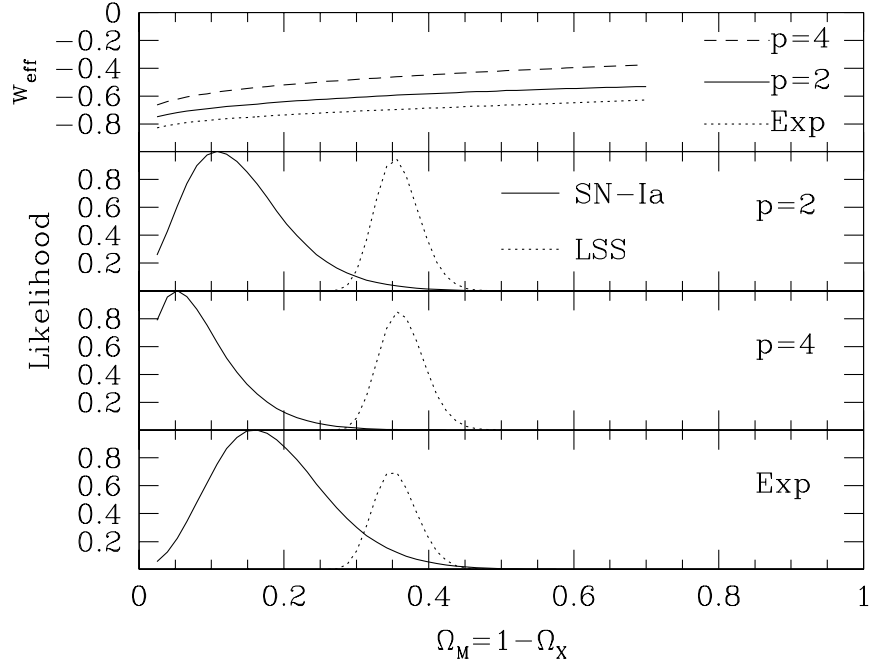}
}
\caption{Upper panel: relation between $w_{eff}$ and $\Omega_m$ for an Inverse Power Law
potential with slope p and an Exponential potential $V\propto$ exp($m_{Pl}$/Q). Lower
panels: One dimensional likelihoods as a function of $\Omega_m$ for CMB+LSS 
(dot) and Sn Ia (solid) data (from \cite{PERL3}).}	
\label{f:fig10}			
\end{figure}

Different dark energy models are usually constrained assuming that the dark energy
behaves as a perfect fluid with a constant equation of state $w_X=p_X/\rho_X\le-1$.
This can take into account for models such as a network of topological defects with
$w_{X}=-n/3$  ($n$ is the dimension of the defect) \cite{VILENKIN1}
and as first approximation time dependent tracker 
quintessence models with $w_Q>-1$. Different values of
$w_X$ will lead to a different expansion rate and consequently
to a different cosmological distance. In particular for larger negative values
of $w_X$ the phase of acceleration starts at earlier times and therefore
cosmological distance indicators will appear farther than in models with smaller
negative values of $w_X$. This can be seen in figure~\ref{f:fig9},
we plot in the solid line the redshifts $z_{acc}$ when the Universe starts accelerating
and in a starred line the redshift $z_{de}$ when the dark energy starts dominating 
as a function of $w_X$ for different values of $\Omega_X$.
For a flat Universe The redshift $z_{acc}$ is given by
\begin{equation}
(1+z_{acc})^3=-(1+3 w_X)\frac{\Omega_X}{1-\Omega_X},
\end{equation}
and the redshift $z_{de}$ is
\begin{equation}
(1+z_{de})^3=\frac{\Omega_X^{-1/w_X}}{1-\Omega_X}.
\end{equation}
Note for $w_X<-0.6$ we have $z_{acc}>z_{de}$.
For a flat Universe the effect on cosmological distances
produced by varying $w_X$ is degenerate with
$\Omega_m=1-\Omega_X$ and with the Hubble parameter $H_0$.
By making use of the Sn Ia data the
authors of \cite{PEL2,GARNAVICH} find the constraint
$w_X<-0.55$ (2$\sigma$) for $\Omega_m>0.1$.
We will comment in the next Chapter about the bias effect introduced in the
analysis of data by assuming a constant equation of state.
The position of the Doppler peaks in the CMB anisotropy power spectrum
is another indicator of the cosmological distance to the last scattering
surface and hence can be used to constrain $w_X$. 
In \cite{PERL3} a constant effective equation of state
\begin{equation}
w_{eff}=\frac{\int w_X(a) \Omega_X(a)da}{\int \Omega_X(a)da},
\label{weff}
\end{equation}
has been constrained with the Sn Ia, CMB and the LSS data. The authors find
$w_{eff}<-0.65$ (2$\sigma$) with $w_{eff}=-1$ being best fit. 
They comment that tracker models can marginally accommodate
the Sn Ia and LSS constraints at the same time (see figure~\ref{f:fig10}).
Such an inconsistency can be explained as follows. The supernova luminosity
distance is sensitive to
the value of $w_{eff}$, while large scale structure data are more sensitive
to the amount of clustered matter $\Omega_m$ through the matter power spectrum. 
This implies that dark energy models with different values of $w_{eff}$ will not
change the likelihood in $\Omega_m$ of the large scale structure data. In fact
assuming the value of the effective equation of state to be $w_{eff}>-1$,
this shifts the Sn Ia likelihood towards smaller values of $\Omega_m$
({\em i.e.} larger values of $\Omega_X$).
As we will show later, extending the likelihood analysis
to the slope of the tracker potential enables us to find models
that simultaneously fit the whole data. Similar constraints from Sn Ia and CMB data
have been found by Efstathiou
\cite{EFST}, namely $w_X \lesssim -0.6$ (2$\sigma$). In a different approach
Saini et al. \cite{SAINI} have
parameterized the luminosity distance and by constraining their parameter space with the
Sn Ia data they have been able to reconstruct the redshift dependence
of the dark energy equation of state.
They find a time dependence with $-1\leq w_X\leq-0.86$ today and
$-1\leq w_X\leq-0.6$ at $z=0.83$. Waga and Frieman have constrained the slope of the 
Inverse Power Law potential $\alpha$ (see Eq.~(\ref{inv}))
by making use of the Sn Ia and lensing statistic data
\cite{WAGA1}. Imposing $\Omega_m>0.3$ they obtain an upper limit on the
present value of the equation of state $w_Q^0<-0.67$ that implies $\alpha<1.8$.
The combined analysis of the CMB power spectrum measured by the Boomerang and MAXIMA
experiments with Sn Ia and large scale structure, limits a constant equation of
state to be $w_X<-0.7$ with $w_X=-1$ being the favoured value \cite{BOND}.

\section{Parameterized quintessence potential}
\begin{figure}[t]
\centerline{
\includegraphics[scale=0.6]{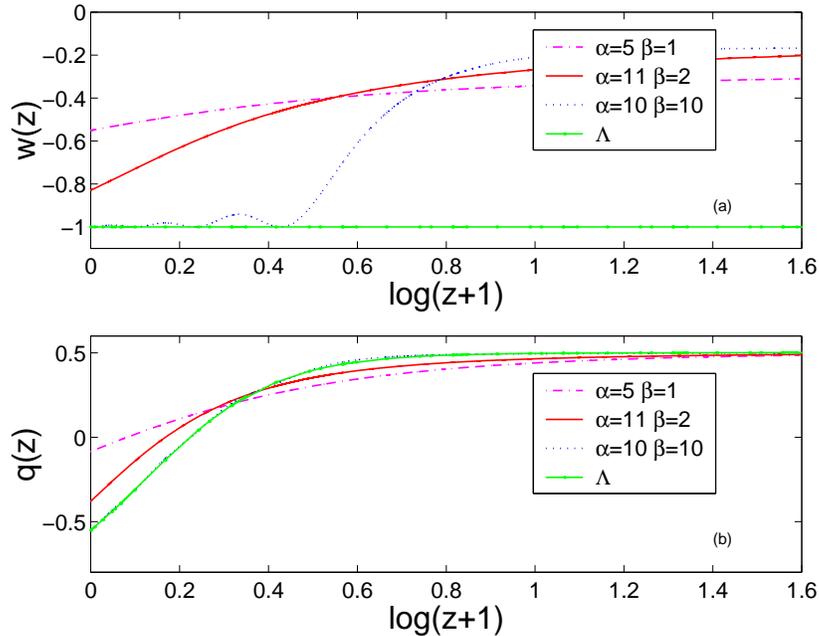}
}
\caption{In (a) the evolution of $w_{Q}$ against the red-shift  is plotted 
for different values of
$\alpha$ and $\beta$. In (b) the behaviour of the deceleration parameter, 
$q$, is plotted against the red-shift. The acceleration starts ($q<0$) earlier 
for models with an equation of state close to that of a true cosmological constant.}
\label{f:fig11}
\end{figure}

We have seen in Chapter~\ref{chap2} that for tracker models the present value
of the equation of state depends of the shape of the scalar field potential.
In particular for SUGRA inspired models \cite{NELSON,ALB,BRAX1,ED}
the equation of state parameter varies in the range $-1\leq w_{Q}^0<-0.8$,
while Inverse Power Law potentials
require larger values.
A general potential which can accommodate for these classes of scenarios is:

\begin{equation}
V(Q)=\frac{M^{4+\alpha}}{Q^{\alpha}}e^{\frac{1}{2}(\kappa Q)^{\beta}},
\label{potenziale}
\end{equation}
where $\kappa =\sqrt{8\pi G}$ and $M$ is fixed in such a way that today $\rho
_{Q}=\rho_{c}\Omega_{Q}$, where $\rho_{c}$ is the critical energy density.
For $\beta=0$ Eq.(\ref{potenziale})
becomes an inverse power law, while for $\beta=2$ we have the 
SUGRA potential proposed by \cite{BRAX1}. For $\alpha=0$, $\beta=1$ and
starting with a large value of $Q$, the Quintessence field evolves in
a pure exponential potential \cite{FERRE}.
We do not consider this case further since it is possible
to have a dark energy dominated universe, but at the expense of fine tuning for
the initial conditions of the scalar field. Larger values of $\beta$ mimic the
late time evolution of the model studied in \cite{ED}. For $\alpha,\beta\neq0$
the potential has a minimum, the dynamics can be summarized as follows:
for small values of $\beta$ and for a large range of initial conditions,
the field does not reach the minimum by the present time and hence $w_{Q}^0>-1$.
For example, if the Quintessence energy density initially dominates over the
radiation, the $Q$ field quickly rolls down the inverse power law part of
the potential eventually resting in the minimum with $w_{Q}\sim-1$ 
after a series of damped oscillations \cite{RIAZ}.
This behaviour however requires fine tuning the initial value of $Q$ to be small. On the other hand, this can be avoided if we consider large 
values of $\alpha$ and $\beta$ (figure~\ref{f:fig11}a).
In these models the fractional energy density of the quintessence field, 
$\Omega_{Q}$, is always negligible during both radiation and matter
dominated eras. In fact, for small initial values of $Q$, $V(Q)$ acts like an inverse 
power law potential, hence as $Q$ enters the scaling regime its energy
density is subdominant compared to that of the background component.
Therefore nucleosynthesis constraints \cite{RACHEL} are always 
satisfied and there are no physical effects on the evolution of the density perturbations. The main consequence is that for a different value of $w_{Q}^0$ the
Universe starts to accelerate at a different redshift (figure~\ref{f:fig11}b).
This implies that  different values of $\alpha$ and $\beta$ lead to a 
different luminosity distance and angular diameter distance.
Consequently by making use of the observed distances we may in principle 
determine an upper limit on $w_{Q}^0$, potentially constraining the allowed shape 
of the quintessence potential \cite{HUT}.

\section{CMB peaks}
The CMB power spectrum provides information on combinations of fundamental
cosmological parameters. The physical processes responsible for the
anisotropy are well understood allowing us to accurately predict the shape
of the anisotropy power spectrum for a given cosmological model.
The presence of acoustic oscillations of the photon-baryon plasma before the 
recombination epoch establishes a multiple peaks pattern in the power spectrum.
For a review of the subject we refer to \cite{HU1}.
During the radiation dominated era the equation 
describing the effective temperature fluctuation
$\Delta T$ of the CMB is of the form \cite{HU2}:
\begin{equation}
\Delta T''+c_s^2 \Delta T =2\Psi'',
\label{dt}
\end{equation}  
where $\Psi$ is the gravitational potential sourced by the energy density
perturbations in the different matter components, 
$c_s^2=1/3(1+3\rho_b/\rho_r)$ is the sound speed of the barotropic fluid and the primes are
derivatives with respect to $k\tau$,
$\tau$ being the conformal time and $k$ the comoving wavenumber.
Eq.~(\ref{dt}) has the form of a perturbed harmonic oscillator, therefore
when the photons decouple from the baryons their
energy carries an imprint of such oscillations.
The characteristic frequency of these oscillations is fixed by the size of 
the sound horizon at the decoupling $r_{sh}=\int c_s d\tau$. Therefore we 
have a series of compressions and rarefactions at scales $k_m r_{sh}=m\pi$.
Today such scales appear at angles that are 
multiple integers of the angular size of the sound horizon at 
the decoupling $\theta_{sh}=r_{sh}/D_{\kappa}(\tau_{ls})$, where
$D_{\kappa}(\tau_{ls})$ is the distance to the last scattering surface
for a space-time with curvature $\kappa$.
As a consequence of these the position of the `Doppler'
peaks in the power spectrum depends on the geometry of the Universe.
For a flat Universe the peaks will appear at the multipoles
\begin{equation}
l_m=m l_{sh}=m \frac{\pi}{\bar{c}_{s}}\left(\frac{\tau_{0}}{\tau_{ls}}-1\right),
\label{lharm}
\end{equation}
where $\bar{c}_{s}$ is the mean sound speed and $\tau_{0}$, $\tau_{ls}$ 
are the conformal time today and at last scattering respectively.
However the acoustic oscillations are perturbed by the evolution of the
gravitational potential $\Psi$ on the right-hand-side of Eq.~(\ref{dt}),
which shifts the position of the peaks by an amount that depends on 
the cosmological parameters that are relevant before the recombination.
This results in a better estimate for the peak positions being given by:
\begin{equation}
l_{m}=l_{sh}(m-\delta l-\delta l_{m}),
\label{peak}
\end{equation}
where $\delta l$ is an overall shift \cite{HU2} and 
$\delta l_{m}$
is the shift of the $m$-th peak. These corrections depend on the 
amount of baryons $\Omega_{b}h^{2}$, on the fractional quintessence energy density
at last scattering ($\Omega_{Q}^{ls}$) and today ($\Omega_{Q}^{0}$),
as well as on the scalar spectral index $n$.
Analytic formulae, valid over a 
large range of the cosmological parameters,
have been provided to good accuracy for 
$\delta l$ and $\delta l_{m}$ \cite{DOR2}.
Of crucial importance is the observation that
the position of the third peak appears to
remain insensitive to other
cosmological quantities. This is because at small scales, well inside the horizon,
$\Psi$ is usually negligible, hence we recover the unperturbed
harmonic oscillator equation for Eq.~(\ref{dt}) that
describes oscillations with zero point offset. 
Hence we can make use of this fact to test dark energy 
models \cite{DOR1,DOR3}. As we shall discuss in more detail in Chapter~\ref{chap6},
the quintessence field can leave a distinctive signature on the shape
of the CMB power spectrum through both the early integrated Sachs-Wolfe effect (ISW)
and the late one \cite{HU3}.
The former is important if
the dark energy contribution at the last scattering surface 
is not negligible \cite{NELSON,ALB} 
or in non-minimally coupled models \cite{LUCA1,PER,BAC2}, whereas
the late ISW is the only effect in models with $\Omega_{Q}^{ls}\sim0$ such as those
described by our parameterized quintessence potential 
\cite{BRAX4}. However the late ISW produces an imprint on the CMB power spectrum of
the order of $10\%$ and therefore is not detectable with the pre-WMAP measurements.   
In such a case an accurate determination of the position of the Doppler peaks is more
sensitive to the actual amount of dark energy.

\section{Likelihood analysis and results}

\subsection{Constraints from supernovae}

\begin{figure}[t]
\centerline{
\includegraphics[scale=0.6]{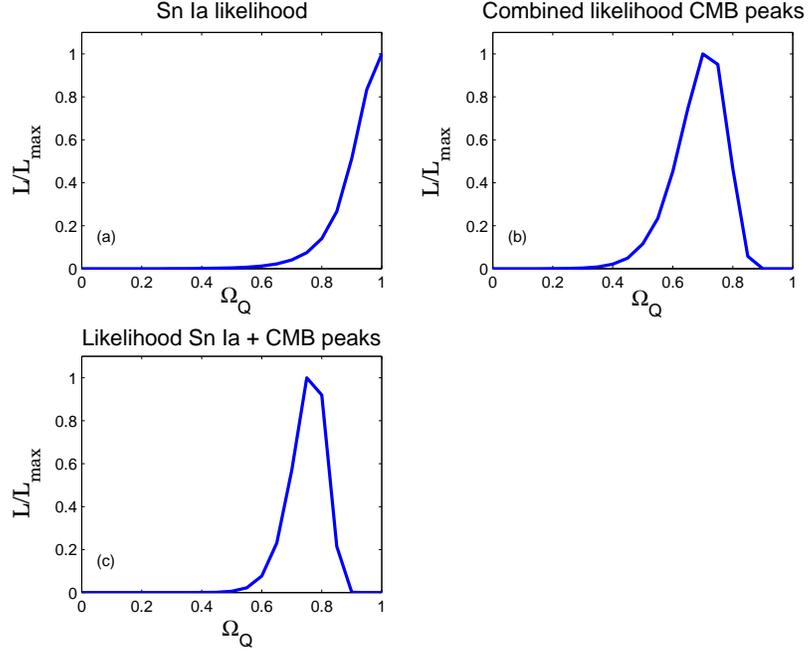}
}
\caption{ Fractional Quintessence energy density likelihoods, (a) for Sn Ia,
(b) for the combined CMB peaks and (c) for the combined 
data sets.}
\label{f:fig13}
\end{figure}

\begin{figure}[ht]
\centerline{
\includegraphics[scale=0.6]{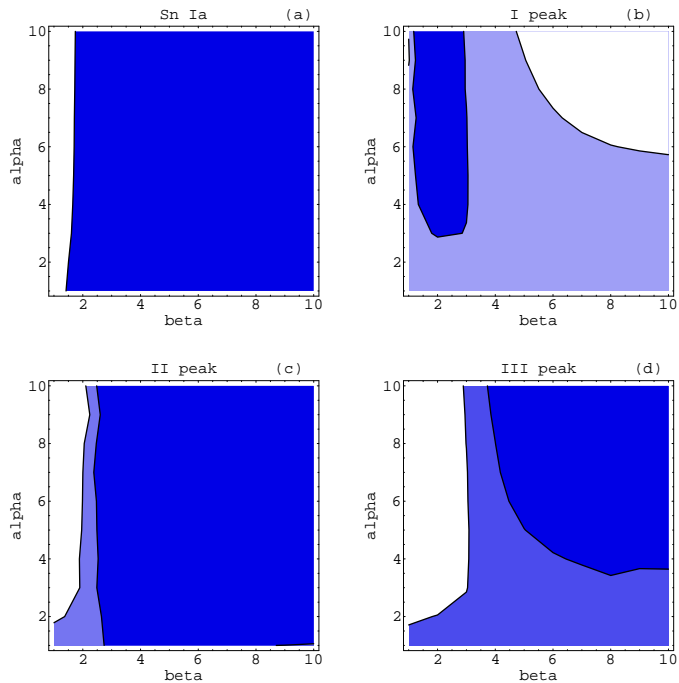}
}
\caption{ Likelihood contour plots  for Sn Ia, I, II and III acoustic peaks.
The blue region is the $68\%$ confidence region while the $90\%$ is
the light blue one. For the Sn Ia the white region correspond to $2\sigma$.
The position of the third CMB acoustic peak strongly constrains the acceptable parameter
space.}
\label{f:fig14}
\end{figure}

We want to constrain the set of parameters $\alpha$, $\beta$ and $\Omega_{Q}$
confined in the range: $\alpha\in(1,10)$, $\beta\in(0,10)$ and $\Omega_{Q}\in(0,1)$,
subject to the assumption of a flat universe. This choice of the parameters
allows us to account for a large number of models.
We use the Sn Ia data fit C of Perlmutter \textsl{et al.} (1999) \cite{PEL2},
that excludes 4 high redshift data points, which are very likely reddened by
their host galaxies. The magnitude-redshift relation is given by:
\begin{equation}
m(z)= 5\log D_{L}(z,\alpha,\beta,\Omega_{Q})+\mathcal{M},
\end{equation}
where $\mathcal{M}$ is the `Hubble constant free' absolute magnitude and
 $ D_{L}(z)=H_{0} d_{L}(z) $ is the free-Hubble constant luminosity
distance. In a flat universe
\begin{equation}
d_{L}(z)=(\tau_{0}-\tau(z))(1+z),
\label{dist}
\end{equation}
where $\tau_{0}$ is the conformal time today and $\tau(z)$ is the conformal
 time at the red-shift $z$ of the observed supernova.
Both of these quantities are calculated solving numerically
Eq.~(\ref{klein}) and Eq.~(\ref{friedmann})
for each value of $\alpha,\beta$ and $\Omega_{Q}$.
In $\mathcal{M}$ we neglect the dependence on a fifth parameter
($\alpha$ in \cite{PEL2}) and assume it to be 0.6, the best fit value in \cite{PEL2}. 
We then obtain a gaussian likelihood
function $\mathcal{L}^{Sn}(\alpha,\beta,\Omega_{Q})$, by 
marginalizing over
$\mathcal{M}$. In figure \ref{f:fig13}a we present the one-dimensional likelihood 
function normalized to
its maximum value for $\Omega_{Q}$. When considering only Sn Ia data,
 there is a maximum at $\Omega_{Q}=1$, such a high
value of $\Omega_{Q}$ is required if we impose the constraint $\Omega_k=0$ and it is
in agreement with the analysis in \cite{EFS2}. In figure \ref{f:fig14}a we present
the likelihood contours in the
$\alpha - \beta$ parameter space, 
obtained after marginalizing over $\Omega_{Q}$. 
Note that all values are allowed at the $2\sigma$ level.
The confidence regions for the Sn Ia data correspond to Quintessence models
with $w_{Q}^0<-0.4$ for $\Omega_{Q}=0.6$, an upper
limit that agrees with those found in \cite{PEL2,EFST}.

\subsection{Constraints from Doppler peaks and Sn Ia}

We now compute the position of the three Doppler peaks
$l_{1},l_{2}$ and $l_{3}$ using Eq.~(\ref{peak}). In addition
to the parameter space used in the supernovae analysis we consider 
the physical baryon density and the scalar spectral index varying respectively in
the range $\Omega_{b}h^2\in(0.018,0.026)$ and $n\in(0.9,1.1)$. 
The Hubble constant is set to $h=0.70$ in agreement with the recent 
HST observations \cite{FREE}. It is worth remarking that the baryons density
and $H_0$ are degenerate with the quintessence parameters.
In fact increasing $\Omega_{b}h^2$ reduces
the value of $c_s$ causing a shift of the CMB peaks towards large multipoles.
The same effect occurs for low values of the Hubble constant, therefore the results
of this analysis will depend on the HST prior.  
The predicted peak multipoles are then compared with
those measured in the Boomerang and DASI spectra \cite{DEB}.
Note, that the third peak has been detected in the Boomerang data
but not in the DASI data. In table~\ref{data} we report the position
of the peaks with $1\sigma$ errors estimated from the
Boomerang, DASI and WMAP data with a model independent
analysis. Since the errors are non Gaussian (see also \cite{ruthd}),
to be conservative we take our $1\sigma$ errors on the data 
to be larger than those reported in \cite{DEB}, so that our analysis
is significant up to $2\sigma$. 
\begin{table}[t]
\centerline{
\begin{tabular}{c c c c}
\hline\hline
Peak & Boomerang & DASI& WMAP \\
\hline \hline
$l_1$  & $213^{+10}_{-13}$ & $202 \pm 15$ & $220.1\pm 0.8$ \\ \\ 
$l_2$  & $541^{+20}_{-32}$ & $548 \pm 10$ & $546\pm 10$\\ \\
$l_3$  & $845^{+12}_{-25}$ & - & -
\end{tabular}
}
\caption{Location of the CMB peaks inferred from the power spectrum of Boomerang and MAXIMA data
using a model independent procedure (from \cite{DEB}).}
\label{data}
\end{table}
We then evaluate a gaussian likelihood function
$\mathcal{L}^{Peaks}(\alpha,\beta,\Omega_{Q},\Omega_{b}h^2,n)$.
The combined one-dimensional likelihood function for the peaks is shown
in figure \ref{f:fig13}b, where we find $\Omega_{Q}=0.69\pm^{0.13}_{0.10}$.
The likelihood for all the data sets combined is shown in figure \ref{f:fig13}c,
where we find $\Omega_{Q}=0.75\pm^{0.09}_{0.08}$. These results are in agreement
with the analysis in \cite{EFS,NETTE,BAC1}.
\begin{figure}[ht]
\centerline{
\includegraphics[scale=0.6]{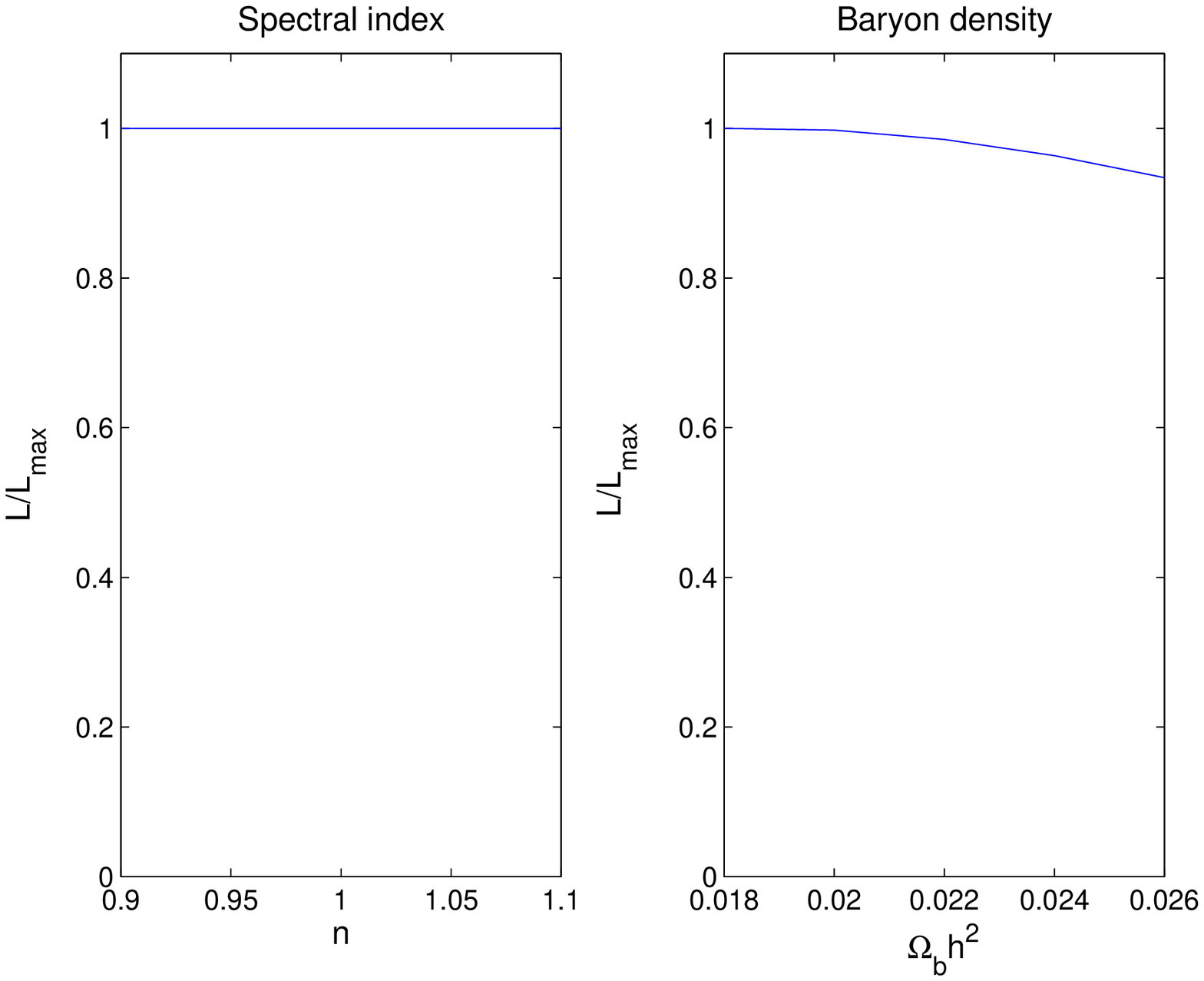}
}
\caption{ One-dimensional likelihood for $n$ and $\Omega_{b}h^2$.}
\label{f:fig15}
\end{figure}
\begin{figure}[ht]
\centerline{
\includegraphics[scale=0.6]{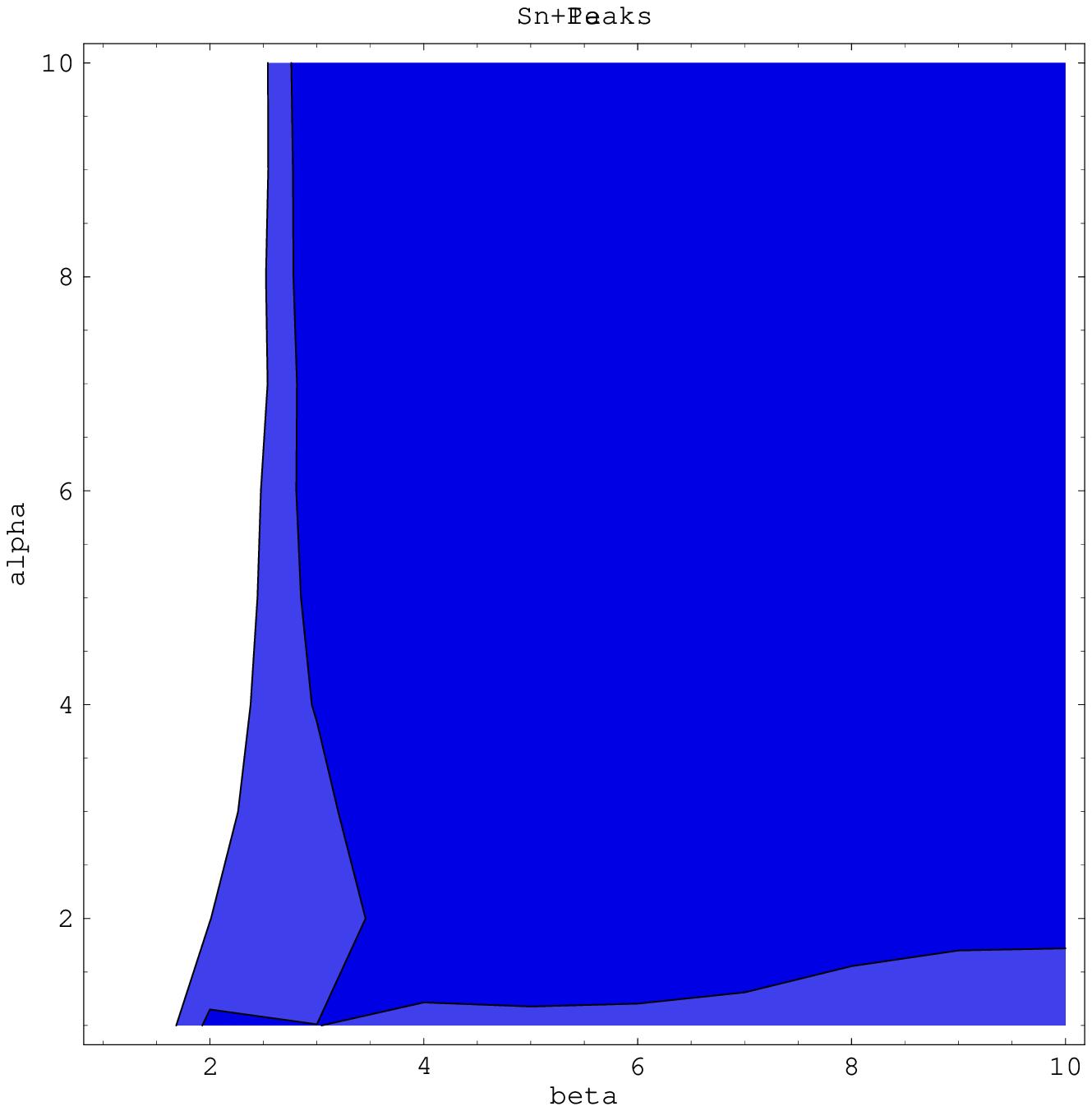}
}
\caption{ Two-dimensional likelihood for Sn Ia and CMB with $1$ (dark blue) and $2\sigma$
 (light blue) contours.}
\label{f:fig16}
\end{figure}
The likelihood functions, combining all the CMB peaks data,
for the scalar spectral index and the physical baryon density 
are shown in figure \ref{f:fig15}.
Since the dependence of the peak multipoles on $\Omega_{b}h^2$
and $n$ is weak, it is not possible to obtain some significant
constraints on these cosmological parameters.
In figure \ref{f:fig14}b-\ref{f:fig14}d we plot the 
two-dimensional likelihood function in the plane $\alpha-\beta$ for each peak,
obtained after having marginalized over $\Omega_{Q}$, $\Omega_{b}h^2$ and $n$. 
Their shape reflects the accuracy in the estimation of the position of the peaks.
Actually the first one is very well resolved, while we are less 
confident with the location of the second and third peak. 
Therefore their likelihoods are more spread and flat in the $\alpha-\beta$ plane. 
The $1\sigma$ confidence contour (figure~\ref{f:fig14}b)
for the first acoustic peak constrains the slopes of our potential in the range:
$3 \leq \alpha \leq 10$ and $1 \leq \beta \leq 3$. In particular the likelihood has
a maximum at $\alpha=9$ and $\beta=2$, corresponding to an equation of state
$w_{Q}^0=-0.8$ for $\Omega_{Q}=0.7$, in agreement with the recent analysis in \cite{BAC1}.
However, the second and third peaks constrain a complementary region where
the equation of state is compatible with the cosmological constant value.
Therefore the effect of including all the data in the likelihood analysis is
to move the constraint from models with $w_{Q}^0 \sim -0.8$ to models with an
equation of state $w_{Q}^0\sim-1$. The constraints on the slopes of the quintessence
potential allows us to infer an upper limit on the present value of the 
equation of state only after having fixed the value of $\Omega_Q$. In fact the 
scalar field potential Eq.~(\ref{potenziale}) is fully specified by three numbers that are 
$M,\alpha,\beta$. However the map $(M,\alpha,\beta)\rightarrow w_Q^0$ is not
one-to-one and therefore it is not possible to transform the likelihood 
$\mathcal{L}(\alpha,\beta,\Omega_{Q})\rightarrow\mathcal{L}(w_Q^0,\Omega_{Q})$.
Therefore there is a substantial difference between inferring the bounds on the dark
energy equation of state by directly constraining its present value
and constraining the scalar field potential. 
As we can see in figure~\ref{f:fig17} the values of $\alpha$ and $\beta$, allowed by
the likelihood including all the data (figure~\ref{f:fig16}), correspond to our models with values
of $w_{Q}^0$ in the range $-1\leq w_{Q}^0 \leq 0.93$ at $2\sigma$ for our
prior probability $\Omega_{Q}=0.75$. Assuming smaller values of $\Omega_Q$ relaxes
the constraints on $w_Q^0$, for instance for $\Omega_Q=0.5$ we have
$-1 \leq w_Q^0 \leq -0.6$ at $2\sigma$. The results of our analysis can be interpreted
as follows.  At the $1\sigma$ level the position of the first peak is 
inconsistent with the position of the other two. A possible explanation of this discrepancy
is that the multipoles $l_{2}$ and $l_{3}$  are less sensitive to small shifts 
induced by the dependence on $\Omega_{b}h^2$ and $n$.
Therefore we can obtain a different constraint
on the dark energy equation of state if we consider the peaks individually.
It is worth remarking that
the position of the first peak in the Boomerang data prefers slightly spatially closed
models. Having assumed a flat geometry affects our upper bounds on the slope
of the potential in a region corresponding to $w_Q^0>-1$.
On the other hand the location of the third peak
inferred in \cite{DEB}, is at $l_{3}=845\pm^{12}_{23}$, for which
values of $w_{Q}^0\sim-1$ fit this multipole better than models with $w_{Q}^0>-1$.
In fact the peaks are
shifted toward larger multipoles as $w_{Q}^0$ approaches the 
cosmological constant value. This is because, in models with $w_{Q}^0\sim-1$
the Universe starts to accelerate earlier than in those with $w_{Q}^0>-1$, consequently the 
distance to the last scattering surface is farther and hence the sound horizon at the 
decoupling is projected onto smaller angular scales. 
\begin{figure}[ht]
\centerline{
\includegraphics[scale=0.6]{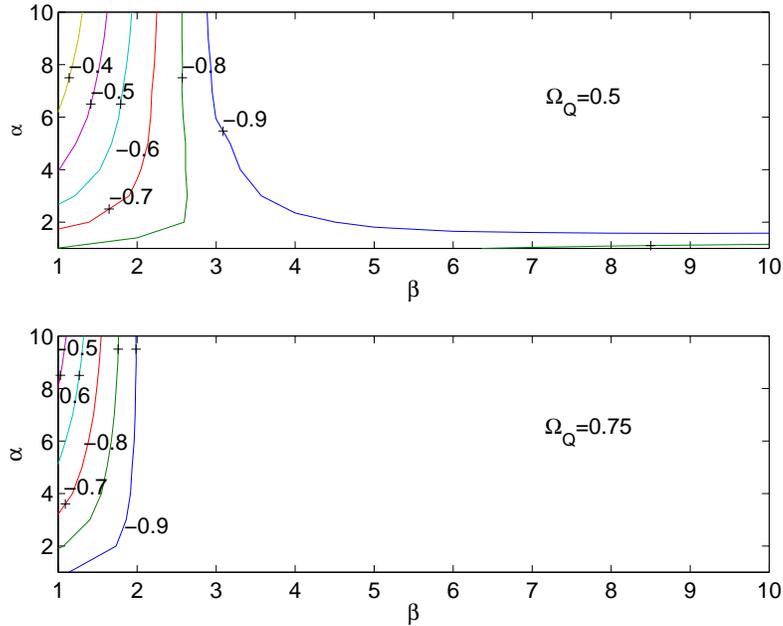}
}
\caption{ Contour values of $w_Q^0$ in the $\alpha-\beta$ plane for $\Omega_Q=0.5$ (upper panel)
and $\Omega_Q=0.75$ (upper panel).}
\label{f:fig17}
\end{figure}
A possible source of bias in our analysis is due to the prior on the value of $H_0$,
in fact assuming $h=0.70$ implies that models with a large negative equation of state
and large value of $\Omega_Q$ are preferred. However our results are consistent with
other studies which assume similar priors. For instance in \cite{BAC1}, using the complete set of 
available CMB measurements the slope of
the Inverse Power Law potential has been constrained to be $\alpha=0.8\pm_{0.5}^{0.6}$
under the prior $h=0.65$, corresponding to $w_Q^0=-0.82\pm_{0.11}^{0.14}$. Similarly in 
\cite{BEAN2}, by making use of the CMB, Sn Ia and large scale structure data,
it has been found $w_Q\leq-0.85$ for a constant equation of state. A time dependent dark energy
equation of state, characterized by a late time transition, has been constrained using
all cosmological data in \cite{BRUCE},
where it has been found $w_Q^0=-1^{+0.2}$.

\section{Discussion}
The analysis of Sn Ia data and the location of the CMB peaks in Boomerang and
DASI data constrain the slope of a general quintessence potential in a range of values
such that the quintessence field is undergoing small damped 
oscillations around a minimum of the potential (see figure~\ref{f:fig11}). 
Such a situation implies that the
value of the quintessence equation of state has to be very close to $-1$. However
this does not imply that the dark energy is the cosmological constant. What our result
means is that the quintessence field today is evolving in a nearly vacuum state.
In fact models where the scalar field is rolling down a very flat region of the potential
can fit the data and pass the constraints even though they are not included in our analysis.
Another important caveat is that this study does not take into account quintessence
scenarios where the contribution of the dark energy is not negligible. In such
a case we would have to take into account physical effects not only on distance 
measurements, but on the structure formation process itself
(see for instance \cite{DOR4,DOR5}). These effects however are not detectable
with the level of accuracy provided by the recent balloon and ground CMB experiments.
A revised analysis of the Boomerang data has been recently published in \cite{RUHL},
the position of the peaks has been determined with a better accuracy, in particular
the third peak is shifted at a smaller multipole than previously detected.
This slightly changes our results, relaxing the lower bounds on the slope $\beta$.
Here we want to stress the fact that it seems likely the present value of the
dark energy equation of state is close to $-1$.
However the possibility that the equation of state was largely different from $-1$
cannot be excluded, as it is shown by the class of models that
best fit the Sn Ia data and the CMB peaks. The new generation of satellite experiments
we will probably be able to detect the specific effects produced by a time varying
dark energy component.

\chapter{A model independent approach to the dark energy equation of state} 
\label{chap5}

The detection of time variation in the dark energy equation of state 
can be considered as the `smoking gun' for the cosmological constant scenario.
The new generation of experiments in cosmology will provide
high precision measurements that in principle can discriminate
between different dark energy candidates.
On the other hand it is unrealistic to assume we can infer
some general information constraining some  particular quintessence model.
In fact a plethora of scalar field potentials have been proposed, all leading 
to similar late time behaviour of the universe, hence to a lack of
predictability. In this Chapter, we first review some
of the methods that have been proposed in the literature
to constrain the time features of a general quintessence component.
We then introduce a model independent approach which
simply involves parameterizing the dark energy equation of
state in terms of known observables \cite{CORAS2}. 
This allows us to analyse the impact dark energy has had on cosmology
without the need to refer to particular scalar field models and opens
up the possibility that future experiments will be able to constrain 
the dark energy equation of state in a model independent manner.

\section{The effective equation of state}
Measuring the present value of the dark energy equation of state
can distinguish between a cosmological constant scenario and a quintessential
component if $w_Q^0>-1$. Using the tracker properties of the quintessence
field, the most reliable method to infer a bound on $w_Q^0$ would
be to constrain directly the quintessence potential. However
this could result in a difficult challenge, for instance
in Table~\ref{potlist} we show a list of proposed tracker
quintessence potentials. Therefore, because there are
no fundamental physical principles that can specify the nature of the
dark energy, we are left with a potentially infinite set of families of 
theoretical models to compare with the data.
\begin{table}
\centerline{
\begin{tabular}{c c c c}
$\Lambda^{4 + \alpha } / Q^\alpha$               & \cite{RATRA1} &
$\Lambda^4 e^{- \lambda Q}$                      & \cite{FERRE} \\ \\ 
$(\Lambda^{4 + \alpha } / Q^\alpha) e^{\frac{\kappa}{2} Q^2}$
                                                 & \cite{BRAX1} &
$\Lambda^4 (\cosh \lambda Q - 1)^p$              & \cite{SW} \\ \\ 
$\Lambda^4 (e^{\alpha \kappa Q} + e^{\beta \kappa Q})$
                                                 & \cite{NELSON} & 
$\Lambda^4 e^{- \lambda Q} (1 + A \sin \nu Q)$   & \cite{DKS} \\ \\ 
$\Lambda^4 [(Q - B)^\alpha + A] e^{- \lambda Q}$ & \cite{ALB} &
$\Lambda^4 [1 + \cos (Q / f)]$                   & \cite{FRIEMAN} \\ \\
$\Lambda^4\left(\frac{1}{Q^{\alpha}}+\frac{1}{Q^{\beta}}\right)$ & \cite{STEIN1} &
$\Lambda^4(e^{1/Q}-1)$                           & \cite{ZLATEV}
\end{tabular}
}
\caption{Quintessence potentials that have been used in the 
litterature.}
\label{potlist}
\end{table}
As a consequence of this it is to be hoped that a more general
way of constraining the dark energy can be developed.
A simple method to take into account a large number of dark energy models is
to consider a constant effective equation of state defined by Eq.~(\ref{weff}).
Using the energy conservation equation, the redshift evolution of
the dark energy density is given by
\begin{equation}
\rho_{de}(z)=\rho_{de}^0 (1+z)^{3(1+w_{eff})}.
\label{rhowe}
\end{equation}   
The effective equation of state appears through the Hubble equation in
the coordinate distance defined by:
\begin{equation}
r(z)=\frac{1}{H_0}\int^{z}_0 \frac{dz'}{\sqrt{\Omega_m(1+z')^3+\Omega_r(1+z')^4+\Omega_{de}(1+z')^{3(1+w_{eff})}}}.
\label{cosmodist}
\end{equation}
Hence the value of $w_{eff}$ can be constrained by measurements
of the luminosity distance and the angular diameter distance, that in a flat
space-time read as $d_L(z)=(1+z)r(z)$ and $d_A(z)=r(z)/(1+z)$ respectively.
There is a general consensus that an accurate determination of 
$d_L$ from high redshift Sn Ia measurements with the proposed SNAP
satellite \cite{LBL} can determine $w_{eff}$ to better than 5 per cent 
\cite{MAOR1}. In such a case if $w_{eff}$ is much larger than $-1$ it will
be possible to rule out the $\Lambda$CDM model. However if 
$w_{eff}\approx-1$ the results of the data analysis have to be
carefully considered. In fact
it has been shown by Maor {\em et al.} \cite{MAOR2} that
assuming a constant $w_{eff}$ introduces a bias that leads to
misleading conclusions about the properties of the dark energy.
In order to make this point clear, let us consider a sample of mock
data generated from a dark energy model with an equation of state
of the form
\begin{equation}
w_Q(z)=w_0+w_1 z,
\label{wz}
\end{equation}
with $w_0<-1/3$ and $w_1>0$.
The constraints on a constant equation of state $w_{eff}$ obtained from
the statistical analysis of such a sample, will push the bounds towards
values of $w_{eff}<w_0$. This effect has been noticed also in other
works, see for instance \cite{WELLA,GERK} and can be explained as follows.
For a model with $w_Q$ given by Eq.~(\ref{wz}) the energy density evolves as:
\begin{equation}
\rho_{de}(z)=\rho_{de}^0 (1+z)^{3(w_0-w_1+1)}exp\left[\frac{3 w_1 z}{1+z}\right].
\label{rhowz}
\end{equation}  
As we may note, comparing Eq.~(\ref{rhowe}) with Eq.~(\ref{rhowz}),
if $w_0<0$ and $w_1>0$, fitting the data with a constant equation of state
roughly requires $w_{eff}\approx w_0-w_1<w_0$. In other words the 
fitting procedure will pick out values of $w_0$ that are more negative
than the fiducial one. Hence a time varying dark energy component
can be confused with a cosmological constant, while
allowing $w_{eff}$ to vary in a range of values more negative
than -1, will shift the best fit models towards $w_{eff}<-1$.
Therefore it is very possible that the claims 
for the existence of {\em phantom} matter (with $w_Q<-1$)
\cite{HANN,TROD} are just a
consequence of assuming a constant equation of state
in the data analysis.

\section{Cosmological distance fitting functions}
Several methods to constrain the time dependence of the dark energy
equation of state have been discussed in the
literature \cite{SAINI,HUT1,EFS,CHIBA,NAKA,ASTIER,WELLA,GOL,WELLB,GERK}.
Since the dark energy became dominant only recently,
it is reasonable  to concentrate on local redshift measurements,
such as the Sn Ia.
The coordinate distance and the Hubble
parameter are uniquely related by $H(z)=(dr/dz)^{-1}$.
Thus we can estimate $r(z)$ from cosmological distance indicators 
and unambiguously calculate $H(z)$ and $dH/dz$. This allows
us to reconstruct $w(z)$ provided the value of $\Omega_m$
is given.
In fact the Hubble equations in the presence of matter and a scalar field are:
\begin{equation}
H^2=\frac{8\pi G}{3}\left[\rho_m+\frac{1}{2}\dot{Q}^2+V(Q) \right]
\end{equation}
and
\begin{equation}
\dot{H}=-4\pi G (\rho_m+\dot{Q}^2),
\end{equation}
where the dot is the derivative with respect to the time.
These equations can be rewritten in the form:
\begin{equation}
\frac{8\pi G}{3 H_0^2}V(x)=\frac{H^2}{H_0^2}-\frac{x}{6H_0^2}
\left(\frac{dH}{dx}\right)^2-\frac{1}{2}\Omega_m x^3,
\end{equation}
\begin{equation}
\frac{8\pi G}{3 H_0^2}\left(\frac{dQ}{dx} \right)^2=\frac{2}{3H_0^2x}-\frac{\Omega_m x}{H^2},
\end{equation}
where $x=1+z$. Using the definition of the scalar field equation of state
we find:
\begin{equation}
w(z)=\frac{\frac{2x}{3}\frac{d ln H}{dx}-1}{1-\frac{H^2}{H_0^2}\Omega_m x^3}.
\label{eqrc}
\end{equation}
The idea is then to introduce a fitting function for the
coordinate distance, so that once its parameters have been determined
from the data analysis, its first and second derivatives 
with respect to the redshift can be 
analytically calculated. The authors of \cite{SAINI} suggest
\begin{equation}
r^{fit}(z)=\frac{2}{H_0}\left[\frac{x-\alpha\sqrt{x}-1+\alpha}{\beta x+\gamma\sqrt{x}+2-\alpha-\beta-\gamma} \right],
\end{equation}
where $\alpha$, $\beta$ and $\gamma$ are fitting parameters, while Huterer $\&$
Turner \cite{HUT1} consider both a polynomial fitting function
\begin{equation}
r^{fit}(z)=\sum_i c_i z^i
\end{equation}
and a Pade' approximate
\begin{equation}
r^{fit}(z)=\frac{z(1+a z)}{1+b z+c z^2}.
\end{equation}
It has been pointed out in \cite{GERK} that such a general method fails
to accurately reproduce the correct behaviour of given scalar field models,
even assuming strong priors on the value
of $\Omega_m$. The difficulty arises because the  
formula Eq.~(\ref{eqrc}) depends on the derivatives of $r(z)$.
Therefore even though
$r^{fit}(z)$ can mimic the $r(z)$ of a specific model, this may
be not true for its derivatives.

\section{{\em Statefinder} method}
A different approach has been developed by Sahni {\em et al.} 
\cite{VARUN1,VARUN2}. They suggest a method to distinguish amongst different dark energy
scenarios by measuring geometrical parameters that are constructed in such
a way they are more sensitive to
the expansion rate of the Universe. Such parameters, called {\em Statefinder}
pair, are defined by:
\begin{equation}
r=\frac{\ddot{a}}{aH^3}, s=\frac{r-1}{3(q-1/2)}.
\end{equation}
For a flat Universe the relation to the dark energy equation
of state $w$ and its time derivative $\dot{w}$ is:
\begin{equation}
r=1+\frac{9}{2}\Omega_{de}w(1+w)-\frac{3}{2}\Omega_{de}\frac{\dot{w}}{H},
\end{equation}
\begin{equation}
s=1+w-\frac{1}{3}\frac{\dot{w}}{w H},
\end{equation}
where $H$ is the Hubble parameter.
In general the value of such parameters evolves with the time, but 
for the cosmological constant ($w=-1$) we have $(r,s)=(1,0)$ at all the times. 
Moreover, since $(r,s)$ are geometrical quantities,
they can naturally take into account 
braneworld models in which the accelerated expansion
is caused by five dimensional gravity effects and which lead to
modifications of the Hubble
equations. 
The Statefinder pair can be estimated using the following ansatz for
the Hubble parameter,
\begin{equation}
H(x)=H_0 \left[\Omega_m x^3+A_1+A_2 x+ A_3 x^2\right]^{\frac{1}{2}},
\end{equation}
where $x=1+z$, and $A_1+A_2+A_3=1-\Omega_m$. In such a case $(r,s)$ are of the
form:
\begin{equation}
r=1-\frac{A_1+\Omega_m x^3}{\Omega_m x^3+A_1+A_2 x+A_3 x^2},
\label{r}
\end{equation}
\begin{equation}
s=\frac{2(A_2 x+A_3 x^2)}{3(3 A_1+2 A_2 x+A_3 x^2)}.
\label{s}
\end{equation}
The value of the fitting parameters $A_i$ can be inferred from 
a likelihood analysis of cosmological distance measurements, such as the Sn Ia.
From Eq.~(\ref{r})
it appears obvious that an accurate estimation of $r$ needs a precise
knowledge of the amount of non-relativistic matter.
In \cite{VARUN2} it has been
shown that imposing a Gaussian prior probability distribution on
$\Omega_m$ with variance $\sigma_{\Omega_m}=0.05$, 
the high redshift Sn Ia measurements provided by
the proposed SNAP satellite will accurately determine the Statefinder pair
offering the chance to distinguish between different dark energy models.
Therefore the Statefinder method seems to be highly sensitive to the nature
of dark energy and avoids the main difficulties of the reconstructing
procedure. In fact it only needs to constrain fitting parameters that
appear in the Hubble equation and does not constrain directly the
equation of state that affects the expansion rate apart from
through a time integrated effect.
A potential limitation of such an approach is that the determined values
of $(r,s)$ do not give us any information about the physical properties
of the dark energy, unless their values are {\em a priori} known for specific
classes of models. 

\section{Low redshift parameterization}
\begin{figure}[t]
\centerline{
\includegraphics[scale=0.6]{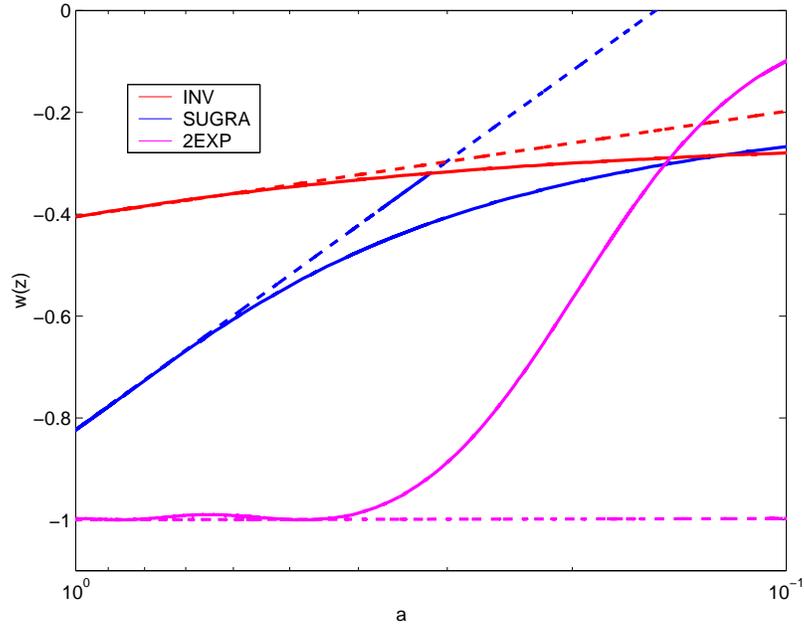}
}
\caption{Low redshift approximation to the quitenssence equation
of state for different types of potentials: Inverse Power Law (red),
Inverse Power Law times Exponential (blue) and Two Exponential (magenta).
The solid line corresponds to the exact numerical solution of the Klein-Gordon
equation, while the dash line is the approximated behaviour provided by Eq.(\ref{efsta}).}	
\label{f:fig18}			
\end{figure}

In the absence of theoretically motivated dark energy models we can
consider a low redshift expansion for $w_Q(z)$ and
the parameters of such an expansion can be interpreted as fundamental 
dark energy parameters. Therefore the constraints on a parameterization
of the equation of state would provide a model
independent determination of the dark energy properties.
Such a parameterization should satisfy the following
requirements:
\begin{itemize}
\item it should depend on a minimal number of parameters $\theta_i$
that specify the physical properties of the dark energy;
\item the functional form of the parameterization has to be such that for a 
given set $(\tilde{\theta_i})$
the behavior of the equation of state $w_Q(z,\tilde{\theta})$ reproduces 
the $w(z)$ predicted by proposed quintessence models and can account
also for more general cases.
\end{itemize}
In \cite{EFS,GERK} it was suggested the use
of a logarithmic expansion,
\begin{equation}
w_Q^p(z)=w_0-\alpha \log(1+z),
\label{efsta}
\end{equation} 
while the authors of \cite{WELLA,ASTIER,WELLB} used a polynomial fitting 
function,
\begin{equation}
w_Q^p=\sum_{i}c_i z^i.
\label{well}
\end{equation}
The general drawback of using a redshift expansion formula for the
dark energy equation of state is that the number of fitting
parameters is now larger than assuming a simple time constant behaviours
and consequently the degeneracy in the dark energy
parameter space is enlarged.
In fact the coordinate distance is related by a multi-integral expression to
the equation of state and therefore widely different $w(z,\theta_i)$
can have the same multi-integral value. There is a general consensus
that assuming strong priors on the value of $\Omega_m$ it will be possible to
tightly constrain $w_0=\theta_1$, but it will be more difficult 
to put tight bounds on the other parameters. The solution would be to
combine different data sets that can break the degeneracy in the
parameter space. The expansions specified by Eq.~(\ref{efsta}) and
Eq.~(\ref{well}) both have limited applicability.
For instance Eq.~(\ref{efsta}) can take into account
only for quintessence models with slowly varying behaviours and breaks
down at large redshifts, as we can see in figure~\ref{f:fig18}.
On the other hand the polynomial expansion Eq.~(\ref{well})
introduces a number of unphysical
parameters whose values are not directly related to the properties of a 
dark energy fluid. 
The consequence is that, unless we pay the cost of considering
very unphysical values, their application is limited to low
redshift measurements and cannot be extended to the
analysis of high redshift data sets such as the CMB data.
An interesting alternative to the fitting
expansion approach, has recently been proposed in \cite{STARK}.
The authors of this study developed a method to reconstruct
the time behavior of the equation of state
from cosmological distance measurements without
assuming the form of its parameterization. In spite of the
efficiency of such an approach, it does not take into account the
effects of the possible clustering properties of dark energy which
become manifest at higher redshifts. Hence its application has to
be limited to the effects dark energy can produce on the expansion
rate of the universe at low redshifts. On the other hand, it has
been argued that dark energy does not leave a detectable imprint
at higher redshifts, since it has only recently become the
dominant component of the universe. Such a statement, however, is
model dependent, on the face of it there is no reason why the dark
energy should be negligible deep in the matter dominated era. For
instance CMB observations constrain the dark energy density at
decoupling to be less then 10 per cent of the critical one
\cite{DOR5}. Such a non negligible contribution can be realized in
a large class of models and therefore cannot be {\em a priori}
excluded. Consequently it is of crucial importance to find an
appropriate parameterization for the dark energy equation of state
that allows us to take into account the full impact dark energy
has on different types of cosmological observations.

\section{An exact parameterization for the dark energy equation of state}
\begin{figure}[t]
\centerline{
\includegraphics[scale=0.6]{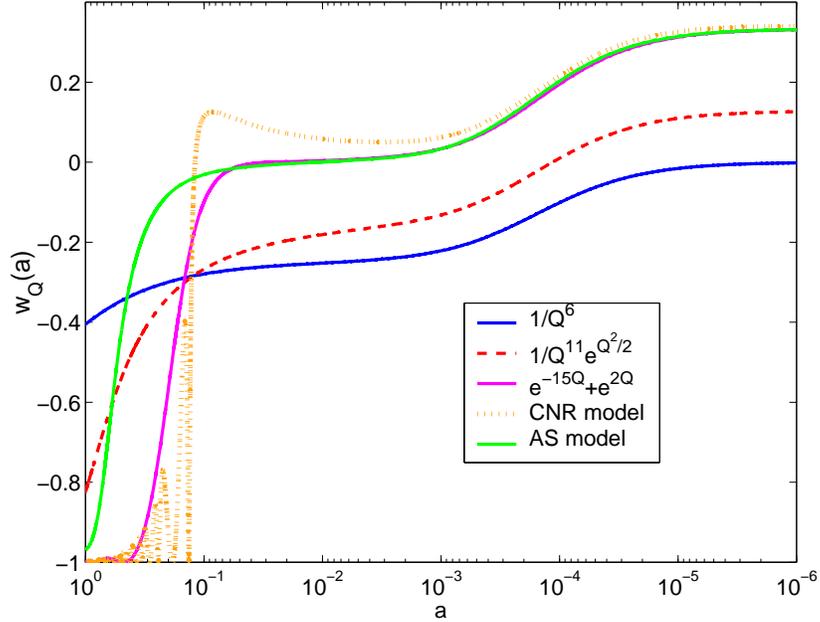}
}
\caption{Evolution of $w_{Q}$ against the scale factor
for an Inverse Power Law model (solid blue line),
SUGRA model (\cite{BRAX1}) (dash red line),
two exponential potential model (\cite{NELSON}) (solid magenta line),
AS model (\cite{ALB}) (solid green line) and
CNR model (\cite{ED}) (dot orange line).
}	
\label{f:fig19}			
\end{figure}

Our goal is to determine an appropriate analytical form of the 
equation of state $w_Q^p(a)$ valid at all redshifts in terms of
physical quantities, so that it can describe a general fluid and
take into account most of the proposed dark energy models. 
In Chapter 2 we have seen that the specific evolution of $w_Q(a)$,
depends on the shape of the potential,
however there are some common features in its behaviour that
can be described in a model independent manner and which allow us
to introduce some physical parameters. As we have seen in Chapter 2
a large number of quintessence models are
characterized by the existence of the 'tracker regime'.
It consists of a period during which the scalar field, while it is
approaching a late time attractor, evolves with an almost constant
equation of state whose value can track that of the background
component. Here we consider
a broad class of tracking potentials.
These include models for which $w_Q(a)$ evolves
smoothly, as with the inverse power law \cite{RATRA1}, $V(Q) \sim
1/Q^{\alpha}$ (INV) and the supergravity inspired potential
\cite{BRAX1}, $V(Q) \sim 1/Q^{\alpha} e^{Q^2/2}$ (SUGRA).
Late time rapidly varying equation of states arise in potentials with two
exponential functions \cite{NELSON}, $V \sim e^{- \alpha
Q}+e^{\beta Q}$ (2EXP), in the so called `Albrecht \& Skordis'
model \cite{ALB} (AS) and in the model proposed by Copeland et al.
\cite{ED} (CNR). To show this in more detail, in figure~\ref{f:fig19} we plot
the equation of state obtained by solving numerically for each of these
potentials the scalar field equation of motion defined by
Eq.~(\ref{klein}) and Eq.~(\ref{friedmann}).

There are some generic features that appear to be present, and
which we can make use of in our attempts to parameterize $w_Q$.
For a large range of initial conditions of the quintessence field,
the tracking phase starts before matter-radiation equality. In
such a scenario $w_Q(a)$ has three distinct phases, separated by
two `phase transitions'. Deep into both the radiation and matter
dominated eras the equation of state, $w_Q(a)$, takes on the
values $w_Q^r$ and $w_Q^m$ respectively, values that are related
to the equation of state of the background component $w_B$ through:
\begin{equation}
\ w_Q \approx \frac{w_B-2(\Gamma-1)}{2\Gamma-1},
\label{track}
\end{equation}
where $\Gamma=V''V/(V')^2$ and $V' \equiv dV/dQ$ etc.
For the case of an exponential potential, $\Gamma=1$,
with $w_Q=w_B$, but
in general $w_Q \neq w_B$. Therefore if we do not specify
the quintessence potential the values of
$w_Q^r$ and $w_Q^m$ should be considered as free parameters.
The two transition phases can each be described by two parameters;
the value of the scale factor $a_c^{r,m}$ when the equation of state $w_Q$
begins to change and the width $\Delta_{r,m}$ of the transition.
Since $\Gamma$ is constant or slowly varying during the tracker regime,
the transition from $w_Q^r$ to $w_Q^m$ is always
smooth and is similar for all the models (see figure~\ref{f:fig19}).
To be more precise, we have found
that $a_c^r\sim 10^{-5}$ and $\Delta_r\sim 10^{-4}$ during this transition,
the former number expected from the time of matter-radiation
equality and the latter from the transition period from radiation to
matter domination.
However, when considering the transition in $w_Q$ from $w_Q^m$ to
the present day value $w_Q^0$, we see from figure~\ref{f:fig19} that this can
be slow ($0 < a_c^{m}/ \Delta_m < 1$)
or rapid ($a_c^{m}/ \Delta_m > 1$) according to the slope of
the quintessence potential. For instance in models with a steep
slope followed by
a flat region or by a minimum, as in the case of the two exponentials,
the AS potential
or the CNR model, the scalar field evolves towards a solution that
approaches the late time attractor, finally deviating from the tracking
regime with the parameter $\Gamma$ rapidly varying.
In contrast the inverse power law potential always has a
slower transition since $\Gamma$ is constant for all times.
Given these general features we conclude that the behavior of $w_Q(a)$ can be
explained in terms of functions, $w_Q^p(a)$, involving the following parameters:
$w_Q^0$, $w_Q^m$, $w_Q^r$, $a_c^{m}$ and $\Delta_m$.
The authors of \cite{BRUCE} have recently used an
expansion in terms of a Fermi-Dirac function in order to constrain a class of
dark energy models with rapid late time
transitions.
\begin{figure}[t]
\centerline{
\includegraphics[scale=0.6]{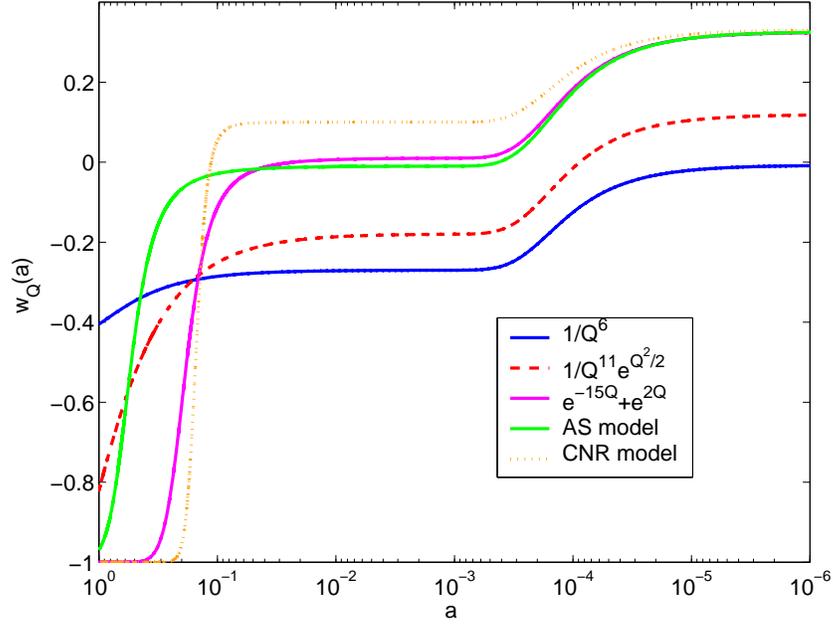}
}
\caption{Plot of $w_Q^{p}(a)$ (Eq.~\ref{exp}) best fit
for different potentials.
}	
\label{f:fig20}			
\end{figure}

In what follows we find that a generalisation of this involving
a linear combination of such functions allows for a wider range of models to be
investigated. To be more precise, we propose the following formula for $w_Q^p(a)$:
\begin{equation}
w_Q^{p}(a)=F_1f_r(a)+F_2f_m(a)+F_3,
\label{exp}
\end{equation}
with
\begin{equation}
f_{r,m}(a)=\frac{1}{1+e^{-\frac{a-a_c^{r,m}}{\Delta_{r,m}}}}.
\end{equation}
The coefficients $F_1$, $F_2$ and $F_3$ are determined by
demanding that $w_Q^{p}(a)$ takes on the respective values
$w_Q^r$, $w_Q^m$, $w_Q^0$ during radiation ($a_r$) and matter
($a_m$) domination as well as today ($a_0$). Solving the algebraic
equations that follow we have:
\begin{eqnarray}
F_1&=&\frac{( w_Q^m-w_Q^r)\left( f_m(a_0)-f_m(a_r)
\right)-
 ( w_Q^0-w_Q^r)\left( f_m(a_m)-f_m(a_r) \right)}
 {\left( f_r(a_m)-f_r(a_r) \right)\left( f_m(a_0)-f_m(a_r) \right)-
\left( f_r(a_0)-f_r(a_r) \right)\left( f_m(a_m)-f_m(a_r)
\right)}, \nonumber \\ \\
F_2&=&\frac{ w_Q^0-w_Q^r }{ f_m(a_0)-f_m(a_r)}-F_1
\frac{f_r(a_0)-f_r(a_r)}{f_m(a_0)-f_m(a_r)}, \\ \nonumber\\
F_3&=&w_Q^r-F_1f_r(a_r)-F_2f_m(a_r),
\end{eqnarray}
where $a_0=1$, and the value of and $a_r$ and $a_m$ can be
arbitrarily chosen in the radiation and matter era because of the
almost constant nature of $w_Q$ during those eras. For example in
our simulations we assumed $a_r=10^{-5}$ and $a_m= 10^{-3}$. In
table~\ref{fitpar} we present the best fit parameters obtained by minimizing a
chi-square for the models we have considered and
in figure~\ref{f:fig20} we plot
the associated fitting functions $w_Q^{p}(a)$. It is encouraging to see
how accurately the Fermi-Dirac functions mimic the exact time
behavior of $w_Q(a)$ for the majority of the potentials.
\begin{table}[t]
\centerline{
\begin{tabular}{cccccccc}
 &$w_Q^0$ &$w_Q^m$ &$w_Q^r$& $a_c^m$ &$\Delta_m$\\
\hline
INV& -0.40 & -0.27 & -0.02 & 0.18 & 0.5 \\
SUGRA& -0.82 & -0.18 & 0.10 & 0.1 & 0.7 \\
2EXP& -1 & 0.01 & 0.31 & 0.19 & 0.043 \\
AS& -0.96 & -0.01 & 0.31 & 0.53 & 0.13 \\
CNR& -1.0 & 0.1 & 0.32 & 0.15 & 0.016 \\
\end{tabular}
}
\caption{\label{tab:table1} Best fit values of the parameters of the
expansion ~(\ref{exp}).}
\label{fitpar}
\end{table} 

\begin{figure}[t]
\centerline{
\includegraphics[scale=0.6]{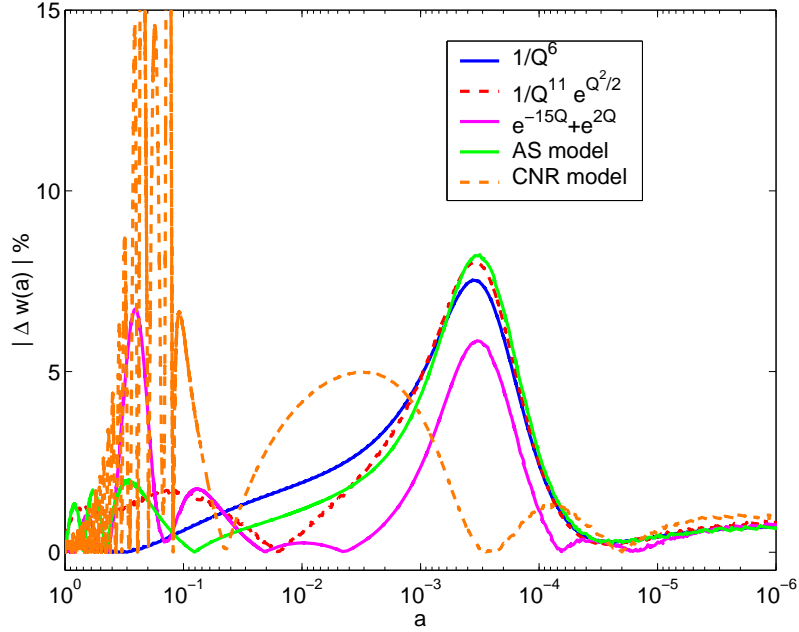}
}
\caption{Absolute value of the difference
 between $w_Q(a)$ and $w_Q^{p}(a)$ for the models of figure~\ref{f:fig19}.}
\label{f:fig21}			
\end{figure}

\begin{figure}[t]
\centerline{
\includegraphics[scale=0.6]{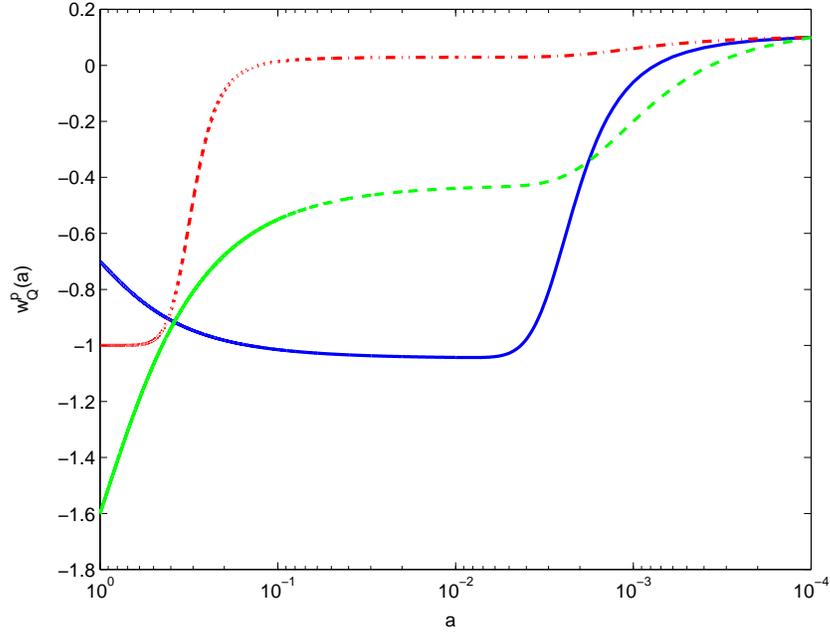}
}
\caption{Time evolution of $w_Q^p(a)$ as in the case
of K-essence (blue solid line), late time transition (red dash-dot line)
end with $w_Q^o<-1$ (green dash line).}	
\label{f:fig22}			
\end{figure}

\begin{figure}[b]
\centerline{
\includegraphics[scale=0.6]{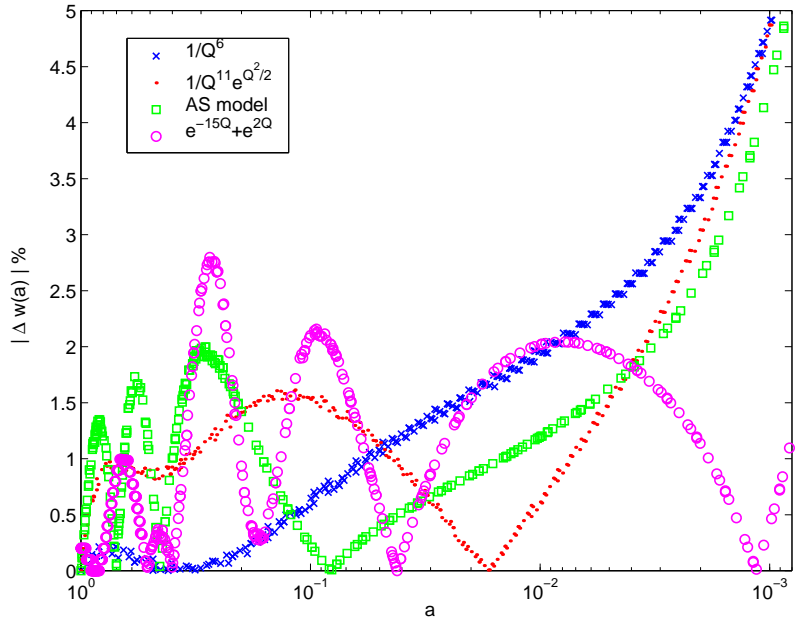}
}
\caption{Absolute value of the difference
 between $w_Q(a)$ and the low redshift formula Eq.~(\ref{lowz})
 for the models of figure~\ref{f:fig19}.}
\label{f:fig23}			
\end{figure}

In figure~\ref{f:fig21} we plot the absolute value of 
the difference $\Delta w(a)$ between
$w_Q(a)$ and $w_Q^{p}(a)$. The discrepancy is less then $1\%$ for
redshifts $z<10$ where the energy density of the dark energy can
produce observable effects in these class of models and it remains
below $9\%$ between decoupling and matter-radiation equality. Only
the CNR case is not accurately described by $w_Q^{p}(a)$ due to
the high frequency oscillations of the scalar field which occur at
low redshift as it fluctuates around the minimum of its potential.
In fact these oscillations are not detectable, rather it is the
time-average of $w_Q(a)$ which is seen in the cosmological
observables, and can be described by the corresponding
$w_Q^{p}(a)$. There are a number of impressive features that can
be associated with the use of $w_Q^p(a)$ in Eq.~(\ref{exp}).
For instance it can mimic the behavior of more general models.
As an example of this in figure~\ref{f:fig22} 
we plot $w_Q^{p}(a)$ that approximately corresponds to
three cases: a K-essence model \cite{MUKA} (blue solid line); a
rapid late time transition \cite{PARKER} (red dash-dot line) and
finally one with an equation of state $w_Q^0 < -1$ (green dash
line). 
The observational constraints on $w_Q^0$, $w_Q^m$,
$w_Q^r$, $a_c^{m}$ and $\Delta_m$ lead to constraints on a large
number of dark energy models, but at the same time it provides us
with model independent information on the evolution of the dark
energy. It could be argued that the five dimensional parameter
space we have introduced is too large to be reliably constrained.
Fortunately this can be further reduced without losing any of the
essential details arising out of tracker solutions in these
Quintessence models. In fact nucleosynthesis places tight
constraints on the allowed energy density of any dark energy
component, generally forcing them to be negligible in the
radiation era \cite{RACHEL,YA}. The real impact of dark energy occurs
after matter-radiation equality, so we can set $w_Q^r=w_Q^m$.
Consequently we end up with four parameters: $w_Q^0$, $w_Q^m$,
$a_c^{m}$ and $\Delta_m$. Although they increase the already large
parameter space of cosmology, they are necessary if we are to
answer fundamental questions about the nature of the dark energy.
The parameters make sense, if $w_Q(a)$ evolves in time, we need to
know when it changed ($a_c^{m}$), how rapidly ($\Delta_m$) and
what its value was when it changed ($w_Q^m$). Neglecting the
effects during the radiation dominated era it proves useful to
provide a shorter version of Eq.~(\ref{exp}), in fact since we can
neglect the transition from radiation to matter dominated eras,
then the linear combination Eq.~(\ref{exp}) can be rewritten as
\footnote[1]{We thank Eric Linder for pointing this out to us.}:
\begin{equation}
w_Q^{p}(a)=w_Q^0+(w_Q^m-w_Q^0)\times\frac{1+e^\frac{a_c^m}{\Delta_m}}
{1+e^{-\frac{a-a_c^m}{\Delta_m}}}
\times\frac{1-e^{-\frac{a-1}{\Delta_m}}}{1-e^\frac{1}{\Delta_m}}.
\label{lowz}
\end{equation}
As we can see in figure~\ref{f:fig23},
the relative difference between the exact
solution $w_Q(a)$ of the Klein-Gordon equation and
Eq.~(\ref{lowz}) is smaller than $5\%$ for redshifts $z<1000$,
therefore it provides a very good approximation for the evolution
of the quintessence equation of state. Both Eq.~(\ref{exp})
and Eq.~(\ref{lowz}) are very useful in that they allow us to take
into account the clustering properties of dark energy (see for
instance \cite{DAVE}) and to combine low redshift measurements
with large scale structure and CMB data.
\\

\chapter{Dark energy effects in the Cosmic Microwave Background Radiation}
\label{chap6}

The physics of the Cosmic Microwave Background radiation has been
deeply studied during the past 30 years. The theoretical developments
in understanding the different processes responsible for the temperature
anisotropies have given to us the possibility to
accurately predict the spectrum of the CMB fluctuations for a given
cosmological model. The existence of these cosmological temperature fluctuations was 
initially confirmed by the COBE satellite \cite{COBE}, but it is only
with the recent generation of balloon and ground experiments, such as 
Boomerang \cite{DEB1}, Maxima \cite{BALBI1} and DASI \cite{PRYKE},
that the observations reached the level of accuracy necessary
for testing different cosmological scenarios. As we have reviewed in 
Chapter~\ref{chap1}, these measurements have detected the multiple peak
structure of the CMB power spectrum, providing an accurate determination
 of the cosmological parameters.
The improvements in the CMB experiments are mainly due to the high
performance of the new high frequency microwave detectors. The recent WMAP
satellite, using this technology has started a new generation of satellite
experiments that will measure the temperature and polarization anisotropies
close to the theoretical `cosmic variance' limit.
It is therefore of crucial importance to study
the imprint dark energy leaves in the anisotropy power spectrum.
In this Chapter we start reviewing the main concepts of the CMB physics.
Using a model independent approach we study the impact of different
dark energy models in the CMB spectrum. In particular we find that features of the
dark energy equation of state can leave a characteristic imprint in 
the Integrated Sachs-Wolfe effect \cite{CORAS3,CORAS4}. 
We will identify the dark energy models that are distinguishable
from the cosmological constant. In conclusion we will show the limits on a class of 
dark energy models that can be obtained by
cosmic variance limited experiments.

\section{A beginner's guide to CMB physics}
The starting point of the CMB anisotropy calculation is the kinetic
theory of photons in a perturbed space-time.
Here we shall briefly review the main concept of the subject and
for more authoritative reviews we refer to \cite{HU3,HU1,LYTH}.

\subsection{Basic equations}
Let us consider a flat pertubed background. In the Newtonian gauge
the metric element is
\begin{equation}
ds^2=a^2(\eta)[-(1+2\Psi)d\eta^2+(1-2\Phi)dx^idx_i],
\end{equation}
where $\Psi$ and $\Phi$ are the metric perturbations and $\eta$ is the conformal time. 
The properties of a gas of photons propagating in this space-time are described
by the distribution function $f(x^i,P_i)$
that depends on the three spatial coordinates $x^i$
and the three conjugate momenta $P_i$. In the Newtonian gauge the relation between $P_i$
and the proper momentum $p_i$ measured by an observer at a fixed spatial coordinate
is given by $P_i=a(1-\Phi)p_i$. In order
to eliminate 
the metric perturbations from the definition of the momenta
it is convenient to replace $P_i$ with $q_i=a p_i$,
where $q_i=q \gamma_i$, $q$ and $\gamma_i$ being the module and 
the cosine directions respectively.
In the early Universe the photons propagate
in a hot ionized medium and interact through the Compton scattering with free electrons
and ions. As a consequence of this
the distribution function is conserved along the perturbed geodesic
apart for a collisional term. The evolution of the distribution function
is described by the Boltzmann equation:
\begin{equation}
\frac{df}{d\eta}\equiv \frac{\partial f}{\partial \eta}+\frac{\partial f}{\partial x^i}\frac{dx^i}{d\eta}
+\frac{\partial f}{\partial q}\frac{dq}{d\eta}+\frac{\partial f}{\partial \gamma^i}\frac{d \gamma^i}{d\eta}=C[f],
\label{Boltz}
\end{equation}
where $C[f]$ accounts for the Compton scattering
and the geodesic equation
\begin{equation}
\frac{1}{q}\frac{dq}{d\eta}=\dot{\Phi}-\gamma_i \frac{\partial \Psi}{\partial x_i}.
\end{equation}
In addition the linearized Einstein equations determine the equations for the 
metric perturbations $\Phi$ and $\Psi$ in term of the perturbations in the multiple
fluids system.
At the zeroth order we can consider the Universe as being unperturbed,
in such a case $f$ is independent
of $x^i$ and $\gamma_i$ and if
the collision between photons and charged particles conserve
energy, then $q=a p$ is time independent and consequently $f$ is as well.
This is a good
approximation, since the free electrons are non-relativistic and the Compton
scattering is primarily Thomson scattering
in which momentum and not energy is transferred. 
The equilibrium distribution function 
$f$ is
\begin{equation}
f_{eq}=\frac{2}{\exp{(\frac{q}{T_0})}-1},
\label{eql} 
\end{equation}
where the factor 2 accounts for the number of polarization states of the photon and
$T_0$ is the temperature of the radiation today. The effects of the perturbed background
can be described by expanding the distribution function $f$ about the equilibrium state,
\begin{equation}
f=f_0(q)+f_1(\eta,x^i,\gamma_i,q),
\label{fex}
\end{equation}
where $f_0(q)$ is given by Eq.~(\ref{eql})
and $f_1$ is the perturbed part of the distribution function.
Substituting Eq.~(\ref{fex}) into Eq.~(\ref{Boltz}) and 
taking only the first order terms we obtain: 
\begin{equation}
\frac{\partial f_1}{\partial \eta}+\gamma^i\frac{\partial f_1}{\partial x^i}+
q\frac{d f_0}{dq}\left(\dot{\Phi}-\gamma^i \frac{\partial \Psi}{\partial x^i} \right)=C[f].
\label{pertf}
\end{equation}
For practical purposes it is convenient
to write the perturbed part of the distribution function in term
of the brightness function $\Theta(\eta,x^i,\gamma_i)$, that is the fractional perturbation
in the effective temperature of the photons, $T=T_0(1+\Theta)$.
Expanding $f$ around $T_0(1+\Theta)$
we have:
\begin{eqnarray}
f &=&\frac{2}{\exp{[\frac{q}{T_0(1+\Theta)}]}-1} \nonumber \\
& &\approx  \frac{2}{\exp{(\frac{q}{T_0})}-1}-
2\frac{q}{\left[\exp{(\frac{q}{T_0})}-1\right]^2}\frac{\exp{(\frac{q}{T_0}})}{T_0}\Theta \nonumber\\
& & =f_0-q\frac{d f_0}{dq}\Theta.
\label{fbri}
\end{eqnarray}
Comparing Eq.~(\ref{fex}) with Eq.~(\ref{fbri}) we obtain 
\begin{equation}
f_1=-q df_0/dq \Theta.
\label{f1t}
\end{equation}
Substituting Eq.~(\ref{f1t}) into Eq.~(\ref{pertf}) and taking the Fourier transform
the Boltzmann equation becomes \cite{PEBYU}:
\begin{equation}
\dot{\Theta}+ik\mu(\Theta+\Psi)=-\dot{\Phi}+S_C,
\label{bz}
\end{equation}
where $\mu=\frac{\gamma^i k_i}{k}$ and
\begin{equation}
S_C=\dot{\tau}\left(\Theta_0-\Theta+\gamma_i V_b^i+
\frac{1}{16}\gamma_i\gamma_j\Pi^{ij}_r \right),
\label{sc}
\end{equation}
where $\dot{\tau}=x_e n_e\sigma_T a$ is the differential optical depth, with $x_e$ the ionization
fraction, $n_e$ the electron number density and $\sigma_T$ the Thomson cross section;
$\Theta_0$ is the isotropic component of $\Theta$, $V_b^i$ is the baryon velocity and
$\Pi^{ij}_r$ is the anisotropic stress perturbation for the photons.
Hence the collisional term couples the evolution of the photon perturbation to
that of the baryons. The equations for the baryon perturbations are obtained by
linearizing the conservation equation of the stress energy tensor:
\begin{eqnarray}
\dot{\delta_b}&=&-k (V_b-V_r)-3\dot{\Phi},\label{db}\\
\dot{V_b}&=&-\frac{\dot{a}}{a}V_b+k\Psi+\dot{\tau}(V_r-V_b)/R,\label{vb}
\end{eqnarray}
where $V_r$ is the velocity perturbation of the photons.
It is useful to expand the brightness function in a Legendre series:
\begin{equation} 
\Theta=\sum_{l=0}^{\infty} (-i)^l (2l+1)\Theta_l P_l({\mu}).
\label{leg}
\end{equation}
Using the definition of the photon stress energy tensor in term of the 
distribution function $f$, it can be shown that $\Theta_0=\delta_r/4$ and $\Theta_1=V_r$.
Substituting Eq.~(\ref{leg})
in Eq.~(\ref{bz}) and using
the explicit form of the Compton scattering term Eq.~(\ref{sc})
we obtain the hierarchy equations \cite{Wilson,Gouda}:
\begin{equation}
\dot{\Theta}_0=-\frac{k}{3}\Theta_1-\dot{\Phi}, \label{t1}\\
\end{equation}
\begin{equation}
\dot{\Theta}_1=k \left[ \Theta_0+\Psi-\frac{2}{5}\Theta_2\right]-\dot{\tau}(\Theta_1-V_b), \label{t2}\\
\end{equation}
\begin{equation}
\dot{\Theta}_2=k \left[ \frac{2}{3}\Theta_1-\frac{3}{7}\Theta_3\right]-
\frac{9}{10}\dot{\tau}\Theta_2, \label{t3}\\
\end{equation}
\begin{equation}
\dot{\Theta}_l=k \left[\frac{l}{2l-1}\Theta_{l-1}-\frac{l+1}{2l+3}\Theta_{l+1} \right]
-\dot{\tau}\Theta_l \;(l>2). \label{tl}
\end{equation}
The anisotropy power spectrum is defined as:
\begin{equation}
\frac{2l+1}{4\pi}C_l=\frac{1}{2\pi^2}\int d\eta \frac{dk}{k} \frac{k^3 |\Theta_l(k,\eta)|^2}{2l+1}. 
\end{equation}
From Eq.~(\ref{bz}) we note there are two sources of anisotropy formation,
the Compton scattering that couples the photons to the baryons and the gravitational effect 
produced by the presence of density fluctuations in all the matter components. 

\subsection{CMB anisotropies}

\subsubsection*{Acoustic Oscillations}

Before recombination the differential optical depth $\dot{\tau}$ is very large and the
scattering between photons and baryons is extremely rapid and efficient. 
In this regime, called the
{\em tight coupling} limit ($\dot{\tau}/k>>1$),
the photons and the baryons
behave as a single fluid. Because of this the baryons and photons velocity
are the same at the zeroth order, $V_b\approx V_r=\Theta_1$. 
In such a case Eq.~(\ref{vb}) becomes:
\begin{equation}
\dot{\Theta}_1=-\frac{\dot{R}}{R}\Theta_1+k\Psi+\frac{\dot{\tau}}{R}(\Theta_1-V_b),
\end{equation}
where $R=3\rho_b/4\rho_r$ is the baryon to photon ratio. Inverting this equation
in terms of $\dot{\tau}(\Theta_1-V_b)$ and substituting in Eq. (\ref{t2}) we obtain
the iterative first order solution,
\begin{eqnarray}
\dot{\Theta}_0=-\frac{k}{3}\Theta_1-\dot{\Phi}, \\
\dot{\Theta}_1=-\frac{\dot{R}}{1+R}\Theta_1+\frac{k\Theta_0}{1+R}+k\Psi.
\end{eqnarray}
These equations can be combined into a single second order equation \cite{HUSUG},
\begin{equation}
\ddot{\Theta}_0+\frac{\dot{a}}{a}\frac{R}{1+R}\dot{\Theta}_0+k^2 c_s^2 \Theta_0= F(\eta),
\label{theta0}
\end{equation}
where $c_s^2=\frac{1}{3}(1+R)^{-1}$.
The source function $F(\eta)$ arises
from the gravitational potentials and is given by
\begin{equation}
F(\eta)=-\ddot{\Phi}-\frac{\dot{a}}{a}\frac{\dot{R}}{1+R}\dot{\Phi}-\frac{k^2}{3}\Psi.
\end{equation}
From Eq.~(\ref{theta0}) we can distinguish three different contributions to the
evolution of the isotropic part of the brightness function during the tight coupling regime:
\begin{itemize}

\item the radiation pressure given by the term $k^2 c_s^2 \Theta_0$ which dominates on
subhorizon scales and supports oscillations of the photon-baryon plasma;
\item the gravitational infall of the photon-baryon fluctuations in the potential
well, $k^2 \Psi$;
\item the $\ddot{\Phi}$ term that sources the oscillations at scales that enter in the horizon between
the epochs of matter-radiation equality and decoupling. The $\dot{\Phi}$ term contribute
as a friction term.

\end{itemize}
Neglecting time variations
in the gravitational potentials,
Eq.~(\ref{theta0}) describes a forced harmonic oscillator
equation. 
On subhorizon scales the photon pressure balances the gravitational collapse of the perturbations
and causes
 the propagation of pressure waves in the photon-baryon plasma.
The characteristic scale upon which these waves can propagate is fixed by the size
of the sound horizon $r_s=\int c_s(\eta)d\eta$. When the photons last scatter,
the imprint of these compression and rarefaction regions will appear as a series of
peaks in the anisotropy power spectrum on the angular scales smaller
than the angle subtended by the sound horizon at the decoupling.
The odd and the even peaks correspond to photons coming
from regions that are respectively 
in compression and expansion phases at last scattering.
However the compression and rarefaction phases are not
completely equivalent. In fact the baryons increase the effective mass of the
barotropic fluid and consequently the gravitational infall generated by the density fluctuations 
leads to a deeper trough during the compression phase ({\em baryon drag}). 
The compression is enhanced over rarefaction resulting in 
a larger amplitude of the odd peaks relative to the even ones. Therefore the
relative heights of the acoustic peaks is extremely sensitive to the baryon density $\Omega_b h^2$.
An additional source of anisotropies is due to the motion of the fluid at the last scattering 
surface due to the Doppler shift of the CMB photons. However due to the large value of $R$ this
effect is subdominant.

\subsubsection*{Sachs-Wolfe effect}
In the large wavelength limit ($k\eta<<1$) the Boltzmann equation Eq.~(\ref{t1})
reduces to the form $\dot{\Theta}_0=-\dot{\Phi}\approx \dot{\Psi}$ ($\Phi=-\Psi$).
The solution of this equation depends on the initial condition of the solution of the 
linear density perturbation
equations. In fact $\Theta_0=\delta_r$ and for adiabatic initial conditions
$\delta_r(0)=-\frac{1}{2}\Psi(0)$, reflecting the fact that
the photons are overdensities inside the potential well. However due
to the decay of the gravitational potential
after matter-radiation equality the photon temperature at decoupling becomes
$\Theta_0(\eta_*)=-\frac{2}{3}\Psi(\eta_*)$, where $\eta_*$ is the conformal time
at decouplings. When the photons last scatter they climb out of the
potential well $\Psi$ so that the effective superhorizon perturbations in the photon temperature is:
\begin{equation}
[\Theta_0+\Psi](\eta_*)=\frac{1}{3}\Psi(\eta_*).
\end{equation} 
This is the so called Sachs-Wolfe effect \cite{SACHS}. It is the dominant source of
anisotropies on large angular scales and is responsible for the plateau at the low
multipoles in the CMB power spectrum. Since these scales are superhorizon at the time of
decoupling the amplitude of this Sachs-Wolfe depends on the primordial spectrum
of the density fluctuations. 

\subsubsection*{The Integrated Sachs-Wolfe effect}
\begin{figure}[t]
\centerline{
\includegraphics[scale=0.28,angle=270]{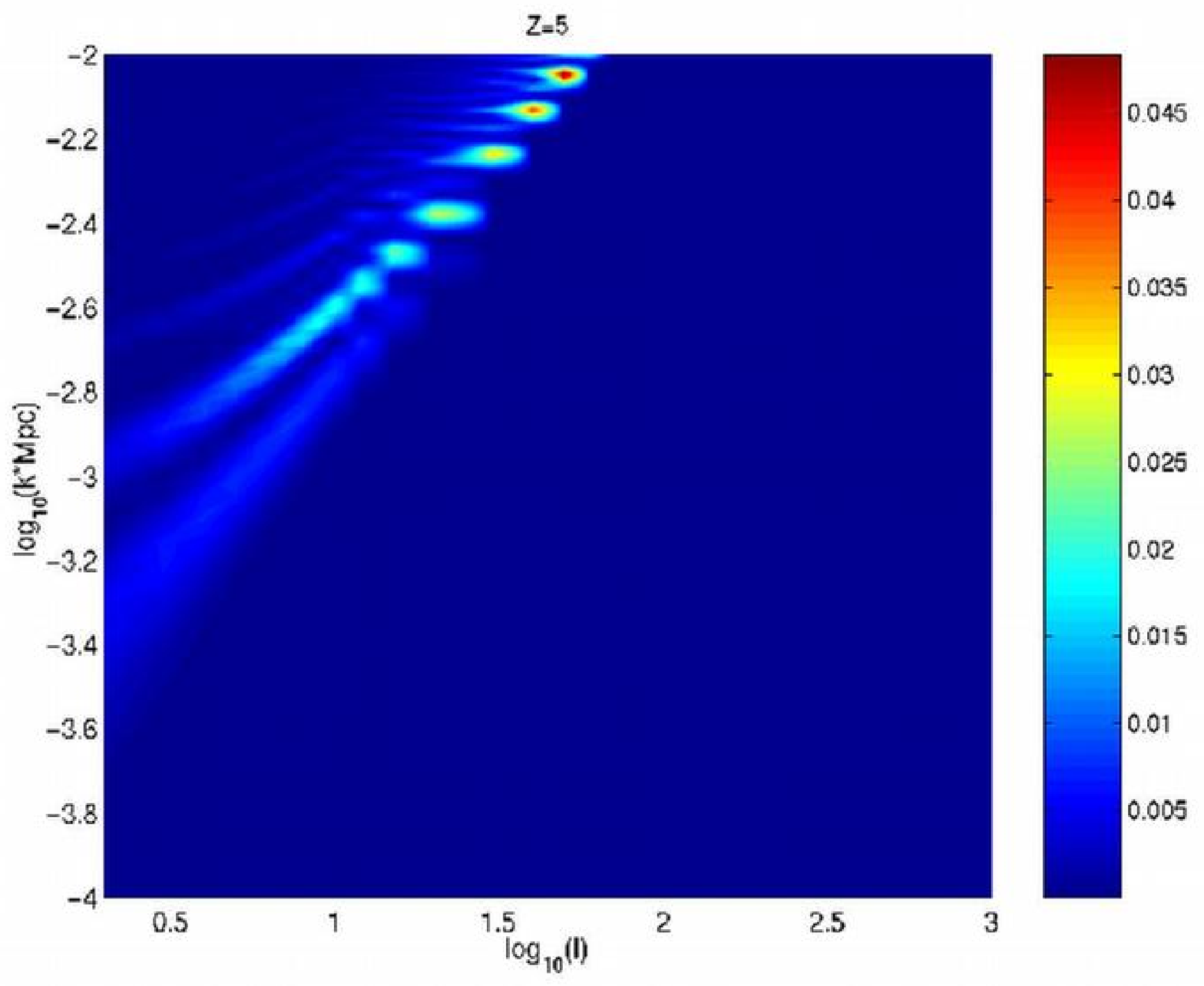}
\includegraphics[scale=0.28,angle=270]{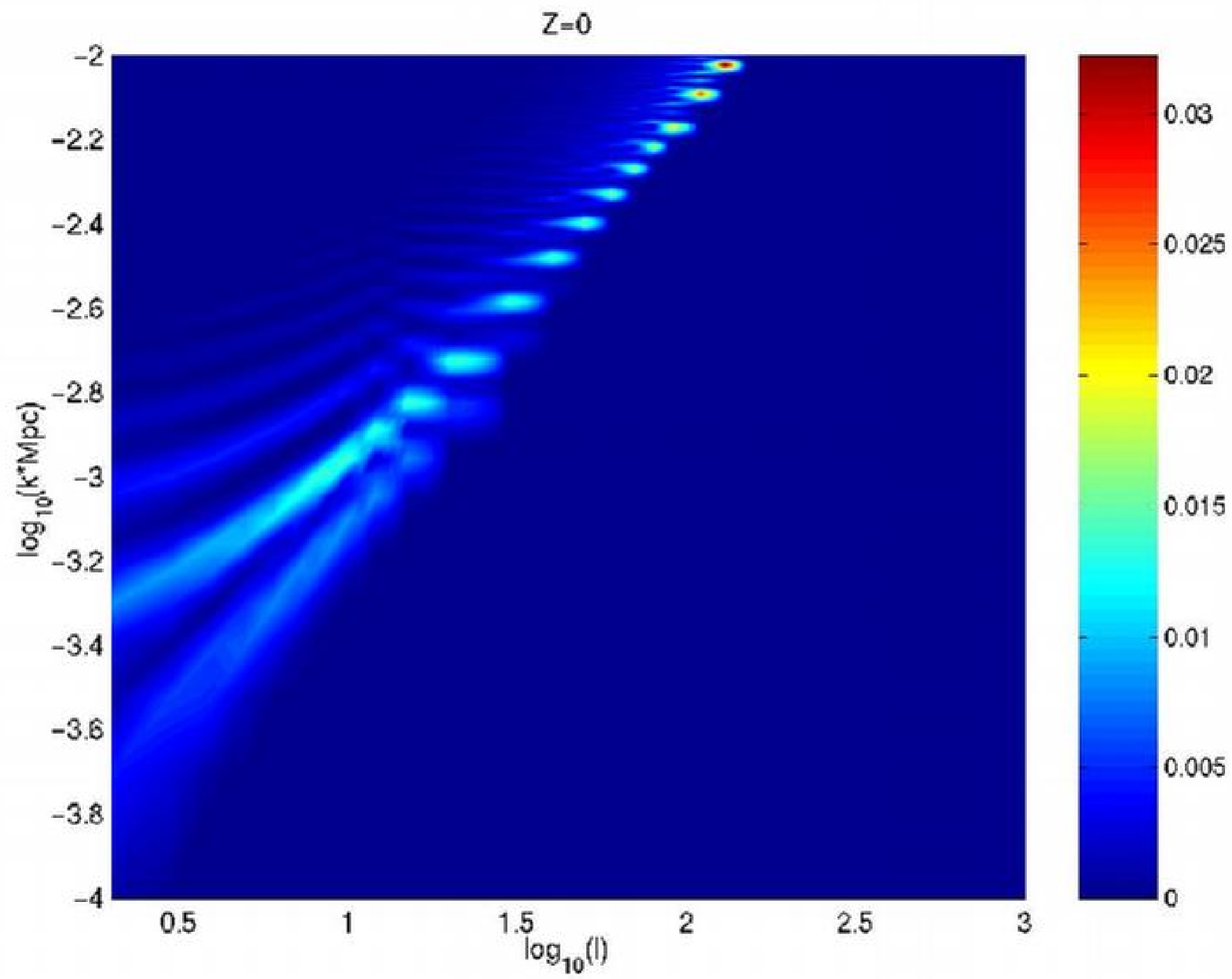}
}
\caption{Density plots of $C_l(k)$ in the $k-l$ plane for a $\Lambda$CDM model.
The modes in the left panel are integrated up to redshift $z=5$ and those in the
right panel are integrated up to $z=0$.}
\label{f:figisw}
\end{figure}

The time evolution of the gravitational potentials is a source of primary anisotropies, in fact
as the photons free stream after decoupling the solution to the Boltzmann equation is given by \cite{HUGSUG1}:
\begin{equation}
\frac{\Theta_l(\eta)}{2l+1}=[\Theta_0+\Psi](\eta_*)j_l(k(\eta-\eta_*))+
\int_{\eta_*}^\eta(\dot{\Psi}-\dot{\Phi})j_l(k(\eta-\eta'))d\eta',
\end{equation}
where $j_l(k\eta)$ are the Bessel functions.
The first term represents the contribution of the Sachs-Wolfe effect,
while the second term is the Integrated Sachs-Wolfe effect (ISW) \cite{REES}.
We can distinguish two contributions to the ISW:
an early and a late ISW.
The former is due to the decays of the gravitational
potentials after horizon crossing at
the end of the radiation dominated era.
The latter occurs in dark energy dominated cosmologies, as the Universe starts
accelerating the rapid expansion causes the decay of the density fluctuations that drive
the decay of the gravitational potentials. 
The combined effect of the SW and the two components of the ISW can be studied 
by analysing the the
anisotropy power spectrum per logarithmic $k$ and $l$ interval defined by
\begin{equation}
C_l(k)=\frac{1}{l(l+1)}\int_{\eta*}^{\eta}|\Theta(k,\eta')|^2 d\eta'.
\label{clk}
\end{equation}
In figure~\ref{f:figisw} we show the density plots of Eq.~(\ref{clk}) integrated 
up to $z=5$ ({\em left panel})
and to $z=0$ ({\em right panel}) for
a $\Lambda$CDM with $\Omega_{\Lambda}=0.7$. The brighter regions correspond to values of
the wavenumbers and multipoles where there is more power. Since for the model we have considered
the acceleration starts at redshift $z=0.7$,
the contribution of the late ISW can be emphasized by comparing the distribution of
power in the left panel
with that in the right panel. As we expect, we note that due to the projection
onto the last scattering surface the wavenumbers $k$ and the multipoles 
$l$ are strongly correlated. As a consequence of this
the power at redshift $z=5$ is shifted toward lower multipoles
than at $z=0$. This is simply because the last scattering surface is closer to the observer 
at redshift $z=5$ and
different scales are projected on larger angles. The filamentary and periodic
structure of the density plots are due to the Bessel functions. We therefore expect a different
structure in the case of an open or closed space-time. Note that the power is
distributed along an upper and a lower ridge respectively.
The latter is the contribution of the Sachs-Wolfe effect caused by modes that are superhorizon
at decoupling ($k\sim10^{-3}$) while the upper ridge is produced by the early ISW.
Since this latter effect arises closer to
the present time, the corresponding anisotropies are projected
on angular scales larger than the SW effect itself.
The bright spot at the top of the panels correspond to the rise of the first Doppler peak,
the other peaks do not appear in these plots since we limited our analysis to scales larger
than the horizon at decoupling ($k<10^{-3}$).
As we can see in the right panel, integrating the modes over the period of time when
the Universe started to accelerate boosts the power at lower multipoles.
In fact the signature of the late ISW can be noticed from the fact that the upper ridge becomes 
brighter than that in the right panel. 
Note that the two plots have slightly different brightness scales, this is because
the late ISW overlaps with the SW leading to a different distribution of the power on the
large angular scales.

\subsubsection*{Damping mechanisms}
The overall amplitude of the CMB power spectrum is suppressed by the effects of
diffusive mechanisms.
The most important of these effects is due to the diffusion
of the photons on scales smaller than their mean free path $\lambda_C$. In fact
due to the Compton scattering the photons randomly walk through the
baryons with a mean free path $\lambda_C\approx\dot{\tau}^{-1}$.
After $N$ collisions the diffusive length is $\lambda_D\sim
\sqrt{N}\lambda_C$. Consequently on scales $\lambda<\lambda_D$ the diffusion
exponentially
suppresses the amplitude of anisotropies.
The baryons suffer a similar effect, in fact from Eq.~(\ref{vb}) we can see that
on scales $k<<\dot{\tau}/R$ the Compton scattering can drag the baryons
in and out of the potential wells leading to a destruction
of the baryonic oscillations ({\em Silk damping} \cite{SILKDAMP}).
Another damping effect arises from the fact that he last scattering process
is not instantaneous. The photons we observe today in a given direction
of the sky may have decoupled
from different points along the line of sight. 
Therefore fluctuations on scales smaller then the thickness of the last scattering
surface have destructively interfered causing a `washing out' of the
anisotropies on very small angular scales. 
The reionization history of the Universe also contributes to the overall damping of
the temperature anisotropy power spectrum. 
In fact when the first stars form they reionize the intergalactic
medium, therefore a fraction of the CMB photons are scattered on scales smaller then size of the
horizon at the epoch of reionization.
This lead to a suppression of the power at large multipoles.
 In light of the recent WMAP measurements of
the temperature-polarization cross power spectrum, it is worth mentioning
that the reionization sources the polarization of the anisotropies.
Consequently a late reionization, $z_{ion}<100$, will produce significant power at low multipoles
in the temperature-polarization spectrum.

\section{Dark energy and the Integrated Sachs-Wolfe effect}

\begin{figure}[t]
\centerline{
\includegraphics[scale=0.6]{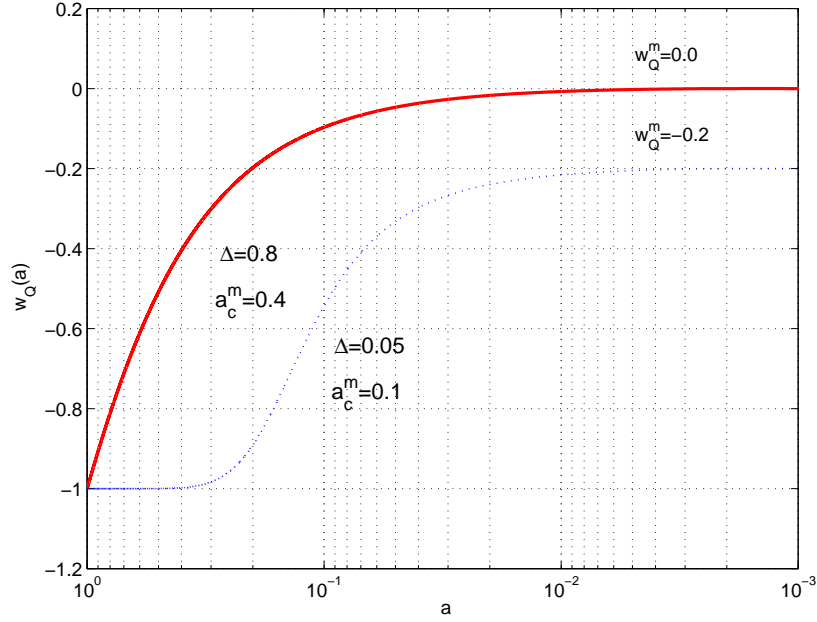}
}
\caption{ Time evolution of the equation of state
for two classes of models, with slow (red solid line) and rapid transition 
(blue dotted line). The dark energy parameters specify the features of $w_Q(a)$}.
\label{f:fig24}
\end{figure}

\begin{figure}[t]
\centerline{
\includegraphics[scale=0.6]{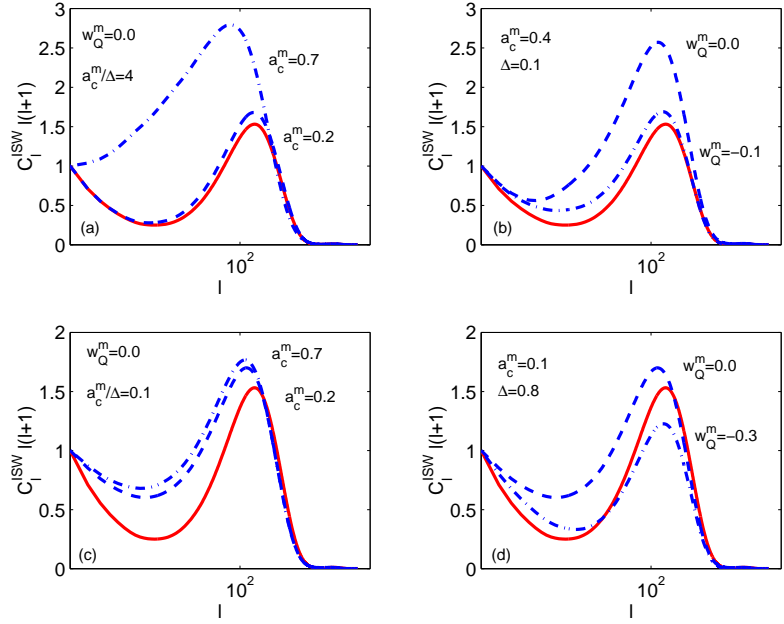}
}
\caption{Power spectrum of the ISW for rapidly varying models (top panels) and
slowly varying ones (bottom panels). The solid red line is the ISW effects produced
in the cosmological constant case. Detailed explanation in the text.} 
\label{f:fig25}
\end{figure}

We have previously discussed the importance of distinguishing between different dark energy
models from the ``concordance'' $\Lambda$CDM model. The ISW is particularly sensitive
to the late time evolution of the Universe. In fact during the matter dominated era the
gravitational potential $\Phi$ associated with the density perturbations is constant and
there is no ISW effect. However, we have seen that in $\Lambda$CDM models $\Phi$ starts
decaying at redshifts when $\Lambda$ starts to dominate, producing large angular scale
anisotropies \cite{CRI}. In dark energy scenarios the cosmic acceleration is not the
only contribution to the decay of $\Phi$:
on large scales the clustering properties alter
the growth rate of matter perturbations \cite{CALDWELL1,MA2}.
It is the signal of this clustering \cite{HU0} that we are hunting 
in as model-independent a way as possible. 
We assume a flat spatial geometry and fix the value of the Hubble constant
$H_o=70$ Km$s^{-1}$$Mpc^{-1}$, and the amount of matter (CDM) $\Omega_{m}=0.3$.
We can usefully distinguish two classes of models, 
\begin{itemize}
\item those with a 
slowly varying equation of state for which $0 < a_c^{m}/ \Delta < 1$, as in
the case of the inverse power law potential \cite{ZLATEV}; 
\item a rapidly
varying $w(z)$, such as the 'Albrecht-Skordis` model \cite{ALB} and
the two exponential potential \cite{NELSON}, with 
$ a_c^m/ \Delta > 1$. This class also includes many interesting 
radical models such as vacuum metamorphosis \cite{PARK}, 
late-time phase transitions \cite{HILL1}, and backreaction-induced 
acceleration \cite{WETT5}.
\end{itemize}
We show these two classes in figure~\ref{f:fig24}.
The red solid line corresponds to dark energy that tracks
the dust during the matter era ($w_Q^m=0.0$) and 
evolves slowly toward $w_Q^o=-1$, and the 
blue dotted line corresponds to a model with a rapid transition in its equation
of state at $a_c^m=0.1$ ($z = 9$). 
Given current data it is worth studying the case with $w_Q^o=-1$, 
(since it is also the most difficult to distinguish from $\Lambda$CDM) 
whilst allowing the other parameters $w_Q^m$ and $a_c^m$, to vary.
We have modified the CMBFAST code \cite{ZALDA} in order 
to include the effects of a dark energy fluid and its perturbations
by specifying the time evolution of the equation of state. 
Figure~\ref{f:fig25} shows the anisotropy power spectrum, $C_l^{ISW}$, produced through
the integrated Sachs-Wolfe effect by  a rapidly evolving ($top$ $panels$) and 
a slowly evolving ($bottom$ $panels$) equation of
state; the red (solid) line corresponds to the $\Lambda$CDM model.
As we can see in the top left panel (figure \ref{f:fig25}a), varying $a_c^m$ can 
produce a strong
ISW. The effect is larger if the transition in the equation of state occurs
at redshifts $z<3$. On the other hand the $C_l^{ISW}$ is the same as in the
cosmological constant regime if $a_c^m<0.2$ ($z>4$). 
In the top right panel (figure \ref{f:fig25}b)
we plot the ISW for two different values of $w_Q^m$, corresponding to
$w_Q^m=0.0$ (dashed line) and $w_Q^m=-0.1$ (dot-dash line).
We note that the signal is larger if the quintessence field is perfectly
tracking the background component. But as $w_Q^m$ diverges from the dust
value the ISW effect approaches that of $\Lambda$CDM. 
This means that even
for rapidly varying $w(z)$ (small $\Delta$), the ISW is distinguishable 
from that in the $\Lambda$CDM scenario only if $w(z)$ 
during matter domination closely mimics the dust value and the 
transition occurs at low redshifts, $z<3$. 
We can see that the amplitude of the integrated Sachs-Wolfe effect is smaller in
slowly varying models ($bottom$ $panels$). As we expect the $C_l^{ISW}$
is independent of $a_c^m$ (figure~\ref{f:fig25}c), since for these models a
different value in the transition redshift does not produce a large 
effect on the evolution of the dark energy density.
In figure~\ref{f:fig25}d, the ISW power spectrum is large for $w_Q^m=0.0$ (dash line)
and becomes smaller than the cosmological constant on horizons
scales as $w_Q^m$ has negative values (dot-dashed line),
and increases toward $\Lambda$ for $w_Q^m$ approaching $-1$. 
This class of models is then more difficult to distinguish 
from the $\Lambda$CDM if the equation of state today is close to 
$w_{\Lambda} = -1$.
This can be qualitatively explained noting that perfect 
tracking between dark energy and CDM causes a delay in the time when the 
gravitational potential starts to decay, compared to the case of $\Lambda$CDM. 
This effect is stronger for models with a rapidly varying equation 
of state since the rapid change in $w_{Q}$ produces a stronger variation in 
the gravitational potential.

\section{Differentiating dark energy models with CMB measurements}

The information in the CMB power spectrum can be encoded in 
the position of the first three Doppler peaks, $l_1$, $l_2$ and $l_3$
and in three parameters, $H_1$, $H_2$ and $H_3$ that account for the amplitude \cite{FU}.
These are the height of the first peak relative to the power at $l=10$,
\begin{equation}
H_1\equiv\frac{l_1(l_1+1)C_{l_1}}{l_{10}(l_{10}+1)C_{l_{10}}},
\end{equation}
the height of the second peak relative to the first,
\begin{equation}
H_2\equiv\frac{l_2(l_2+1)C_{l_2}}{l_{1}(l_{1}+1)C_{l_{1}}},
\end{equation}
and the height of the third peak relative to the first,
\begin{equation}
H_3\equiv\frac{l_3(l_3+1)C_{l_3}}{l_{1}(l_{1}+1)C_{l_{1}}}.
\end{equation}
In principle the position of the second and third peaks is not necessary to characterize  
the CMB power spectrum, since their value is set by the position of the
first peak through the harmonic relation
Eq.~(\ref{lharm}).
However, as we have previously discussed in Chapter \ref{chap4} this relation
is affected by pre-recombination effects and therefore the value of $l_2$ and $l_3$
carry information about a combination of the cosmological parameters. 
Since $H_1$, $H_2$ and $H_3$ quantify the amplitude of the power spectrum on different
multipoles we expect they are sensitive in different ways to the cosmological parameters.
For instance since $H_1$ depends on the ratio between $C_{l_1}$ and $C_{l_{10}}$
it mainly depends
on the scalar spectral index $n$, the physical baryon density $\Omega_b h^2$ due to
the baryon drag that changes $C_{l_1}$, $\Omega_{\Lambda}$ through the ISW and
the reionization since increasing $\tau$ will lower the first peak.
On the other hand $H_2$ is sensitive only to $\Omega_b h^2$ and $n$.
Since the baryons affect in the same way the height of the third and first
peak, $H_3$ depends only on the matter density $\Omega_m$ and $n$.
This implies that if the amplitude of the first three
peaks is accurately determined the degeneracy between $\Omega_b h^2$ and $n$ can
be broken and their value can be
measured with high precision. We use $H_1$, $H_2$ and $H_3$ 
to determine the overall effect
of the dark energy in the CMB spectrum. 
In particular we expect $H_1$ to be extremely sensitive to a quintessential 
effect through the ISW, while 
$H_2$ and $H_3$ are weak indicators of such a signal. 
We have computed the CMB spectra for the class of models with $w_Q^o=-1$
described in the previous section and we have inferred the values of 
$H_i$ and compared them with those of the $\Lambda$CDM 
model for the same values of the cosmological parameters.
\begin{figure}[t]
\centerline{
\includegraphics[scale=0.6]{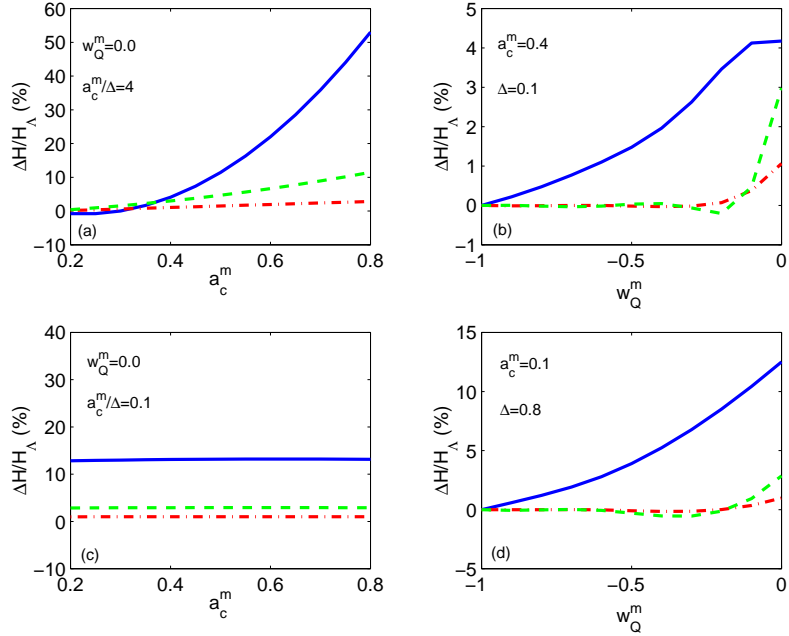}
}
\caption{Relative difference of $H_1$ (blue solid line), $H_2$ (green dash
line) and $H_3$ (red dash-dot line) to the $\Lambda$CDM model, for rapidly
varying models ($top$ $panels$), and with slow transition ($bottom$ $panels$).
For these models the present value of the equation of state is $w_Q^o=-1$.} 
\label{f:fig26}
\end{figure}
\begin{figure}[h]
\centerline{
\includegraphics[scale=0.6]{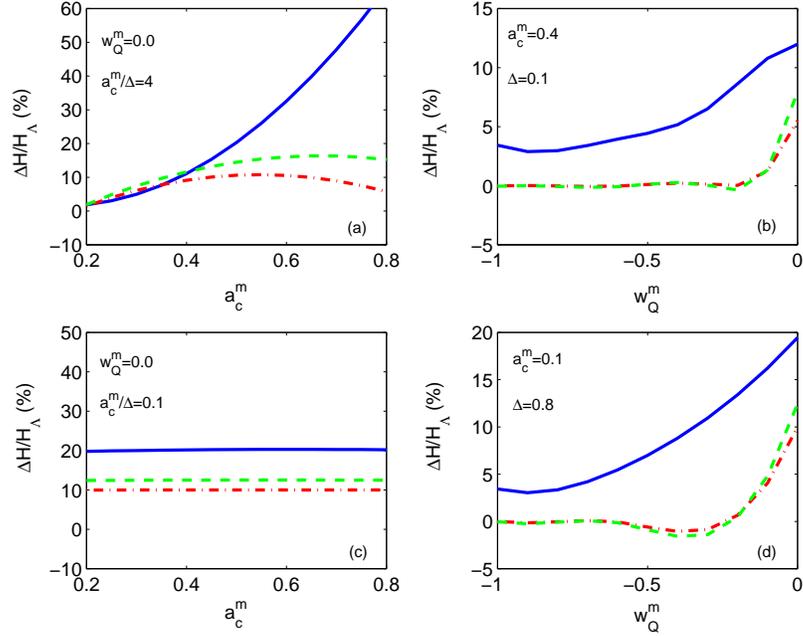}
}
\caption{As in figure~\ref{f:fig26} for $w_Q^o=-0.88$.} 
\label{f:fig27}
\end{figure}
In figure~\ref{f:fig26} we plot the absolute value of the relative difference of $H_1$,
$H_2$ and $H_3$ to the $\Lambda$CDM model.
The rapidly varying models are shown in the top panels. 
We can see the strong ISW effects produced by changing $a_c^m$ are now
evident in the large discrepancy between $H_1$ and $H_1^{\Lambda}$
(blue line) (figure~\ref{f:fig26}a): it can be larger then 20 per cent for $a_c^m>0.6$.
The effect on $H_2$ and $H_3$ is smaller. However varying $w_Q^m$ (figure~\ref{f:fig26}b)
produces a discrepancy of only order $4$ per cent on $H_1$, while
$H_2$ and $H_3$ remain the same as in $\Lambda$CDM.
For a slowly varying equation of state, $H_1$, $H_2$ and $H_3$ are 
independent of $a_c^m$ (figure~\ref{f:fig26}c). The dark energy imprint
is only on $H_1$ for which the discrepancy to the $\Lambda$ case
is about 10 per cent. This discrepancy decreases
when changing the value of the quintessence equation of
state during matter from $w_Q^m=0.0$ to $w_Q^m=-1$ (figure~\ref{f:fig26}d). 
Values of the equation of state
today $w_Q^o > -1$ imply a stronger ISW effect.
Consequently the curves of fig.3 are shifted upwards. For instance in figure~\ref{f:fig27}
we plot the class of models previously analysed, with $w_Q^o=-0.88$. We note
the same behavior as we vary the dark energy parameters, but the discrepancy
with the $\Lambda$CDM model is now larger. 
In figure~\ref{f:fig27}d it is worth noticing
the case $w_Q^m=-1$, that corresponds to a model very similar to 
a `k-essence' model \cite{MUKA}.
We can see that the relative difference with the $\Lambda$CDM case 
is of the order of a few percent, in agreement with \cite{ERIC} 
for the same value of $w_Q^o=-0.88$.
At this point we ask the key question whether such differences are observable. 
We have shown that $H_1$ is a good estimator of the ISW effect, and that it
is a tracer of the dark energy imprint on the
CMB. However its estimation from the data
will be affected by cosmic variance at $l=10$. Hence with even perfect 
measurements of the first acoustic peak the
uncertainty on $H_1$ will be dominated by the  
$30$ per cent uncertainty due to cosmic variance. 
With the plots of figures~\ref{f:fig26}-\ref{f:fig27} in mind, 
this means that if the present value
of the equation of state is close to $-1$,
slowly varying dark energy models are hardly distinguishable from $\Lambda$CDM,
while rapidly 
varying ones can produce a detectable signature only if the transition in the
equation occurred at $a_c^m>0.7$, but in any case it will be difficult to 
constrain $w_Q^m$. Moreover it should be taken into account that $H_1$
is degenerate with other cosmological parameters, such as the
scalar spectral index $n$, the optical depth $\tau$ and the scalar to tensor
ratio $r$. 
Hence only an accurate determination of the angular
diameter distance, inferred from the location of the acoustic peaks,
would allow detection of such deviations from the cosmological
constant model.
\begin{figure}[t]
\centerline{
\includegraphics[scale=0.6]{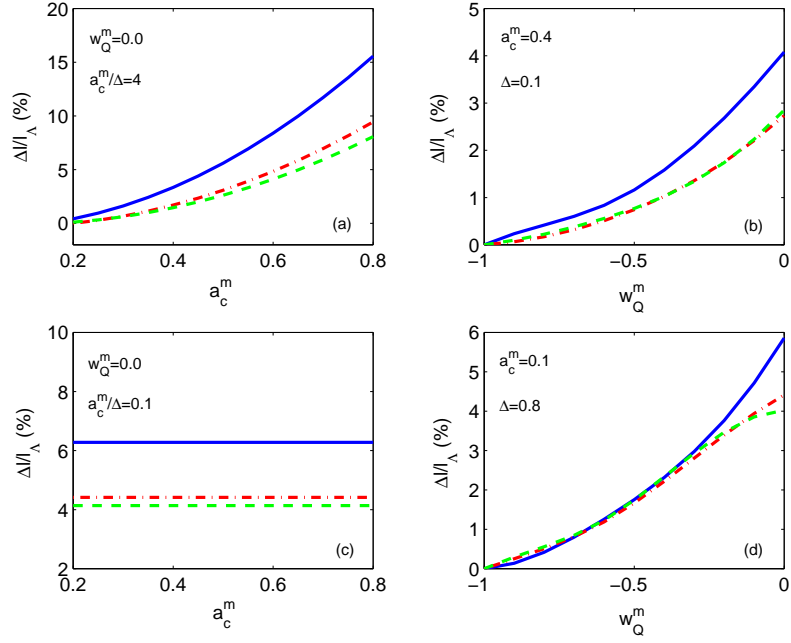}
}
\caption{Relative difference of $l_1$ (blue solid line), $l_2$ (green dashed
line) and $l_3$ (red dot-dashed line) to the $\Lambda$CDM model, for rapidly
varying models ($top$ $panels$), and with slow transition ($bottom$ $panels$).
For these models the present value of the equation of state is $w_Q^o=-1$.} 
\label{f:fig28}
\end{figure}
The shift of the multipole 
positions ($\ell_i$) of the acoustic peaks caused by the 
evolution of the dark energy in the class of models analysed in figure~\ref{f:fig25} can be
seen in figure~\ref{f:fig28}, where we plot the 
relative difference of $l_1, l_2$ and $l_3$ to the $\Lambda$ case.
We note that due to the additional shift induced on the first acoustic peak
by the ISW effect the difference with the $\Lambda$CDM model 
for the first peak is generally larger than for the second and third peaks. 
As with the comparison of the amplitude of the CMB spectrum,
the largest effect is produced by models with a rapid transition occurring at
small redshifts.
However the degeneracy of the angular diameter distance,
in particular with the value of
Hubble constant and the amount of dark energy density, will limit our ability
to put tight constraints on the
dark energy parameters.
There are alternative ways in which these problems can be alleviated,
for instance cross-correlating
the ISW effect with the large scale structure of
the local universe \cite{HUW,CO,BOUGHN}.
An efficient approach would be to
combine different observations in order to break the
degeneracies with the cosmological parameters \cite{WAS,FRIEMAN02}.

\section{Testing dark energy with ideal CMB experiments}
\begin{figure}[t]
\centerline{
\includegraphics[scale=0.6]{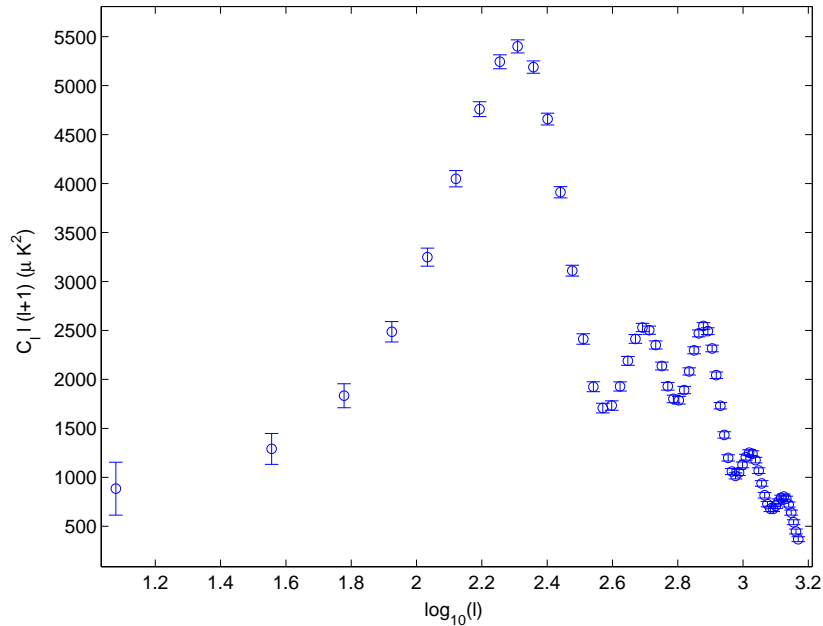}
}
\caption{CMB power spectrum for a fiducial dark energy models, the errorbars are the cosmic
variance errors.} 
\label{f:fig29}
\end{figure}
The new generation of CMB satellite experiments is going to provide
an estimation of the anisotropy power spectrum close to the 
ideal case, where the only source of indetermination is due to the cosmic
variance. It is therefore interesting to test the sensitivity of such
CMB
measurements to the dark energy effects. As we have described in the previous 
section, the class of models most difficult to distinguish from the $\Lambda$CDM
case corresponds to a dark energy fluid with a slowly varying equation of state
characterized by $w_Q^0=-1$. This can be considered as the most
pessimistic situation since it can
prevent us from understanding the nature of the dark energy.
We have studied
the information on this class of models that can be inferred from
cosmic variance limited measurements \cite{CORAS4}
and generated a sample of ideal CMB power spectrum data assuming a fiducial model specified by
the following values of the cosmological parameters: $\Omega_{Q}=0.68$,
$H_o=70$ Km $s^{-1}$ $Mpc^{-1}$, $w_Q^o=-1$, $w_Q^m=-0.4$, $a_c^m=0.2$, $\Delta=0.3$. 
Assuming a flat geometry, no tensor contribution, 
the scalar spectral index $n=1$ and the baryon density $\Omega_b h^2=0.021$,
we have binned the input power spectrum in $62$ data points  plotted in figure~\ref{f:fig29}.
A library of CMB spectra has then been generated using a modified version of CMBFAST~\cite{ZALDA}
assuming the following uniform priors: $\Omega_Q\in(0.5,0.8)$,
$H_0\in(0.5,0.8)$, $w_Q^0\in(-0.1,-0.4)$, $w_Q^m\in(-1.0,0.0)$ and $a_c^m\in(0.01,1.0)$.
As extra priors we have assumed $\Delta=0.5$ (slowly varying equation of state) and the remaining 
parameters have been set as follows: $n=1$, $\Omega_b=0.05$ and $\tau=0$.
Since we are not considering the effect of systematics, but
only cosmic variance errors we evaluate a simple
likelihood defined as
\begin{equation}
L(\alpha_j)=\mathcal N \exp{\left[-\sum_i \frac{C_l^t(l_i;\alpha_j)-C_l^d(l_i)}{\sigma_i^2}\right]}, 
\end{equation}
where $\alpha_j$ are the likelihood parameters, $\mathcal N$ is a normalization constant and
$\sigma_i$ is the cosmic variance at $l=l_i$.
The superscripts $t$ and $d$ refer to the theoretical quantity and to the real data respectively.
\begin{figure}[t]
\centerline{
\includegraphics[scale=0.6]{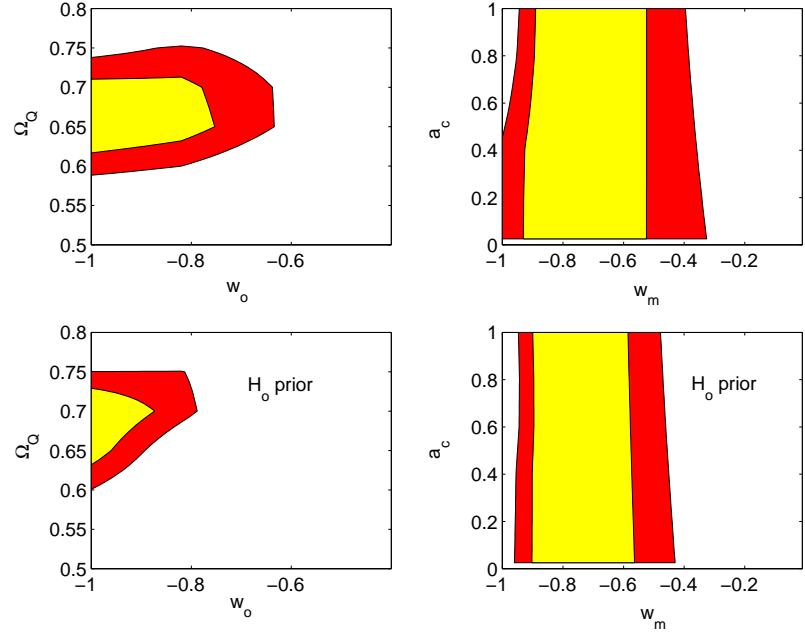}
}
\caption{Simulated 2-D likelihood contour plots for an ideal experiment,
with no priors ($top$ $panels$) and $H_o$ prior ($bottom$ $panels$). 
The yellow and red contours correspond to $1$ and $2\sigma$ respectively.
}
\label{f:fig30}
\end{figure}
The results are shown in figure~\ref{f:fig30}, where we plot the two dimensional likelihood
contours in the $w_Q^0-\Omega_Q$ (left panels) and $w_Q^m-a_c^m$ (right panels) planes respectively.
We find $\Omega_Q=0.68\pm_{0.08}^{0.05}$, $w_Q^0\leq-0.78$, whereas assuming a prior on $H_0$
improves the constraints on $w_Q^0$ to $w_Q^0\leq-0.85$ at $1\sigma$.
The likelihood plot
in the $w_Q^m-a_c^m$ plane shows that
$a_c^m$ is undetermined. This result is expected since for this class of
models the value of $a_c^m$ does not affect the evolution of the dark energy.
On the contrary we find $w_Q^m=-0.79\pm_{0.1}^{0.2}$, but the $\Lambda$ case cannot
be excluded at $2\sigma$. The constraint does not improve assuming the $H_o$ prior.
This is because $w_Q^o$ and $w_Q^m$
are degenerate, 
hence marginalizing the likelihood over $w_Q^o$ shifts
the best fit value of $w_Q^m$ towards more negative values.
However it is remarkable that there is still some sensitivity to the value of $w_Q^m$,
such that the best fit is not for $w_Q^m=-1$. Therefore
we can conclude that if $w_Q^0 \approx-1$, a large class of models will not be distinguished
from a $\Lambda$CDM scenario even with ideal CMB measurements. 
It is possible that by combining different cosmological data, as Sn Ia, large
scale structure and quasar clustering the degeneracy between $w_Q^m$ and $w_Q^0$ can be
broken and
more information on this class
of dark energy models can be inferred \cite{FRIEMAN02}.

\chapter{Alternative cosmological test with higher order statistics}
\label{chap7}

In the recent past the number of papers devoted to the analysis of high order
statistics of the CMB anisotropy has dramatically increased.
In fact the simplest inflationary models predict to first order 
a Gaussian distribution of 
temperature fluctuations, and deviations from gaussianity could be
the signature of other phenomena occurring in the anisotropy formation
process. On top of that different mechanisms can also be a source of non-Gaussian signals
at different angular scales. This is the case of non-linear
effects during the inflationary epoch
or the presence of topological defects. 
The present CMB data strongly constrains the level of non-gaussianity on the scales
so far probed by the experiments. In this Chapter we will briefly review the 
higher order statistics of the CMB anisotropies. We will focus on the use of the
angular bispectrum as an estimator of non-gaussianity, and
introduce a formalism that allows us
to analytically calculate the spectrum and bispectrum in the case of a random distribution
of localized anisotropies in the CMB sky. We will argue that applying this analytical
approach to the analysis of localized anisotropies such as the Sunyaev-Zeldovich
effect or radio point sources, it is possible to constrain the clustering properties
of these objects and determine the cosmological parameters.

\section{Higher order statistics}
Let us expand the temperature fluctuation field in the direction
$\hat{\gamma}$ of the sky into spherical harmonics:
\begin{equation}
\frac{\Delta T(\hat{\gamma})}{T}=\sum_{l,m}a_{lm} Y_l^m(\hat{\gamma}),
\end{equation}
the $a_{lm}$ are the multipole coefficients that contain all the statistical information
of the anisotropy field. For $m\neq0$ these are complex numbers satisfying the condition 
$a_{l-m}=(-1^m)a_{lm}^*$. 
The statistics of the CMB anisotropies depends on the
physical process responsible for the generation of the initial
density perturbations. Simple inflationary models predicts a Gaussian
spectrum of fluctuations \cite{GUTH,BARDEEN}, consequently we expect 
the anisotropy field to be Gaussian. 
In such a case the $a_{lm}$ are random Gaussian variables with Gaussian distributed 
amplitudes and with uniformly distributed phases. As consequence of this
the statistical distribution of the CMB anisotropies
is entirely specified by its second order moment ({\em i.e.} the power spectrum), 
\begin{equation}
C_l=\sum_{m=-l}^l |a_{lm}|^2.
\end{equation}
In this Gaussian case the higher odd moments of the distribution vanish, while the even
moments can be expressed in term of the variance $C_l$.
Therefore any deviation from gaussianity will inevitably produce non vanishing
high order statistics. Since there is a potentially infinite number of higher moments,
the non-Gaussian hypothesis cannot be disproved. Several methods have been proposed
in the literature to measure statistical estimators of the skewness (third moment)
and (kurtosis) from the analysis of CMB maps (for a list of these methods we refer to \cite{KGT}). 
A simple method is to measure the correlation between the temperature fluctuations
in different directions of the sky. In this case the higher order moments are estimated 
by the angular correlation functions. For instance the
$m-point$ angular correlation function is defined as:
\begin{equation}
C_m(\hat{\gamma}_1,\hat{\gamma}_2,...,\hat{\gamma}_m)\equiv \left \langle
\frac{\Delta T}{T}(\hat{\gamma}_1)\frac{\Delta T}{T}(\hat{\gamma}_2)...\frac{\Delta T}{T}(\hat{\gamma}_m)
\right \rangle,
\end{equation}
where $\hat{\gamma}_1,..,\hat{\gamma}_m$ are unitary vectors pointing at $1,..,m$
directions of the sky and the average $\langle..\rangle$ is taken over the whole sky.
In particular an estimate of the skewness is provided by the collapsed three-point
correlation functions $C_3(\alpha)$, that is a specific configuration of 
the three-point correlation function between
two points of the sky separated by an angle $\alpha$ and defined as:
\begin{equation}
C_3(\alpha)\equiv
\left \langle
\frac{\Delta T}{T}(\hat{\gamma}_1)\frac{\Delta T}{T}(\hat{\gamma}_1)
\frac{\Delta T}{T}(\hat{\gamma}_2) \right \rangle,
\end{equation}
where $\cos(\alpha)=\hat{\gamma}_1\cdot\hat{\gamma}_2$.
In terms of the multipoles it reads as:
\begin{equation}
C_3(\alpha)=\sum_{l_1,l_2,l_3}\sum_{m_1,m_2,m_3}P_{l_1}(cos\alpha)
a_{l_1 m_1}a_{l_2 m_2}a_{l_3 m_3}\mathcal{W}_{l_1}\mathcal{W}_{l_2}\mathcal{W}_{l_3}
\mathcal{H}_{l_1 l_2 l_3}^{m_1 m_2 m_3},
\label{c3a}
\end{equation}
where $P_{l_1}(\cos{\alpha})$ is the Legendre polynomial of degree $l_1$, $\mathcal{W}_l$ is
the experimental window function in the multipole space 
and 
\begin{equation}
\label{goodh}
\bar {\cal H}_{\ell_1\ell_2\ell_3}^{m_1 m_2 m_3} =
\sqrt{ \frac{(2\ell_1 +1)(2\ell_2 +1)(2\ell_3 +1)}{4\pi} }
\left(^{\ell_1~\ell_2~\ell_3}_{0~~0~~0}\right)
\left(^{\ell_1~\,\;\ell_2~\,\;\ell_3}_{m_1~m_2~m_3}\right) ~.
\end{equation}
are a combination of the Wigner-3J symbols.
The $C_3(\alpha)$ measured from the COBE-DMR sky maps has been found
to be
consistent with the Gaussian hypothesis within the cosmic variance errors \cite{KGT}. 
These results strongly limit the allowed
amount of non-gaussianity in the CMB anisotropies at large angular scales,
in particular they constrain the parameter space of a class of 
non-Gaussian models \cite{GANGUI2,GANGUI1,GANGUI,CORAS5}.
Similarly the collapsed three-point correlation
function inferred from the WMAP data is also consistent with Gaussian expectations \cite{GAZTA}.
Any non-gaussian analysis can be carried out in the multipole space as well, in this
case the equivalent of the three-point correlation function
is the angular bispectrum defined by:
\begin{eqnarray}
B_{l_{1}l_{2}l_{3}} =\sum_{m_{1},m_{2},m_{3}}
\left(^{\ell_1~\,\;\ell_2~\,\;\ell_3}_{m_1~m_2~m_3}\right)
a_{l_{1}}^{m_{1}}a_{l_{2}}^{m_{2}}a_{l_{3}}^{m_{3}}.
\label{bispe}
\end{eqnarray}
This estimator
is rotationally invariant \cite{MAG,FERMAG} and satisfies geometrical conditions such
that the only non vanishing $B_{l_1 l_2 l_3}$ are those with $|l_i-l_j|\leq l_k \leq l_i+l_j$
($i\neq j\neq k$) for all permutations of indices and $l_1+l_2+l_3=even$.
A normalized version of the angular bispectrum has been applied to
the analysis of COBE-DMR maps for the configuration $l_1=l_2=l_3$ \cite{FERGORSK,BANDAY,PHIL,KOM1}.
More recently the bispectrum has been estimated from the WMAP map and has been found
to be
consistent with the Gaussian hypothesis \cite{KOM2}. 

\section{Frequentist approach and estimation of higher moments}
The frequentist approach is the usual procedure adopted to test gaussianity. 
This implies the measured value of a higher order statistical estimator is compared
against the probability distribution function obtained from random Gaussian simulations
of the data sample. If there is a low probability that the measured value is 
consistent with the Gaussian simulation, then the Gaussian hypothesis is ruled out.
For instance this approach allowed the authors of \cite{FERGORSK}
to rule out the gaussianity of the
anisotropies at multipole $l=16$ in the COBE-DMR data. 
However it is worth mentioning that so far only the diagonal component of the bispectrum
has been measured and in order to be statistically significant the data analysis
should be extended to the estimation of the non-diagonal term of the bispectrum. 
If this is not a problem at low multipoles it could be a computationaly challenge
at higher orders. On the other hand it has been pointed out in a number of papers \cite{PHIL,CONMA}
that within the frequentist approach some non-Gaussian theories will be indistinguishable from
the Gaussian one. We will try to make this point more clear with a specific example
and we refer to the fundamental statistics textbook \cite{KEND} for a more detailed derivation
of the formula used in what follows.

Let us consider a sample of data $\{x_i\}$ ($i=1,..,N$)
generated from a random Gaussian process $f(x)$
such that each $x_i$ is an independent Gaussian random variable
with zero mean $\mu_1=0$ and the second order moment $\mu_2=\sigma^/2$,
\begin{equation}
f(x)=\frac{1}{\sqrt{2\pi}\sigma}e^{-\frac{x^2}{2\sigma^2}}.
\label{gau}
\end{equation}
In order to avoid confusion we will use Greek letters for the moments of the generating
function $f(x)$, $\mu_r$ its $r$-th moment and Roman letters for the moments obtained from the
statistics of the data sample.
In particular the $r$-th moment $m_r$ is given by:
\begin{equation}
m_r=\frac{1}{N}\sum_{i=1}^n(x_i-\langle x \rangle)^r,
\end{equation}
where $\langle x \rangle=m_1$ is the mean value of the sample. Making
$K$ realizations of this data sample, $\{x_i\}_j$ ($j=1,..,K$), we can infer
the distribution function $P_{gauss}(m_r)$ of the $r-th$ moment statistic $m_r$, which is
the frequency (number of times to the total number of realizations) of
a certain value $m_r$ appearing in the sample ${x_i}_j$. The expectation value
of this distribution $E(m_r)$ (the mean value of $m_r$ estimated from the K realizations)
will be indicative of the $r$-th moment $\mu_r$ of the generating 
Gaussian distribution function. The expectation
value $E(m_r)$ and the variance $var(m_r)$ can be related to the moments $\mu_r$ of the generating
function through approximate relations valid up to $(K \cdot N)^{-1/2}$ \cite{KEND}.
It can be shown that:
\begin{equation}
E(m_r)=\mu_r.
\label{expect}
\end{equation}
\begin{equation}
var(m_r)=\frac{1}{N\cdot K}(\mu_{2r}-\mu_r^2+r^2\mu_2\mu_{r-1}^2-2r\mu_{r-1}\mu_{r+1}).
\label{var}
\end{equation} 
For practical purposes we prefer to work in terms of cumulants $k_r$, that
are an equivalent set of numbers characterizing the generating function $f(x)$.
The cumulants $k_r$ and the moments $\mu_r$ are related through 
 Eq.~(3.33) in \cite{KEND}.
Let us consider the third order moment-statistic $m_3$, from
Eq.~(\ref{expect}) and Eq.~(\ref{var}) and using the relation between moments and cumulants we have:
\begin{equation}
E(m_3)=k_3,
\end{equation}
\begin{equation}
var(m_3)=\frac{1}{N\cdot K}(k_6+9k_2 k_4+6 k_2^3+9k_3^2).
\end{equation}
For the Gaussian generating function we have considered $k_2=\mu_2=\sigma^2/2$, $k_3=0$,
$k_4=0$ and $k_6=45\sigma^6$,
consequently we obtain:
\begin{equation}
E(m_3)=0,
\end{equation}
\begin{equation}
var(m_3)=\frac{183}{4N\cdot K}\sigma^6.
\end{equation}
On the other hand let us consider a sample of data generated from a non-Gaussian random
process. A general non-Gaussian random generating function can be constructed by using
an Edgeworth expansion around the Gaussian distribution defined by
\begin{equation}
g(x)=f(x)(1+\frac{1}{6}k_3 H_3(x)+\frac{1}{24}k_4 H_4(x)+...),
\label{edg}
\end{equation}
where $f(x)$ is given by Eq.~(\ref{gau}), the $H_i$ are the Hermite polynomials 
and the cumulants $k_r$ are free parameters.
It can be shown that Eq.~(\ref{edg}) gives a good approximation to any distribution function
provided all moments are defined and the higher order terms do not dominate over
the Gaussian one. 
We can limit the non-gaussianity to 
first order by imposing only the presence of a non vanishing skewness $k_3$ and 
kurtosis $k_4$. In this case Eq.~(\ref{expect}) and
Eq.~(\ref{var}) becomes:
\begin{equation}
E(m_3)=k_3,
\end{equation}
\begin{equation}
var(m_3)=\frac{1}{N\cdot K}(\frac{9}{2}k_4 \sigma^2+\frac{3}{4}\sigma^6+9k_3^2).
\label{varedg}
\end{equation}
Note from Eq.~(\ref{varedg}) the kurtosis and the skewness can sum up in such a way that
they cancel each other. In such a case the
frequency distributions of the third order moment-statistic $m_3$ inferred
from the $K$ realizations of the Gaussian and non-Gaussian process, 
$P_{gauss}(m_3)$ and $P_{n-gauss}(m_3)$ respectively, will have the same variance.
This makes difficult to establish whether the population of the data had been
generated from a Gaussian or a non-Gaussian process by estimating $m_3$.
\begin{figure}
\centerline{
\includegraphics[scale=0.8]{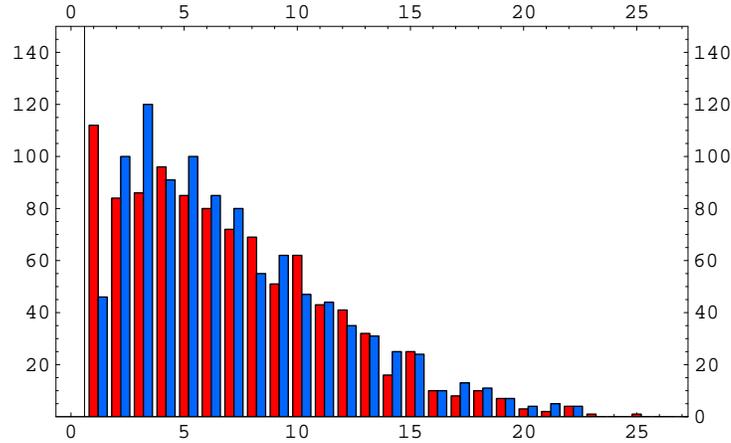}
}
\caption{Probability distribution functions of the third order moment $m_3$
estimated from Monte-Carlo simulation of randomly generated data for a Gaussian
(red bars) and a non-Gaussian process (blue bars).}	
\label{f:fig31}			
\end{figure}
This analysis is confirmed by the numerical simulations in which we simulated two
populations of data from two different random processes.
The Gaussian sample has been generated assuming the generating function Eq.~(\ref{gau})
with $k_1=0$, $k_2=1$, while the non-Gaussian population has been obtained assuming 
Eq.~(\ref{edg}) with $k_1=0$, $k_2=1$, $k_3=3$ and $k_4=-153/18$.
From these two populations we inferred the distribution functions of the third moment-statistic
$m_3$. The results are shown 
in figure~\ref{f:fig31}, where we plot
$P_{gauss}(m_3)$ ({\em red bars}) and $P_{n-gauss}(m_3)$ ({\em blue bars}).
Note that $P_{n-gauss}(m_3)$ is peaked at $m_3\neq0$, while $P_{gauss}(m_3)$
has a maximum at $m_3=0$. 
Even so the variances of the two distributions are the same.
This naive argument shows that for some non-Gaussian models
little information can be obtained using a frequentist
approach when analysing the third order statistics unless higher order moments 
of the CMB anisotropies fields are measured as well. Such simulations 
are necessary to take into account the systematic experimental sources of non-gaussianity.
However they make these tests computationally expensive. Therefore 
it is important to have
a theoretical prediction of the amplitude of non-Gaussianity
at least for known non-Gaussian anisotropies such as the secondary ones.

\section{Modelling localized non-Gaussian anisotropies}

Localized patterns of anisotropies produce a non-Gaussian signature in the CMB statistics. 
This is the case of secondary sources of anisotropies such as radio point sources or the
Sunyaev-Zeldovich (SZ) effect caused by the interaction of the CMB photons with
the hot gas associated with clusters of galaxies \cite{SUN}.
Active sources of non-Gaussianity include topological
defects such as cosmic strings, domain walls (see for a general review \cite{SHEL}) or
primordial bubble relics of a first order phase transition during the inflationary era
\cite{OCC,BAM}. For these classes of models analytical formula of the higher order
correlation functions have been calculated in a number of papers
\cite{MAGUEJO1,GANGUI,CORAS5}. As far as the non-Gaussianity arising from
secondary source of anisotropies is concerned, an analytical estimate of the three point correlation
function has been obtained in the case of the Rees-Sciama effect \cite{MOL},
while the value of the bispectrum produced
by the S-Z effect and extragalactic radio sources has been obtained in \cite{GOLDB}, whilst
the case of the Vishniac effect has been considered in \cite{CASTRO}.
In what follows we determine the spectrum and the bispectrum for the case of a distribution
of localized anisotropies in the CMB sky, using a formalism introduced in \cite{MAGUEJO1}.

\subsubsection*{Power Spectrum}

Let us consider a nearly circular spot on the sky and perform the spherical harmonic
decomposition in the frame where the $z$-axis coincides with the center of the spot.
The brightness of the temperature fluctuation of the spot is $b$, and $f(\theta)$
is
its angular profile. In this frame, we can write the temperature fluctuation as
$\Delta_s(\theta)=b\cdot f(\theta)$, expanding it in spherical harmonics we find \cite{MAGUEJO1}:
\begin{equation}
\tilde{a}_{lm}=2\pi b \sqrt{\frac{2l+1}{4\pi}} \mathcal{I}_l\cdot \delta_{m0}+\epsilon_{ml},
\end{equation}
with 
\begin{equation}
\mathcal{I}_l=\int_{-1}^1 d(\cos{\theta}) f(\theta) P_l(cos{\theta}),
\label{I}
\end{equation}
where $P_l(cos{\theta})$ is the Legendre polynomial
and $\epsilon_{lm}$ is a perturbation induced by the irregularity of the spot. As a first
approximation it can be neglected, as the CMB measurements do not have the enough
resolution to resolve its $\phi$-angular structure. For a distribution of $N$ spots,
performing a rotation to a general frame where the $n$-th spot is at 
the angle $\hat{\gamma}_n=(\theta_n,\phi_n)$ and summing over
all the spots we obtain:
\begin{equation}
a_{lm}=\sum_n b_n \mathcal{I}_n^l Y_l^{m*} (\hat{\gamma_n}),
\label{almn}
\end{equation}
where $b_n$ is the brightness of the $n$-th spot and 
\begin{equation}
\mathcal{I}_n^l=2\pi \int d(\cos{\theta}) f_n(\theta) P_l(cos{\theta})
\label{il}
\end{equation}
is the Legendre transform of its temperature profile.
We can now calculate the power spectrum,
\begin{eqnarray}
C_l & = &\frac{1}{2l+1}\sum_m|a_{lm}|^2 \nonumber \\
& = &\sum_n \sum_{n'} b_n b_{n'} \mathcal{I}_n^l \mathcal{I}_{n'}^l C_{n n'} \cdot 
\sum_m \frac{Y_l^{m*}(\hat{\gamma_{n}}) Y_l^{m*}(\hat{\gamma_{n'}})}{2l+1} \nonumber \\
& = &\sum_n \sum_{n'} b_n b_{n'} \mathcal{I}_n^l \mathcal{I}_{n'}^l C_{n n'}
\cdot \frac{P_l(\hat{\gamma_{n}}\hat{\gamma_{n'}})}{4\pi},
\end{eqnarray}
where we have used the completeness relation of the spherical harmonics and
$C_{n n'}=\delta_{n n'}+w_{n n'}$, with $w_{n n'}$
taking into account the possibility that the spots are uncorrelated ($\delta_{n n'}$),
as for a Poisson distribution, or that are correlated two by two ($w_{n n'}$). 
Therefore by expanding the sums we finally obtain:
\begin{equation}
C_l=\frac{1}{4\pi} \left [\sum_n b_n^2 (\mathcal{I}_n)^2+\frac{1}{2} \sum_{n\not=n'} b_n b_{n'}
w_{n n'}  \mathcal{I}_n^l  \mathcal{I}_{n'}^l P_l(\hat{\gamma_{n}}\hat{\gamma_{n'}}) \right].
\label{clnn}
\end{equation}
We assume now that the spots have about the same brightness and the same temperature profile:
$b=\langle b_n\rangle$ and $f=\langle f_n \rangle$
 where the average is taken over the ensemble of spots. 
Because the number of spots in a circular ring centered on a single spot
is proportional to the angular extension of the ring  
we can substitute the sums over the number of spots
with an integral over the whole sky. Hence we can substitute the discrete
correlation coefficient $w_{n n'}$ with the two-point angular correlation of the spots, $w(\alpha)$.
If the spots correspond to radio sources or clusters of galaxies $w(\alpha)$ would be
the angular correlation function estimated from large scale structure observations.
The mean value over the spot distribution of Eq.~(\ref{clnn}) becomes:
\begin{equation}
\langle C_l \rangle= N b^2 \left[\mathcal{I}_l^2+ \mathcal{I}_l \cdot \mathcal{G}_l \right],
\label{pwspec}
\end{equation}
where $N$ is a normalization constant and
\begin{equation}
\mathcal{G}_l=\int_{-1}^1 d(\cos{\alpha}) \mathcal{I}_l(\alpha)w(\alpha)P_l(\cos{\alpha}),
\label{gl}
\end{equation}
with
\begin{equation}
\mathcal{I}_l(\alpha)=\int_{-1}^1 d(\cos{\theta}) f(\theta+\alpha)P_l(\cos{\theta}).
\label{ilalpha}
\end{equation}
From Eq.~(\ref{pwspec}) we may note that the contribution to the anisotropy power spectrum from
a localized distribution of spots increases with the brightness $b$. If the spots are uncorrelated
the second term $\mathcal{I}_l \cdot \mathcal{G}_l$ drops from the equation and the only 
contribution is due to the projection of the spot's signal in the multipole space $\mathcal{I}_l$.
In particular the largest contribution will occur in the range of multipoles that correspond to the
effective angular size of the spots. If the spots are correlated, the overall contribution
to the power spectrum will depend on the sign of $\mathcal{I}_l \cdot \mathcal{G}_l$.

\subsubsection*{Bispectrum}
Following the same procedure as just described we calculate the angular bispectrum.
 Substituting 
Eq.~(\ref{almn}) in Eq.~(\ref{bispe}) we have:
\begin{eqnarray}
B_{l_{1}l_{2}l_{3}} & = & \sum_{m_{1},m_{2},m_{3}}
\left(^{\ell_1~\,\;\ell_2~\,\;\ell_3}_{m_1~m_2~m_3}\right)
a_{l_{1}}^{m_{1}}a_{l_{2}}^{m_{2}}a_{l_{3}}^{m_{3}} \nonumber \\
& = & \sum_{n_{1},n_{2},n_{3}}
b_{n1}b_{n_2}b_{n_3} \mathcal{I}_{n_1}^{l_1} \mathcal{I}_{n_2}^{l_2}\mathcal{I}_{n_3}^{l_3}
C_{n_1 n_2 n_3} \nonumber \\
& & \times \sum_{m_{1},m_{2},m_{3}} \left(^{\ell_1~\,\;\ell_2~\,\;\ell_3}_{m_1~m_2~m_3}\right)
Y_{l_1}^{{m_1}*} (\hat{\gamma}_{n_1})Y_{l_2}^{{m_2}*} (\hat{\gamma}_{n_2})
Y_{l_3}^{{m_3}*} (\hat{\gamma}_{n_3}),
\label{b1}
\end{eqnarray}
where $C_{n_1 n_2 n_3}$ takes into account all the possible correlations between the spots
up to third order,
\begin{equation}
C_{n_1 n_2 n_3}=\delta_{n_1 n_2}\delta_{n_2 n_3}+w_{n_1 n_2}\delta_{n_3 n_3}
+w_{n_1 n_3}\delta_{n_2 n_2}+w_{n_2 n_3}\delta_{n_1 n_1}+w_{n_1 n_2 n_3},
\end{equation}
with $w_{n_1 n_2 n_3}$ the correlation between three different spots.
Using the definition of the Wigner 3J symbols in terms of an angular integral of 
three spherical harmonics and the completeness relation of the
spherical harmonics Eq.~(\ref{b1}) becomes:
\begin{eqnarray}
B_{l_{1}l_{2}l_{3}} = &\mathcal{M}_{l_1 l_2 l_3} \left[ \sum_{n_1}b_{n_1}^3\mathcal{I}_{n_1}^{l_1}\mathcal{I}_{n_1}^{l_2}
\mathcal{I}_{n_1}^{l_3}\cdot \mathcal{R}^{l_1 l_2 l_3}_{\hat{\gamma}_{n_1}}
+\frac{3}{2}\sum_{n_1\not=n_2} b_{n_1}^2 b_{n_2}\mathcal{I}_{n_1}^{l_1}\mathcal{I}_{n_1}^{l_2}
 \mathcal{I}_{n_2}^{l_3}w_{n_1 n_2}\cdot
\mathcal{R}^{l_1 l_2 l_3}_{\hat{\gamma}_{n_1}\hat{\gamma}_{n_2}}\right.  \nonumber \\
 & \left. +\frac{1}{3}\sum_{n_1\not=n_2\not=n_3}\mathcal{I}_{n_1}^{l_1}\mathcal{I}_{n_3}^{l_2}
 \mathcal{I}_{n_3}^{l_3}w_{n_1 n_2 n_3}\cdot
\mathcal{R}^{l_1 l_2 l_3}_{\hat{\gamma}_{n_1}\hat{\gamma}_{n_2}\hat{\gamma}_{n_3}} \right],
\end{eqnarray}
where
\begin{eqnarray}
\mathcal{M}_{l_1 l_2 l_3}=
\frac{(4\pi)^{3/2}}{\sqrt{(2l_1+1)(2l_2+1)(2l_3+1)}}\left(^{\ell_1~\ell_2~\ell_3}_{0~~0~~0}\right)^{-1},\\
\mathcal{R}^{l_1 l_2 l_3}_{\hat{\gamma}_{n_1}}=\int d\Omega_{\gamma}
P_{l_1}(\hat{\gamma}\cdot\hat{\gamma}_{n_1})P_{l_2}(\hat{\gamma}\cdot\hat{\gamma}_{n_1})
P_{l_3}(\hat{\gamma}\cdot\hat{\gamma}_{n_1}),\label{r1} \\
\mathcal{R}^{l_1 l_2 l_3}_{\hat{\gamma}_{n_1}\hat{\gamma}_{n_2}}=\int d\Omega_{\gamma}
P_{l_1}(\hat{\gamma}\cdot\hat{\gamma_{n_1}})P_{l_2}(\hat{\gamma}\cdot\hat{\gamma}_{n_1})
P_{l_3}(\hat{\gamma}\cdot\hat{\gamma}_{n_2}),\label{r2} \\
\mathcal{R}^{l_1 l_2 l_3}_{\hat{\gamma}_{n_1}\hat{\gamma}_{n_2}\hat{\gamma}_{n_3}}=\int d\Omega_{\gamma}
P_{l_1}(\hat{\gamma}\cdot\hat{\gamma_{n_1}})P_{l_2}(\hat{\gamma}\cdot\hat{\gamma_{n_2}})
P_{l_3}(\hat{\gamma}\cdot\hat{\gamma_{n_3}}). \label{r3}
\end{eqnarray}
Due to the background isotropy of the space,
 we are free to choose $\hat{\gamma}=\hat{\gamma}_{n_1}$. As consequence
the integrals Eq.~(\ref{r1}-\ref{r3}) become:
\begin{eqnarray}
\mathcal{R}^{l_1 l_2 l_3}_{\hat{\gamma}_{n_1}}&=&4\pi, \\
\mathcal{R}^{l_1 l_2 l_3}_{\hat{\gamma}_{n_1}\hat{\gamma}_{n_2}}&=&\int d\Omega_{\gamma_{n_1}}
P_{l_3}(\hat{\gamma}\cdot\hat{\gamma_{n_2}}), \\
\mathcal{R}^{l_1 l_2 l_3}_{\hat{\gamma}_{n_1}\hat{\gamma}_{n_2}\hat{\gamma}_{n_3}}&=&\int d\Omega_{\gamma_{n_1}}
P_{l_2}(\hat{\gamma}\cdot\hat{\gamma}_{n_2})P_{l_3}(\hat{\gamma}\cdot\hat{\gamma}_{n_3}).
\end{eqnarray}
As for the power spectrum, we can average over the ensemble of all the spots and substitute the
sums over the spots with integrals over the whole sky. In this case the discrete correlation
coefficient $w_{n_i n_j}$ and $w_{n_i n_j n_k}$ are replaced with the corresponding angular
correlation function $w(\alpha)$ and $w(\alpha,\beta)$. 
After tedious calculations we obtain:
\begin{eqnarray}
\langle B_{l_{1}l_{2}l_{3}}\rangle =4\pi^2\mathcal{M}_{l_1 l_2 l_3} b^3 \left[\mathcal{B}^{(0)}+\frac{3}{2}\mathcal{B}^{(1)}
+\frac{1}{3}\mathcal{B}^{(3)}\right ],
\label{bispespot}
\end{eqnarray}
where
\begin{equation}
\mathcal{B}^{(0)}=\mathcal{I}_{l_1} \mathcal{I}_{l_2} \mathcal{I}_{l_3},
\end{equation}
\begin{equation}
\mathcal{B}^{(1)}=\mathcal{I}_{l_1} \mathcal{I}_{l_2} \mathcal{G}_{l_3},
\end{equation}
\begin{equation}
\mathcal{B}^{(2)}=\mathcal{I}_{l_1}\int_{-1}^1 d(\cos{\alpha}) \int_{-1}^1 d(\cos{\beta})
\mathcal{I}_{l_2}(\alpha)\mathcal{I}_{l_3}(\beta)w(\alpha,\beta)P_{l_2}(\cos{\alpha})P_{l_3}(\cos{\beta}),
\end{equation}
with $\mathcal{I}_{l}$ and $\mathcal{G}_{l_3}$ defined by Eq.~(\ref{il}) and Eq.~(\ref{gl}) 
respectively and $\mathcal{I}_l(\alpha)$ defined by Eq.~(\ref{ilalpha}).
As we can see from Eq.~(\ref{bispespot}) for a Poisson distribution $\mathcal{B}^{(1)}$
and $\mathcal{B}^{(2)}$ vanish, but the bispectrum remains non-vanishing due to the term 
$\mathcal{B}^{(0)}$ that account for the localized structure of the anisotropies.

\section{Discussion}
In the previous section we presented a general formalism to calculate the contribution
to the power spectrum and the bispectrum
of a distribution of spots in the CMB sky.
The formulae Eq.~(\ref{pwspec}) and Eq.~(\ref{bispespot}) have to be considered as a starting
point for further investigation.  
They take into account several effects, for instance the brightness
and the angular size of the signals contribute at zeroth order to the spectrum and 
 the bispectrum. On the other hand the presence of internal correlations 
in the spot distribution, which are described by the angular
correlation functions $w(\alpha)$ and $w(\alpha,\beta)$, contribute
 as first and second order effects respectively. As a specific application, 
Eq.~(\ref{pwspec}) and Eq.~(\ref{bispespot}) can be computed in the case of a distribution 
of spots caused by the SZ effect of a cluster of galaxies. Approximating the shape of this
signal with a Gaussian profile characterized by a given width, it will be possible to
numerically compute the integrals Eq.~(\ref{il}). Moreover without loss of generality
the angular correlation function $w(\alpha)$ can be assumed to be a power law. In such a case 
the integrals Eq.~(\ref{gl}) can also be numerically computed. The resulting power spectrum can be
compared with the prediction of numerical simulations.
This will allow us to test how crucial is the assumption that $b=\langle b_n\rangle$ and 
$f=\langle f_n \rangle$.
The next generation of CMB measurements
will measure the bispectrum at very high multipoles, therefore using
Eq.~(\ref{bispespot}) in a specific case such 
as the SZ effect offers an alternative way of inferring cosmological information.
\chapter*{Conclusion and prospects}
\label{conclu}

In this thesis we have discussed various aspects of dark energy dominated
cosmologies. In Chapter~1 we have reviewed the observational evidence
of the dark energy. We have seen that different cosmological measurements
are consistent only if the dark energy accounts for most of the matter content
of the Universe. In Chapter~2 we have discussed some of the proposed dark
energy candidates and we have focused on the quintessence scenario. 
In spite of the theoretical difficulties of a viable quintessence model
building, this scenario has a number of interesting features
that can be tested with cosmological measurements. In particular
in Chapter~3 we reviewed the dynamics of scalar field perturbations for
two different class of minimally coupled quintessence models.
We have learnt that quintessence perturbations have no active role during
the structure formation. However their presence can lead to time integrated
effects in the evolution of the gravitational potential. Therefore
they can leave a characteristic imprint in the Cosmic Microwave Background
anisotropy power spectrum through the Integrated Sachs-Wolfe effect.
In Chapter~4 we have constrained with Sn Ia data and the position of the CMB
peaks a parameterized quintessence potential that accounts for a large
class of quintessence models. Using the properties of the tracker regime
we have been able to put upper limits on the present value of the quintessence
equation of state. We have found that by the present time the scalar field
is evolving in flat region or close to a minimum of its potential.
However the results of this analysis clearly indicate that 
the possibility the quintessence equation 
of state was largely different from its present
value cannot be excluded. In this direction a lot of effort has been made
to detect time variation of the quintessence with cosmological distance 
measurements. In Chapter 5 we reviewed some of the proposed methods and we 
pointed out a number of potential problems. In particular the use of a constant
equation of state to parameterize the dark energy leads to misleading conclusions.
Using a very general argument we showed that if the dark energy is time varying
the constraints on a constant equation of state will be pushed towards large
negative values. Therefore all the results obtained using this approach have
to be carefully interpreted. In particular the fact that several
data analysis found the equation of state best fit value to be $w\lesssim-1$
can be just a bias effect. On the contrary we have proposed a time parameterization
of the dark energy equation of state in terms of physical parameters.
This accounts for most of the proposed dark energy models and moreover is
valid at all the redshifts. Hence this approach allows us
to take into account in a model independent way not only the effects
dark energy has had on the expansion rate of the Universe but also on the 
structure formation. In Chapter~5 we have applied this parameterization
to study the dark energy effects in the CMB power spectrum. As conjectured
in Chapter~3 we have found that the dark energy leaves a characteristic imprint
through the ISW effect. The amplitude of this signature selects
the class of models which are distinguishable from the cosmological constant
scenario. In particular we have shown that using ideal CMB data, only models
characterized by a rapid transition of the equation of state can be distinguished
from the $\Lambda$ case. In Chapter~7 we have introduced an alternative
cosmological test using higher order statistics of the CMB anisotropies.
We have computed the spectrum and the bispectrum for a distribution of localized
non-Gaussian anisotropies. These can be applied to specific cases such as 
the SZ effect to constrain cosmological parameters through the non-gaussianity
produced by the imprint of cluster of galaxies.
We can find a number of directions where the work so far reviewed can be further
extended. Under some general assumptions it would be
interesting to test the formulae developed in Chapter~7 with
the predictions of the SZ effect from numerical simulations of cluster galaxies
\cite{ANTONIO}. A complete likelihood analysis of the parameterized
dark energy equation of state is currently in progress. We make use of the
full cosmological data so far available. We intend to extend this analysis to
the quasar clustering, that
is a good candidate for testing the dark energy \cite{WAGA}.
An interesting issue arises from the recent WMAP data. In fact it has
been found that the quadrupole and octupole are suppressed in contrast with
the prediction of $\Lambda$CDM cosmologies \cite{Spergel}. The possibility that this is caused
by a cancellation mechanism between the SW and ISW effects due to clustering
properties of the dark energy needs further investigation \cite{Contaldi}.
At moment there are no final conclusions about the nature of the dark energy,
luckily the upcoming and future cosmological data will provide
a new insight of the dark energy phenomenology.
It has to be hoped that such measurements will help us to formulate
the new paradigm of Cosmology that will allows us to correctly address
the dark energy problem. The history of science shows that no scientific
activity is possible in subjects where no paradigms have been found.
Therefore we should ask ourself what direction will the cosmological
investigation take if the dark energy problem remains unsolved.
\bibliographystyle{myhep}
\bibliography{thesis}


\end{document}